\documentclass[a4paper,11pt]{article}

\usepackage{jcappub}
\usepackage[utf8]{inputenc}
\usepackage{bigfoot}
\usepackage{bm}
\usepackage{lmodern}
\usepackage[usenames,dvipsnames,svgnames,table]{xcolor}
\usepackage{orcidlink}
\usepackage{enumitem}
\usepackage{mathrsfs}
\usepackage{graphicx} 
\usepackage{ctable}
\usepackage{array}
\usepackage{dcolumn}
\usepackage{placeins}
\usepackage{enumitem}
\usepackage{tensor}
\usepackage[export]{adjustbox}
\usepackage{tcolorbox}

\tcbuselibrary{skins}
\tcbuselibrary{breakable}

\newenvironment{grouppanel}{\begin{tcolorbox}[enhanced,colback=gray!10,frame hidden]}{\end{tcolorbox}}

\newcommand{\fNL}{f_\text{NL}}
\newcommand{\fNLeq}{f_\text{NL}^\text{eq}}
\newcommand{\fNLfold}{f_\text{NL}^\text{fold}}
\newcommand{\fNLsq}{f_\text{NL}^\text{sq}}

\newcommand{\ShiftEqFold}{\Delta(\text{eq}\rightarrow\text{fold})}
\newcommand{\ShiftSqEq}{\Delta(\text{sq}\rightarrow\text{eq})}

\newcommand{\Clm}{\ensuremath{C_{LM}}}

\newcommand{\ellI}{\ensuremath{\ell_1}}
\newcommand{\ellII}{\ensuremath{\ell_2}}

\newcommand{\J}[1]{\ensuremath{J_{\ell_{#1}, m_{#1}, R} (\theta_{#1}) }}
\newcommand{\JBare}{{J_{\ell, m, R}(\theta)}}

\newcommand{\Jup}[1]{\ensuremath{J^\Upsilon_{\ell_{#1}, m_{#1}, R} (\theta_{#1}) }}
\newcommand{\JupBare}{\ensuremath{J^\Upsilon_{\ell, m, R}}}
\newcommand{\Jom}[1]{\ensuremath{J^\Omega_{\ell_{#1}, m_{#1}, R} (\theta_{#1}) }}
\newcommand{\JomBare}{\ensuremath{J^\Omega_{\ell, m, R}}}

\newcommand{\Eharmonic}{\mathcal{E}}
\newcommand{\GreenFunction}{\mathfrak{Gr}}
\newcommand{\BetaFunction}{\mathrm{B}}

\newcommand{\suml}{\mathscr{L}}
\newcommand{\summ}{\mathscr{M}}
\newcommand{\sumR}{\mathscr{R}}
\newcommand{\sumL}{\mathbb{L}}
\newcommand{\sumM}{\mathbb{M}}
\newcommand{\lik}{\mathcal{L}}

\newcommand{\As}{\ensuremath{A_s}}
\newcommand{\ns}{\ensuremath{n_s}}
\newcommand{\At}{\ensuremath{A_t}}
\newcommand{\nt}{\ensuremath{n_t}}
\newcommand{\dimP}{\mathcal{P}}

\newcommand{\vect}[1]{\bm{\mathrm{{#1}}}}
\newcommand{\Mp}{M_{\mathrm{P}}}
\renewcommand{\d}{\mathrm{d}}
\newcommand{\e}[1]{\mathrm{e}^{{#1}}}
\newcommand{\grad}{\nabla}
\newcommand{\im}{\mathrm{i}}

\newcommand{\rep}[1]{{(#1)}}

\newcommand{\wavefn}{w}
\newcommand{\pidx}[1]{\mathbb{{#1}}}
\newcommand{\pcoord}[1]{\mathcal{{#1}}}

\newcommand{\Hend}{H_{\text{end}}}

\newcommand{\Ntot}{N_{\text{tot}}}
\newcommand{\Nsubh}{N_{\text{subh}}}

\newcommand{\brane}[1]{\mathrm{D}{{#1}}}
\newcommand{\antibrane}[1]{\overline{\mathrm{D}{{#1}}}}

\newcommand{\rIR}{r_{\mathrm{IR}}}
\newcommand{\rUV}{r_{\mathrm{UV}}}
\newcommand{\xIR}{x_{\mathrm{IR}}}
\newcommand{\phiUV}{\phi_{\mathrm{UV}}}

\newcommand{\PhiPI}{\Phi_{\mathscr{F}}}
\newcommand{\PhiCF}{\Phi_{\mathscr{H}}}

\newcommand{\VCoulomb}{V_{\mathscr{C}}}
\newcommand{\Vmass}{V_{\mathscr{M}}}
\newcommand{\VPI}{V_{\mathscr{F}}}
\newcommand{\VCF}{V_{\mathscr{H}}}

\newcommand{\gstring}{g_s}

\renewcommand{\geq}{\geqslant}
\renewcommand{\leq}{\leqslant}

\newcommand{\Mpc}{\text{Mpc}}
\newcommand{\GeV}{\text{GeV}}
\newcommand{\seconds}{\text{s}}
\newcommand{\hours}{\text{hr}}

\newcommand{\kpiv}{k_\star}
\newcommand{\Npiv}{N_\star}
\newcommand{\fNLpiv}{\fNL^\star}
\newcommand{\DeltaMax}{\Delta_{\text{max}}}

\newcommand{\numparams}{$1,212$}
\newcommand{\numterms}{$3,881$}
\newcommand{\rawcataloguesize}{$450,000,000$}
\newcommand{\precataloguesize}{$90,000$}
\newcommand{\cataloguesize}{$55,000$}
\newcommand{\precatalogueexact}{$90,039$}

\newcommand{\minicataloguesize}{$18,000$}
\newcommand{\compareprecatalogueexact}{$22,453$}

\newcommand{\squeezedcataloguesize}{$3,000$}

\newcommand{\TrajNofit}{${\#}88,167$}
\newcommand{\TrajBestfit}{${\#}43,942$}

\newcommand{\Ascut}{10^{-4}}
\newcommand{\fNLeqcut}{3 \times 10^{-2}}
\newcommand{\WMAPsigma}{3\sigma}

\newcommand{\nshape}{n_{\text{shape}}}

\usepackage{relsize}
\newcommand\Cpp{C\nolinebreak\hspace{-.05em}\raisebox{.4ex}{\relsize{-3}{\textbf{+}}}\nolinebreak\hspace{-.10em}\raisebox{.4ex}{\relsize{-3}{\textbf{+}}}}

\newcommand{\semibold}[1]{{\fontseries{b}\selectfont{#1}}}
\newcommand{\sansbold}[1]{{\sffamily\fontseries{sbc}\selectfont{#1}}}

\newcommand{\para}[1]{\par\vspace{2mm}\noindent\semibold{{#1.}---}\ignorespaces} 
\newcommand{\stephead}[2]{\vspace{3mm}\par\noindent\colorbox{gray!20}{\begin{minipage}{\linewidth}\textsc{Step {#1} --- {#2}}\end{minipage}}\vspace{1mm}\par\noindent\ignorespaces}
\newcommand{\stepsidehead}[1]{\par\vspace{1mm}\noindent\textsf{{#1}}:}

\newcommand{\CppTransport}{\sansbold{CppTransport}}

\newcommand{\PyTransport}{\sansbold{PyTransport}}
\newcommand{\CLASS}{\sansbold{CLASS}}
\newcommand{\Cosmosis}{\sansbold{CosmoSIS}}
\newcommand{\Cobaya}{\sansbold{Cobaya}}

\newcommand{\GiNaC}{\sansbold{GiNaC}}
\newcommand{\SymPy}{\sansbold{SymPy}}
\newcommand{\Mathematica}{\sansbold{Mathematica}}
\newcommand{\apriori}{\sansbold{apriori}}
\newcommand{\branch}[1]{\texttt{{#1}}}
\newcommand{\commit}[1]{\texttt{{#1}}}

\DeclareMathOperator{\RePart}{Re}
\DeclareMathOperator{\ImPart}{Im}

\renewcommand{\Re}{\RePart}
\renewcommand{\Im}{\ImPart}

\expandafter\let\csname c@tblerows\endcsname\rownum

\definecolor{SussexCobaltBlue}{HTML}{1d4289}
\definecolor{SussexDeepAquamarine}{HTML}{007a78}
\definecolor{SussexPowderBlue}{HTML}{7da1c4}
\definecolor{SussexCornYellow}{HTML}{f2c75c}
\definecolor{SussexChinaRose}{HTML}{be84a3}
\definecolor{SussexBurntOrange}{HTML}{dc582a}

\hypersetup{colorlinks=true,citecolor=SussexDeepAquamarine,linkcolor=SussexCobaltBlue,urlcolor=SussexBurntOrange}

\newcommand{\StandardTable}{\small
    	\heavyrulewidth=.08em
    	\lightrulewidth=.05em
    	\cmidrulewidth=.03em
    	\belowrulesep=.65ex
    	\belowbottomsep=0pt
    	\aboverulesep=.4ex
    	\abovetopsep=0pt
    	\cmidrulesep=\doublerulesep
    	\cmidrulekern=.5em
    	\defaultaddspace=.5em
}

\setupctable{mincapwidth=\textwidth,doinside={\StandardTable\renewcommand{\arraystretch}{1.5}\rowcolors{2}{SussexPowderBlue!20}{white}},footerwidth,captionskip=3ex}
\newcolumntype{C}{>{$}c<{$}}
\newcolumntype{L}{>{$}l<{$}}
\newcolumntype{R}{>{$}r<{$}}
\newcolumntype{S}{D{.}{.}{-1}}
\newcolumntype{Y}{>{\raggedleft\arraybackslash}X}

\DeclareMathOperator{\Or}{O}
\DeclareMathOperator{\tr}{tr}
\DeclareMathOperator{\Prob}{\bm{\mathrm{P}}}

\newcommand{\glossarypage}[1]{p.~\pageref{#1}}

\title{Non-Gaussianity in D3-brane inflation}

\author[1]{Kareem Marzouk,\,$^{\text{\orcidlink{0000-0003-2060-8956}}}$}
\author[1,2]{Alessandro Maraio,\,$^{\text{\orcidlink{0000-0003-2383-3462}}}$%
\note{The first two authors made equal contributions to the work reported in this paper.}}
\author[1]{David Seery\,$^{\text{\orcidlink{0000-0003-3421-6080}}}$}

\affiliation[1]{Astronomy Centre, University of Sussex, Falmer, Brighton, BN1 9QH, UK}
\affiliation[2]{Institute for Astronomy, Royal Observatory, Blackford Hill, Edinburgh, EH9 3HJ, UK}

\emailAdd{B.K.Marzouk@sussex.ac.uk}
\emailAdd{Maraio@roe.ac.uk}
\emailAdd{D.Seery@sussex.ac.uk}

\abstract{We update predictions for observables
in the `delicate' $\brane{3}$/$\antibrane{3}$
inflationary model on the conifold.
We use a full CMB likelihood calculation to assess
goodness-of-fit, which is necessary because
in this model the
$\zeta$
power spectrum often cannot be approximated as a power-law over
observable scales.
For the first time we are able to provide
accurate forecasts for
the amplitude of three-point
correlations.
In a significant portion of its parameter
space the model follows Maldacena's single-field
prediction $\fNL \approx -(5/12)(\ns-1)$
if
$|\nt| \ll 1$.
Therefore
$|\fNL|$ is usually small when
the power spectrum satisfies observational
constraints.
In a small number of cases
the bispectrum is instead dominated by
effects from rapid switching between angular minima.
The resulting amplitudes are larger,
but mostly with unacceptable spectral behaviour.
In the most extreme case
we obtain $|\fNLeq| \sim 75$
at $k_t/3 = 0.002 \, \Mpc^{-1}$.
It has been suggested that the quasi-single
field inflation (`QSFI') mechanism could
produce significant 3-point correlations in this model.
We do observe rare shifts in amplitude between
equilateral and squeezed configurations that could
possibly be associated with QSFI effects, but
more investigation is needed to establish the full
bispectrum shape.
There is evidence of `shape' running between equilateral
and squeezed configurations that may
be inherited
from the scale dependence of the spectrum.
We explore the dependence of observables on discrete choices
such as the truncation point of the potential.
Our analysis illustrates the advantages of a standard
format for information exchange within the
inflationary model-building and testing community.}

\begin{document}

\maketitle

\newpage
\listoffigures

\newpage
\section{Introduction}
\label{sec:introduction}
Inflation continues to retain its favoured
position as the leading scenario for the
origin of structure in the universe---%
but
there has been little progress towards
identifying the degrees of freedom that were active during the
inflationary era, or the manner in which they interacted.
Among the major reasons for this slow progress are
well-rehearsed arguments showing that inflation is sensitive to
small, nonrenormalizable interactions suppressed by Planck-scale
masses, and therefore may depend on the precise way in which
the inflationary sector is embedded within its ultraviolet
completion.

This phenomenon is not a failure of decoupling in the
technical sense~\cite{Appelquist:1974tg}, but shares its
double-edged character.
On the one hand,
if successful inflation can depend on physics
at or near the Planck scale, we are encouraged to believe
that it may be possible to discover details of quantum
gravity by studying inflationary observables.
On the other, dependence on high-scale physics means that
inflation can \emph{not} be studied on its own:
assumptions about physics at higher energies
are required, even if they
are not made explicit.
The predictivity of the scenario is therefore reduced.

This situation has encouraged development
of approaches
in which inflationary model-building
takes place
in the context of a concrete proposal for
its ultraviolet completion. The most well-developed
of these use string theory as the ultraviolet
model, paired with a variety of suggestions for the microscopic
origin of the low-energy fields that populate the inflationary
sector. The field was surveyed at length in a recent monograph
by Baumann \& McAllister~\cite{Baumann:2014nda}.

One proposal is that inflation is driven
by the dynamics of a $\brane{3}$/$\antibrane{3}$ brane
pair within a warped deformed conifold.
The attraction of this scenario is not that we think it
more realistic than any other model,
but that it is highly computable.
In particular, the functional form of the
low-energy effective action
can be computed reliably, even accounting for contributions
from moduli stabilization and supersymmetry breaking.
This high degree of computability is remarkable.

The most significant drawback is the complexity of the
resulting effective theory:
the potential we describe in~{\S}\ref{sec:construct-potential}
below has {\numparams} independent parameters and is a sum
of {\numterms} terms,
many of which are themselves complicated.
Numerical methods are necessary:
it is impractical
to extract observational predictions
from such complex potentials
using analytical
techniques.
Progress therefore becomes
dependent on compute resource and
the availability of suitable
software tools.
It is arguable that analysis of this model---and others of
comparable
complexity---has been hampered by
\emph{both}
the paucity of powerful, general-purpose software tools
for
inflationary model analysis,
\emph{and}
an accepted means for exchanging
the specification of models within the community.
In the sister discipline of collider
phenomenology these roles are played by
the FeynRules system and its
\href{https://feynrules.irmp.ucl.ac.be/wiki/ModelDatabaseMainPage}{online model database}~\cite{Degrande:2011ua,Alloul:2013bka}.

These disadvantages notwithstanding,
its unusual theoretical control has made the
$\brane{3}$/$\antibrane{3}$ model
an interesting
laboratory in which to study the likelihood of inflation,
the distribution of observables
such as the primordial spectral index,
and the prospects for accommodating fine-tuning
issues such as the well-known `$\eta$-problem'
(that is, light scalar fields in a quasi-de Sitter
spacetime typically acquire masses of
order $H$).
For these reasons the model has developed
its own literature, which we
review briefly beginning on
p.~\pageref{page:previous-results}.

\para{Non-Gaussianity and observables}
In this paper we return to the
$\brane{3}$/$\antibrane{3}$ model and re-analyse it
using updated numerical methods.
Our principal aim is an accurate characterization
of the
primordial non-Gaussianity it produces,
for which reliable estimates have
not yet been reported.
To achieve this
we leverage
new versions of
the {\CppTransport}~\cite{Seery:2016lko,Butchers:2018hds}
and {\PyTransport}~\cite{Mulryne:2016mzv,Ronayne:2017qzn}
codes that automate evaluation of inflationary
correlation functions directly from
low-energy effective action.
Such automated methods are the most practical
way to handle models
whose
numerical implementation is
otherwise too laborious
or error-prone,
especially for
calculation of three-point statistics.
Some details of the improvements in the new versions of
these codes are described below, but they will be
discussed more completely in a forthcoming
publication.

The underlying technology is an evolution of the `transport'
method already used to analyse the
$\brane{3}$/$\antibrane{3}$ model by Dias et
al.~\cite{Dias:2012nf}.
It has already been described in the
literature~\cite{Mulryne:2009kh,Mulryne:2010rp,
Dias:2011xy,Anderson:2012em,Dias:2015rca,Dias:2016rjq},
and our implementation introduces no significant novelties
compared to these treatments.
Therefore we recapitulate
only those properties
relevant to our analysis or its interpretation.
First,
neither implementation makes use of
the slow-roll approximation
and therefore time dependence is treated exactly.
However,
in common with all other
general-purpose
frameworks for calculation of inflationary
correlation functions,
the {\CppTransport} and {\PyTransport}
implementations
are valid only to tree-level.
Here, `tree-level' has its usual meaning
in which a term at $n^{\text{th}}$ order in the
loop expansion involves $n$ unrestricted momentum
integrals.
There are two types of
loop in the Schwinger (or `in--in')
formalism appropriate for cosmological correlation
functions~\cite{Weinberg:2005vy,Weinberg:2006ac,Seery:2010kh}.
The first type represent the
familiar averages over
virtual quanta
that appear in `in--out' amplitudes,
and can be absorbed into a renormalization
of masses, coupling constants and field
amplitudes.%
    \footnote{A major advantage of
    the AdS/CFT computation used to obtain the $\brane{3}$-brane
    potential is that it accounts
    automatically for mixing between scales
    that is usually generated by averaging over virtual quanta.}
The second type can be regarded as averages over
unobserved physical particles,
which may include decay products or particles
generated from
non-adiabatic evolution, including
resonance~\cite{Seery:2008ms,Flauger:2016idt}.
Momentum integrals of this type
are a measure of
back-reaction from these particle production
processes.
Any tree-level framework, including the transport
method, is blind to this back-reaction.
In this paper we simply assume there is no
problematic back-reaction from particle production.
However, see footnote~\ref{footnote:tachyons}
on p.~\pageref{footnote:tachyons}.

What \emph{is} included?
In both two- and three-point functions
we capture all
effects
from quadratic mixing between modes on superhorizon scales
where momenta are soft compared to $H$ in the sense
$k/(aH) \ll 1$, and even
small off-diagonal terms in the mass matrix are relevant.
In the traditional language of inflationary
phenomenology these effects describe transfer of
power between adiabatic
and isocurvature modes.
Meanwhile,
in the three-point function we capture
the effect of three-body interactions.
These can loosely be regarded as describing
processes in which a pair of particles
are produced from the gravitational field,
before one member of the pair decays into two
daughter
particles~\cite{Seery:2009hs,Arkani-Hamed:2015bza}.
At later times the three resulting particles are
correlated due to their shared history.

We capture effects from any nontrivial mass
spectrum, including modes that are
much lighter
($m \ll H$),
much heavier ($m \gg H$),
or comparable to the Hubble scale ($m \sim H$).
At horizon exit these effects can
reduce the amplitude of fluctuations,
or change the subtle interference effects
imprinted in the three-point function---%
and higher-order correlations---%
from interaction between growing and decaying modes.
They may also induce significant correlations,
or anti-correlations, between the field
degrees of freedom at horizon exit.

To detect the emergence of an adiabatic limit before the end of
inflation
we employ a technique based on tracking eigenvalues of the
mass matrix~\cite{Elliston:2011dr,Seery:2012vj}.
If an adiabatic limit is reached this implies the model is
predictive without the need to specify details of a later
reheating
phase~\cite{GarciaBellido:1995qq,Weinberg:2003sw,Weinberg:2004kf,Weinberg:2004kr,
Rigopoulos:2003ak,Lyth:2004gb,
Elliston:2011dr,Elliston:2011et,Elliston:2014zea}.

Because our interest lies in computation of
observables, the {\CppTransport} 
and {\PyTransport} codes
are only the front-end of a longer pipeline.
Once the inflationary computation is complete,
the two-point function
is used as an
initial condition for the {\CLASS}
Boltzmann code~\cite{Blas:2011rf}.%
    \footnote{We
    use only the primordial
    two-point function (evaluated at the end of inflation)
    as an initial condition for the subsequent CMB
    calculation. We could equally
    well use this as an initial condition for the matter
    or galaxy power spectra,
    but
    we do not do so here because
    constraining the model from data is not
    the primary purpose of this paper
    and significant extra complexity is needed
    to describe galaxy bias.
    However, if desired,
    the flexibility of the {\Cosmosis} framework makes
    it simple to include more datasets in the likelihood calculation.
    
    We will see below that computations of the primordial
    three-point function are still sufficiently expensive that
    we cannot routinely compute the \emph{shape} of the bispectrum,
    for example to accurately model scale-dependent bias in the power spectrum,
    or to compute a full CMB bispectrum.
    We comment on the computational challenges
    in~{\S}\ref{sec:computational-issues}
    and Appendix~\ref{sec:run_time}
    below.}
This
enables us to generate custom predictions
for the
CMB angular power spectra $C_\ell$
and hence the likelihood function.
This is particularly important
for the $\brane{3}$/$\antibrane{3}$ model
because it frequently produces power spectra
with significant scale dependence~\cite{Dias:2012nf}.
Summary statistics evaluated at a single scale---%
such as the scalar amplitude $\As$
and spectral index $\ns$---%
are therefore misleading,
and
use of the full primordial power
spectrum is required.
The entire pipeline is controlled by the
{\Cosmosis} parameter estimation framework~\cite{Zuntz:2014csq},
allowing it to be attached to a number of efficient
general-purpose sampling algorithms.
We collect values for observables built from the
two- and three-point functions and use these to
estimate distribution functions.
Our focus is on general properties of
perturbations produced by the model,
whether or not their statistical character
falls in an observationally viable window.

Using this pipeline
we are able to study the distribution of
the three-point correlation amplitude on
representative `equilateral' and `folded'
configurations (where $k_1 \sim k_2 \sim k_3$
and $k_1 \sim k_2 \sim k_3/2$, respectively,
if $\vect{k}_1$, $\vect{k}_2$, $\vect{k}_3$ 
are the 3-momenta appearing
in the three-point correlator).
`Squeezed' configurations, where one of the $k_i$
becomes significantly smaller than the other two,
are substantially more expensive to simulate
and we are not able to compute these for every realization.
This is unfortunate because it is the squeezed
correlation amplitude that can be measured
most cleanly~\cite{Meerburg:2019qqi}.
Instead, we study how the squeezed
amplitudes correlate with the equilateral
and folded ones by constructing
a separate, smaller sample.
From this we infer the behaviour of
squeezed configurations in our full catalogue.

\label{page:previous-results}
\para{Previous results}
The brane inflation paradigm was introduced by
Dvali \& Tye~\cite{Dvali:1998pa},
and elaborated into the concrete
$\brane{3}$/$\antibrane{3}$ scenario by
Burgess et al.~\cite{Burgess:2001fx}
and
Dvali et al.~\cite{Dvali:2001fw}.
The branes carry opposite charges,
and in early work the resulting Coulomb attraction
was identified with the inflationary potential.
Unfortunately, this proposal was not viable
due to phenomenological difficulties.

Kachru et al.\ (`KKLMMT'~\cite{Kachru:2003sx},
following earlier work by `KKLT'~\cite{Kachru:2003aw}),
showed that the Coulomb potential would be flattened
by warping of the metric in the extradimensional space. 
Such warped-product geometries were already familiar from the Randall--Sundrum
scenario~\cite{Randall:1999ee,Randall:1999vf}.
Although
the flattened potential relieved most phenomenological
problems, Kachru et al.\ demonstrated
that it would receive significant corrections
from effects due to
moduli stabilization~\cite{Kachru:2003sx}
and in particular that these would
lift the inflaton mass to be of order $H$.
This is a manifestation of the familiar $\eta$-problem of inflation.
The conclusion is that successful
$\brane{3}$/$\antibrane{3}$ inflation would have to be regarded
as a `delicate' accident caused by partial cancellation
to produce a mass smaller than $H$.

An explicit computation of these corrections was given by
Baumann et al.~\cite{Baumann:2006th} for the case
where stabilization is due to $\brane{7}$-branes wrapping
four-cycles of the extradimensional space.
The implications
for $\brane{3}$/$\antibrane{3}$ inflation were
summarized in Refs.~\cite{Baumann:2007np,Baumann:2007ah}.
Their calculation showed that,
by fine-tuning the cancellation between different effects,
a small window existed for inflation to occur near an
inflexion point of the $\brane{3}$-brane potential.

The approach used in these papers left open the question of
what happens when moduli stabilization occurs by a more
general mechanism.
This issue was taken up by Baumann
et al., first in a linearized analysis~\cite{Baumann:2008kq}
and later including nonlinear effects~\cite{Baumann:2009qx,Baumann:2010sx}.
Their most developed
method made use of the AdS/CFT correspondence to map operators
in the $\brane{3}$-brane potential to the spectrum of
non-normalizable perturbations
of the warped conifold.
These perturbations can be determined by harmonic analysis
of the conifold base space, the coset space $T^{1,1}$,
for which
the necessary tools had already been
assembled by
Gubser~\cite{Gubser:1998vd}
and
Ceresole et al.~\cite{Ceresole:1999zs,Ceresole:1999ht}.
We review this method of constructing the brane potential
in~\S\ref{sec:construct-potential}.

\para{Inflationary analysis}
Each of these computations yields a prediction for 
the functional form
of the $\brane{3}$-brane potential,
parametrized by a number of mass scales $M_i$.
These scales may be regarded as encoding
ultraviolet information about the compactification
that has been integrated out to produce the
low-energy description.
In our present state of ignorance they cannot be
computed and must be estimated from
observations.

If the low-energy effective action
were to be expanded
in a basis of local operators,
the Wilson coefficient for each operator
would be determined by
the mass scales
$M_i$.
These coefficients would all have been regarded
as independent by the methods of traditional
effective field theory (`EFT').
For the $\brane{3}$ model, however,
relationships inherited from the functional form
of the brane potential
imply that
certain Wilson coefficients are correlated
or even absent.
It is these correlations that represent the gain in
information from using
an explicit ultraviolet completion
in contrast to a traditional analysis using
an ultraviolet-agnostic EFT.
We will see below that
the likelihood of inflation and its detailed predictions
can depend on this pattern of correlations.
This clearly illustrates
the weak decoupling between
inflation and
the assumed ultraviolet model.

Agarwal et al.\ assessed these importance of these
correlations between the low-energy
coefficients by repeatedly
drawing values for the $M_i$ from a
specified distribution~\cite{Agarwal:2011wm}---%
a strategy that had been introduced earlier by
Easther, Peiris and collaborators~\cite{Mortonson:2010er,Easther:2011yq}.
Assuming inflation would always begin from the same initial
field configuration,
Agarwal et al.\ were able to determine its likelihood
as a function of the number of e-folds
achieved.
They also determined the distribution
of the  single-scale summary statistics
$\As$, $\ns$ and $r$
derived from the two-point function.
Because the angular directions are typically heavy they argued
that an effectively unique inflationary trajectory would often
emerge,
and used a single-field approximation based on this trajectory
to estimate observables.
This approximation does not
capture multiple-field effects that transfer power between
entropic modes and the curvature perturbation.
It also does not account for the contribution
of fields that are not light compared to the Hubble scale.

Agarwal et al.\ used an ensemble of $> 70,000,000$ realizations
to study the homogeneous background, finding that
the probability of more than 60 e-folds of inflation was small, of order $10^{-5}$
to $\Ascut$.
Two further ensembles were used to study observables:
one with
$4,900,000$ realizations, of which $8,301$ yielded more than 60 e-folds;
and a second with
$500,000$ realizations, of which only $750$ yielded more than 60 e-folds.
The two ensembles differed in their truncation of the $\brane{3}$-brane
potential, to be described in~\S\ref{sec:construct-potential}.
Agarwal et al.\ concluded
that the tensor--scalar ratio $r$ would typically be unobservable,
and that the scalar spectral index $\ns$
fell roughly in the range
$0.94 \lesssim \ns \lesssim 1.10$,
with values in the WMAP7 range $\ns = 0.963 \pm 0.014$
(at $k=0.002 \, \Mpc^{-1}$~\cite{Larson:2010gs,Komatsu:2010fb})
coming from cases where the single-field treatment was likely to be
acceptable.
Their results did not depend strongly on the truncation.

In a small fraction of cases, Agarwal et al.\ observed abrupt transitions
between different angular minima and speculated that these might generate
significant non-Gaussianity from multiple-field
effects~\cite{Elliston:2011dr,Seery:2012vj}.
Our results demonstrate that this suggestion is essentially correct.
Indeed, the amplitude of three-point correlations generated in this way
can be surprisingly large, although trajectories that exhibit the effect
are rare within our ensemble.

Dias et al.\ used a more sophisticated numerical scheme to compute
observables~\cite{Dias:2012nf},
based on superhorizon `transport' of the
inflationary
correlation functions~\cite{Mulryne:2009kh,Mulryne:2010rp,Seery:2012vj,
Elliston:2012ab,Mulryne:2013uka}.
This approach was a precursor of the
technology we
deploy in this paper.
(The approach used here
is more complete because it
correctly accounts for subhorizon effects.)
Their method
correctly tracked transfer of power on superhorizon
scales, including contributions from fields that
were not light at the time of horizon exit.
Like the scheme of Agarwal et al.,
it applied only to the two-point function.%
    \footnote{Strictly this applies to v2 of the arXiv
    version of this paper, which includes
    an erratum to the published version (matching arXiv v1).
    Originally this paper contained an error from
    omission of the conifold
    metric, pointed out in Ref.~\cite{McAllister:2012am},
    which caused all fields to be light at horizon exit.
    Based on this error
    the published version included a discussion of the $\fNL$ observable,
    but its conclusions were invalidated when the correct
    conifold metric was introduced in the erratum.
	We would like to thank
    Mafalda Dias and Jonathan Frazer for helpful
    correspondence,
    and for kindly sharing their Mathematica code.}
As part of their analysis, Dias et al.\ attempted to quantify how many
instances of inflation converged to an adiabatic limit.
As explained above, when this occurs it implies that the model is
predictive without specifying the details of a later reheating phase.
Conversely,
if an adiabatic limit is not reached, the final value of each observable
may depend on the details of reheating~\cite{Elliston:2014zea}.

Dias et al.\ used an ensemble with 564 realizations giving more than
60 e-folds of inflation.
Within the statistical limits of their sample size,
these results confirmed the conclusion of Agarwal et al.\ that $r$
would be negligible, and yielded a comparable
distribution for $\ns$.
In most cases, they found that an adiabatic limit would be reached
during the inflationary phase.

Later, McAllister, Renaux-Petel \& Xu studied the same model using
a different numerical technique,
finding 18,731 realizations that yielded
at least 66 e-folds of inflation~\cite{McAllister:2012am}.
They found that 21\% of their realizations were consistent
with WMAP7 constraints on $\ns$
(see above). In agreement with Agarwal et al.,
these realizations typically exhibited
a unique inflationary trajectory
over the final 60 e-folds of inflation, making a single-field
treatment sufficient.
This usually occurs when inflation is of long duration,
with multiple-field effects appearing only as transients
at early times.
Finally, based on an analytic approximation
for the `quasi single-field'
regime~\cite{Chen:2009we,Chen:2009zp,Gong:2013sma},
they suggested that three-point correlations on squeezed
configurations could
occasionally become large,
with $|\fNL| \gtrsim 10$ in perhaps 0.07\% of realizations.
If it occurs, this form of non-Gaussianity has a very different
origin to the rapid shifting between angular minima
suggested by Agarwal et al.
Unfortunately,
detecting the presence of `quasi single-field'
effects is numerically expensive, and
in this paper we are not yet
able to form a definitive judgement regarding
their occurrence.

A different approach was pursued by
Hertog \& Janssen~\cite{Hertog:2015zwh},
who studied the possibility of eternal inflation
near the flat inflexion point that characterizes
inflating potentials in the
$\brane{3}$/$\antibrane{3}$ model.
This is very similar to the proposal of topological
inflation~\cite{Vilenkin:1994pv,Linde:1994wt};
see Ref.~\cite{Kim:2010ud} for an analysis of
non-Gaussianity in related models.
Hertog \& Janssen computed observables
in their scenario
using a prior based on the no-boundary wavefunction
proposal.
Accordingly the observable distributions reported by these
authors cannot be compared with those
given here, although their suggestion
$|\fNL| < 10^{-4}$ within their
ensemble is notable.

\para{Organization of this paper}
To build 
our primary catalogue required
sampling more than {\rawcataloguesize}
trajectories, of which
over {\precataloguesize} yielded
more than 60 e-folds of inflation.
It will be explained in~\S\ref{sec:code-comparison}
that some of these are excluded
due to concerns about representative sampling,
leaving roughly {\cataloguesize}
`safe' trajectories for which observables can be computed.
This is nearly
three times the number of trajectories
used by McAllister et al.,
and nearly 100 times the number used by Dias et al.
The large sample size
means
we are able to characterize the distribution of
each observable with reasonable accuracy.
(However, we will see that there is evidence we
still
undersample away from the central region
for some distributions.)
The main obstruction to generating even larger
ensembles is compute time.
Running on a Haswell-era
compute cluster, our production code
required
$\sim 95,000$ CPU hours
to build the primary catalogue,
and
a further
$\sim 40,000$ CPU hours
to compute observables
for each trajectory.
We comment further on the resource requirements
for the computation
in Appendix~\ref{sec:run_time}.

This primary catalogue is complemented by a number of `small'
catalogues,
each comprising roughly {\minicataloguesize} inflating trajectories,
that are used to study the dependence of observables on
various arbitrary choices made during construction of the model.
These include the way the potential is truncated,
the initial conditions for inflating trajectories,
and the treatment of contributions to the brane potential from
bulk fluxes (see~\S\ref{sec:flux-contributions}). Each of these
`small' catalogues has similar size to the current
best-in-class analysis reported in Ref.~\cite{McAllister:2012am},
giving us considerable statistical power when comparing
distributions.

In~\S\ref{sec:construct-potential}
we review the construction of the $\brane{3}$-brane potential, paying
particular attention to the harmonic
modes on $T^{1,1}$.
The elements of this discussion have all been given before, but are
scattered across a number of papers.
We collect the relevant formulae in a unified notation.

During the course of this work we discovered
instances
where inadvertent omissions
meant that
previous analyses of this model
had not been described in sufficient detail to allow replication.
These relate to minor technical choices in the
construction of the $\brane{3}$-brane potential
or in specifying priors for the sampling procedure.
To assist authors who wish to replicate our own analysis
we have attempted to document the construction of
the potential in sufficient detail to allow
replication if desired.
Our trajectory catalogues are available
for download from the Zenodo open-access
repository,
and may be 
re-used under a permissive
Creative Commons
licence.
(See Appendix~\ref{sec:data-availability}.)
Further,
our computational pipelines are open source
and published on GitHub.
We would like to thank the authors of the previous studies
for their ready assistance in relating our analysis to theirs.

The reader whose interest lies solely with the prediction of inflationary
observables may wish to skip~\S\ref{sec:construct-potential},
which involves ideas from extradimensional compactifications
in string theory,
and return to it only
to understand the relationship between parameters.
For this purpose Tables~\ref{Table:ParameterGlossary}
and~\ref{Table:NotationGlossary} may be helpful.
Conversely, readers who are already familiar with the detailed
constructions of Refs.~\cite{Baumann:2009qx,Baumann:2010sx}
will not find any new material and may also wish to proceed
directly to~\S\ref{sec:experimental-procedure}.

In~\S\ref{sec:experimental-procedure}
we describe our software stack and the
numerical method used to compute observables.
We document our choice of priors and
initial conditions, and the precise sampling strategy
we apply to build both the primary catalogue
and the `small' catalogues
used for comparison.
In~\S\ref{sec:cpptransport-pipeline}
and~\S\ref{sec:pytransport-pipeline}
we explain how observables are computed
within each pipeline,
and in~\S\ref{sec:computational-issues}
we
discuss general
computational issues that arise within
the transport framework
irrespective of implementation.
In~\S\ref{sec:adiabatic-limit}
we describe our procedure for detecting an adiabatic limit
at the end of inflation.
Finally, in~\S\ref{sec:code-comparison}
we compare the distributions reported by
each pipeline
and develop a choice of cuts intended to ensure
the integrity of our analysis.
Imposing these cuts reduces our primary
catalogue from (roughly) {\precataloguesize}
to {\cataloguesize} trajectories, as explained above.

In~\S\ref{sec:Results}
we study the distribution for each observable over the
catalogues
constructed in~\S\ref{sec:experimental-procedure}.
\S\ref{sec:results-background}
discusses the behaviour of
trajectories at the level of the background.
In~\S\ref{sec:twopf-observables}
we consider observables derived from the two-point
function, and in~\S\ref{sec:threepf-observables}
we extend this to 
include information on three-point correlations on
equilateral, folded, and (via a separate catalogue)
squeezed configurations.
We compare our distributions with results previously
given in the literature.
In~\S\ref{sec:abrupt-change-minima}
we discuss a population of rare trajectories that
exhibit the abrupt transitions between angular minima
observed by Agarwal et al., and show that these
yield very large three-point correlations
of `local' shape.
Finally, we conclude in~\S\ref{sec:discussion}.
Three appendices summarize information tangential
to the main discussion.
Appendix~\ref{sec:data-availability}
provides information about the data deposit
accompanying this paper.
Appendix~\ref{sec:run_time}
gives more details on computational resource
requirements.
Appendix~\ref{sec:transport_spectral}
summarizes the `transport' computation of
spectral indices, including new subleading terms
intended to accelerate convergence.

\para{Notation and conventions}
We work in natural units where $c = \hbar = 1$.
The reduced Planck mass is defined by $\Mp = (8\pi G)^{-1/2}$
and is numerically equal to $2.435 \times 10^{18} \, \GeV$.
Latin indices $a$, $b$, {\ldots}, label coordinates
in four-dimensional space,
and indices $A$, $B$, {\ldots}, label coordinates
in the six-dimensional compact space.
The distributions $U(a,b)$ and $N(\mu, \sigma)$
are the uniform distribution with lower limit $a$
and upper limit $b$,
and the normal distribution of mean $\mu$ and
standard deviation $\sigma$, respectively.

The $\brane{3}$/$\antibrane{3}$ model is very complicated---%
indeed, it may be the most
complicated model for which inflationary observables have
yet been computed. Its description entails a heavy overhead of
notation. To assist readers who are unfamiliar
with this maze of definitions we
provide a glossary
in Tables~\ref{Table:ParameterGlossary}--\ref{Table:NotationGlossary}.
Table~\ref{Table:ParameterGlossary} lists the
parameters of the potential, together with their mass dimension,
point of definition, and whether they are fixed, sampled,
or derived quantities in our sampling procedure.
Table~\ref{Table:NotationGlossary} provides similar
information for other
relevant
quantities
that do not parametrize the potential.

In~{\S}\ref{sec:brane-potential}
and Appendix~\ref{sec:transport_spectral}
we discuss the effective field theory for the
$\brane{3}$-brane system and its fluctuations,
for which 
we briefly summarize our
conventions.
The effective field theory
comprises six scalar fields $X^A$ inherited
from the extradimensional coordinates,
with momenta $\pi^A = \d X^A / \d N$,
where $N = \int^t H \, \d t$
represents the accumulated e-folds of expansion.
We collect the fields and momenta into
a phase space coordinate
$\pcoord{X}^{\pidx{A}}$
with index $\pidx{A}$.
The corresponding fluctuations
are $\delta \pcoord{X}^{\pidx{A}}$.
\ctable[
    label = Table:ParameterGlossary,
    caption = {\small Glossary of parameters for the $\brane{3}$/$\antibrane{3}$ model.},
    doinside = {\StandardTable\renewcommand{\arraystretch}{1.2}\rowcolors{2}{gray!20}{white}},
    center, sideways, botcap
]{LCXccR}{
\tnote[a]{The initial conditions for the fields $\phi^A$ and their velocities
$\dot{\phi}^A$ do not vary in our primary ensemble and do not appear in this
table. See the discussion in~\S\ref{sec:experimental-procedure}.}
\tnote[b]{The $Q$ parameter was introduced by Agarwal et al.~\cite{Agarwal:2011wm}
[see Eq.(10) of this reference]. It effectively sets the normalization for the Wilson
coefficients $C_{LM}$ and $\mathscr{C}_{\sumL \sumM}$.}
\tnote[c]{This prior for $Q$ is selected with foreknowledge of the $Q$
posterior and a view to sampling efficiency. We will see in~\S\ref{sec:Results}
that $Q$-values which frequently allow $N > 60$ e-folds of inflation
are rather tightly clustered. We have verified that our final distributions for
observables are unchanged if we select a weaker, broad prior for $Q$, except that
the sampling process spends more time in regions of parameter space where
$N > 60$ e-folds are unlikely. We use the uniform prior quoted here only
in the {\CppTransport} pipeline.
The {\PyTransport} pipeline instead uses a `beta-prime' distribution
$\beta'(\alpha, \beta)$
with shape
parameters $\alpha = 4.16$, $\beta = 4.94$.
See the discussion of the $Q$-prior in~\S\ref{sec:pytransport-pipeline}.
Note that this $\alpha$ parameter is not the same as $\alpha = V_0/D_0$
appearing in the table.}}
{
    \toprule
    \multicolumn{1}{l}{parameter\tmark[a]} &
    \multicolumn{1}{c}{mass dim.} &
    &
    definition &
    type & \NN
    \midrule
    \rIR = \xIR \rUV & -1 & IR limit of throat & \glossarypage{glossary:rIR} & derived & \NN
    \rUV = \phiUV/T_3^{1/2} & -1 & UV limit of throat & \glossarypage{glossary:rUV} & derived & \Mp^{-1} \NN
    T_3 & 4 & $\brane{3}$-brane tension & \glossarypage{eq:nambu-goto} & fixed & 10^{-2} \Mp^4 \NN
    \xIR & 0 & $\rIR$ in dimensionless $x$ coordinate & \glossarypage{glossary:xIR} & fixed & 0.02 \NN
    \phiUV & 1 & UV limit of throat in $\phi$ coordinate & \glossarypage{glossary:phiUV} & fixed & 0.1 \Mp \NN
    Z = (2\Mp / \phiUV)^2 & 0 & $\brane{3}$-brane charge & \glossarypage{glossary:Z} & derived & 400 \NN
    a_0 = \e{A(\rIR)} & 0 & value of warp factor at IR limit of throat & \glossarypage{glossary:a0} & fixed & 10{-3} \NN
    \gstring = \e{\phi} & 0 & string coupling & \glossarypage{glossary:gstring} & not required & \text{---} \NN
    D_0 = 2 T_3 a_0^4 & 4 & normalization of Coulomb potential & \glossarypage{glossary:D0} & derived & (3.76 \times 10^{-4} \Mp)^4 \NN 
    V_0 = \alpha D_0 & 4 & energy density from distant supersymmetry breaking & \glossarypage{glossary:V0} & derived & \NN
    \mu = (V_0 + D_0)^{1/4} (\phiUV / \Mp)^{1/2} & 1 & mass scale in $\VCF$ and $\VPI$ & \glossarypage{glossary:mu} & derived & 1.19 \times 10^{-4} (1 + \alpha)^{1/4} \Mp \NN
    C_{LM} & 0 & Wilson coefficients for zero-model labelled by $LM$ & \glossarypage{glossary:CLM} & sampled & Q \times \big[ N(0,1) + \im N(0,1) \big] \NN
    D_{LM} = C_{LM} + C^\ast_{-L-M} & 0 & combined Wilson coefficient for mode $LM$ if $L>0$ & \glossarypage{glossary:DLM} & derived & Q \times \big[ N(0,\sqrt{2}) + \im N(0,\sqrt{2}) \big]\NN
    \mathscr{C}_{\sumL \sumM} & 0 & Wilson coefficient for flux mode $\sumL \sumM$ & \glossarypage{glossary:fluxCLM} & sampled & Q \times \big[ N(0,1) + \im N(0,1) \big] \NN
    \midrule
    \alpha & 0 & relative magnitude of $V_0$ and $D_0$ & --- & sampled & U(-1,1) \NN
    Q & 0 & typical distance to nearest $\brane{7}$-brane stack\tmark[b] & --- & sampled\tmark[c] & U(0,0.04) \NN
    \bottomrule
}

\ctable[
    label = Table:NotationGlossary,
    caption = {\small Glossary of notation for the $\brane{3}$/$\antibrane{3}$ model.},
    width = \textwidth,
    doinside = {\StandardTable\renewcommand{\arraystretch}{1.2}\rowcolors{2}{gray!20}{white}}
]{LXl}{
\tnote[a]{The $\alpha$ appearing in $C_4$ and $\Phi_-$ is the same,
but not the same $\alpha$ that appears in
Table~\ref{Table:ParameterGlossary}.}
}
{
    \toprule
    \multicolumn{1}{l}{quantity} &
    &
    definition \NN
    \midrule
    T^{1,1} & Romans manifold $SU(2) \times SU(2) /U(1)$; base of conifold & \glossarypage{glossary:T11} \NN
    G_{AB} & conifold metric & \glossarypage{glossary:GAB} \NN
    G^{T^{1,1}}_{AB}, \d s^2_{T^{1,1}} & metric on $T^{1,1}$ & \glossarypage{glossay:GABT11} \NN
    A(r) & Klebanov--Strassler warp factor & \glossarypage{glossary:Ar} \NN
    H & generator of $U(1)$ divisor in $T^{1,1}$ & \glossarypage{glossary:H} \NN
    K & generator of $U(1)$ $R$-symmetry of $T^{1,1}$ & \glossarypage{glossary:K} \NN
    T_i & generators of $SU(2)$ & \glossarypage{glossary:Ti} \NN
    \Psi = (\theta_1, \theta_2, \phi_1, \phi_2, \psi) & coordinates on $T^{1,1}$ & \glossarypage{metric_line} \NN
    r & radial coordinate on the conifold & \glossarypage{eq:cone-metric} \NN
    C_4 = \alpha \, \omega & Chern--Simons form coupling to $\brane{3}$-brane\tmark[a] & \glossarypage{glossary:C4} \NN
    \omega & volume form on $\brane{3}$-brane; $\omega = \star 1$ & \glossarypage{glossary:omega} \NN
    \gamma_{ab} & induced metric on $\brane{3}$-brane & \glossarypage{eq:nambu-goto} \NN
    X^A & embedding coordinates $X^A = X^A(x^a)$ of $\brane{3}$-brane & \glossarypage{glossary:phiA} \NN
    \Phi_- = \e{4A} - \alpha & supergravity field; AdS/CFT dual to $\brane{3}$-brane potential\tmark[a] & \glossarypage{glossary:PhiMinus} \NN
    x = r/\rUV & dimensionless $\brane{3}$-brane radial coordinate & \glossarypage{glossary:x} \NN
    R_4 & 4-dimensional Ricci scalar on $\brane{3}$-brane & \glossarypage{glossary:ricci4} \NN
    \Lambda & 3-form flux derived from $G_3$ & \glossarypage{glossary:Lambda} \NN
    G_3 & 3-form field of Type IIB supergravity & \glossarypage{glossary:G3} \NN
    \VCoulomb & Coulomb contribution to potential & \glossarypage{glossary:VCoulomb} \NN
    \Vmass & mass term in potential from $R_4$ & \glossarypage{glossary:Vmass} \NN
    \VCF & `complementary function' part of potential & \glossarypage{glossary:VCF} \NN
    \VPI & `particular integral' part of potential & \glossarypage{glossary:VPI} \NN
    \rep{\mu} & label for representations of a Lie group $G$ & \glossarypage{glossary:replabel} \NN
    D^\rep{\mu}_{mm'}(\beta) & Wigner's $D$-matrix in representation $\rep{\mu}$ & \glossarypage{glossary:WignerD} \NN
    d^\rep{\mu}_{mm'}(\beta) & Wigner's little $d$ matrix in representation $\rep{\mu}$ & \glossarypage{glossary:Wignerd} \NN
    L(g) & coset representative of group element $g$ & \glossarypage{glossary:cosetL} \NN
    L = (\ell_1, \ell_2, R) & labels representation of isometry group $SU(2) \times SU(2) \times U(1)$ & \glossarypage{glossary:L} \NN
    M = (m_1, m_2) & runs over representation space for representation $L$ & \glossarypage{glossary:M} \NN
    \Eharmonic_{LM}(\Psi) & harmonic on $T^{1,1}$ & \glossarypage{glossary:Eharmonic} \NN
    \Lambda_L & eigenvalue of $\Eharmonic_{LM}$ in Laplacian $\grad^2_{T^{1,1}}$ & \glossarypage{glossary:LambdaL} \NN
    {\Jup{}}, {\Jom{}} & normalized $\theta$ modes on $T^{1,1}$ & \glossarypage{glossary:Jup} \NN
    \Delta(L) = -2 + \sqrt{4 + \Lambda_L} & scaling dimension of zero-mode for representation $L$ & \glossarypage{glossary:DeltaL} \NN
    \Delta, \delta & scaling dimensions \emph{unrelated} to $L$ & \glossarypage{glossary:DeltaNoL} \NN
    \GreenFunction(x,x'), g_L(r,r') & Green's function on conifold and eigensum decomposition $g_L$ & \glossarypage{glossary:GreenFunction} \NN
    \suml, \summ, \sumR & quantum numbers in Clebsch-Gordan expansion & \glossarypage{glossary:suml} \NN
    \sumL, \sumM & $\sumL = (\suml_1, \suml_2, \sumR)$, $\sumM = (\summ_1, \summ_2)$ & \glossarypage{glossary:sumL} \NN
    \mathscr{A}(L, L', \sumL) & unknown amplitude in Clebsch--Gordan expansion & \glossarypage{glossary:Ascr} \NN
    \bottomrule
}

\FloatBarrier
\section{The $\brane{3}$/$\antibrane{3}$ model of inflation}
\label{sec:construct-model}
The $\brane{3}$/$\antibrane{3}$ model
has been studied extensively in the literature,
to which we refer for a more complete account of its
construction~\cite{Dvali:1998pa,Burgess:2001fx,
Kachru:2003sx,Baumann:2006cd,Baumann:2006th,Baumann:2007ah,
Baumann:2007np,Baumann:2008kq,Baumann:2009qx,Baumann:2010sx};
see also Ref.~\cite{Baumann:2014nda} for a textbook description.
In this section we recall only those details needed to
fix notation or
make our account self-contained.
Our primary intention is to give an unambiguous specification
of the $\brane{3}$-brane potential and its parametrization.

In~\S\ref{sec:conifold-geometry} we recall the geometry of
the conifold and explain how its four-dimensional low energy
description can support an inflationary phase.
In~\S\ref{sec:brane-potential}
we summarize the procedure of Baumann et al.\ for constructing
the $\brane{3}$-brane potential~\cite{Baumann:2009qx,Baumann:2010sx}.
This depends on the details of harmonic analysis
on the conifold, originally discussed by Gubser~\cite{Gubser:1998vd}
and Ceresole et al.~\cite{Ceresole:1999zs,Ceresole:1999ht},
and summarized here from a different perspective
in~\S\ref{sec:harmonic-analysis}.
These details are used to construct zero modes of the
conifold Laplacian,
which represent a class of contributions to the brane potential
from `unsourced' deformations of the conifold geometry.
Finally, in~\S\ref{sec:flux-contributions}
we give our prescription for a second class of `sourced'
deformations. Although these have been included in previous analyses,
certain arbitrary choices are required to
completely specify their
contribution.
In this section we document our choices in
sufficient detail
(at least in intention)
to allow replication of our analysis.

\label{sec:construct-potential}
\subsection{The conifold geometry}
\label{sec:conifold-geometry}

In the model, a mobile $\brane{3}$-brane
moves in an extradimensional space
due to its mutual Coulomb attraction
with a distant $\antibrane{3}$-brane.
From the perspective of
a four-dimensional observer, the displacement between the
branes in each coordinate can be regarded as a
four-dimensional scalar field.
These fields will be nearly homogeneous if the branes
are nearly parallel.
The forces experienced by the brane as it moves in the
extradimensional space can be summarized
as an effective potential for the displacement in each
coordinate.

\para{The Klebanov--Strassler throat}
It was explained in~{\S}\ref{sec:introduction} that
the Coulomb force generates a potential that is too
steep to support inflation if the extradimensional
geometry is flat.
To flatten the potential requires warping in the extradimensional
space. A candidate geometry is the singular warped
conifold studied by Klebanov \& Strassler~\cite{Klebanov:2000hb}.
The ten-dimensional metric is a warped product,
\label{glossary:Ar}
\begin{equation}
    \d s^2 = \e{2A} g_{ab} \, \d x^a \d x^b
    + \e{-2A} G_{AB} \, \d X^A \d X^B ,
\end{equation}
where $g_{ab}$ is the four-dimensional spacetime metric
with coordinates $x^a$
and
$G_{AB}$ is the metric on the
transverse
extradimensional space with coordinates $X^A$.
We take this to be a cone over the Romans space
$T^{1,1} = SU(2) \times SU(2) / U(1)$~\cite{Romans:1984an},
\label{glossary:T11}
\label{glossary:GAB}
\begin{equation}
    \label{eq:cone-metric}
    G_{AB} \, \d X^A \d X^B
    =
    \d r^2 + r^2 \, \d s_{T^{1,1}}^2 . 
\end{equation}
The geometry is supposed to be supported by a stack of
$Z \gg 1$
$\brane{3}$-branes positioned at the tip.
The base space $T^{1,1}$ is five-dimensional%
    \footnote{\label{footnote:T11}For a description of coordinates on $T^{1,1}$, see
    Refs.~\cite{Romans:1984an,Candelas:1989js,Witten:1993yc,Klebanov:1998hh,Ceresole:1999zs,Ceresole:1999ht}.
    We follow
    Candelas \& de la Ossa~\cite{Candelas:1989js},
    especially {\S}2 and Appendix A.
    $SU(2)$ is isomorphic to the three-sphere $S^3$,
    which itself is a Hopf fibration
    of $U(1)$ over $S^2$.
    The $U(1)$ can be regarded as parametrizing
    motion along a great circle of $S^3$.
    In the product $SU(2) \times SU(2)$
    there are two $U(1)$ fibres corresponding to motion
    along great circles of the left and right $S^3$s.
    We define the linear combinations
    $H = T_3 + \hat{T}_3$
    \label{glossary:H}
    \label{glossary:Ti}
    and
    $K = T_3 - \hat{T}_3$,
    \label{glossary:K}
    where the $T_i$ are generators of the left-hand copy of $SU(2)$
    and $\hat{T}_i$ are corresponding generators
    of the right-hand copy,
    and $T_3$, $\hat{T}_3$ generate the $U(1)$ factors.
    
    $T^{1,1}$ is the quotient
    $SU(2) \times SU(2) / U(1)_H$, obtained by identifying
    points that can be reached by a rotation generated by $H$.
    (It is part of a family of spaces $T^{p,q}$ obtained by
    generalizing $H$ to $H = p T_3 + q \hat{T}_3$. To preserve supersymmetry we must
    choose $p=q=1$~\cite{Romans:1984an}.)
    Therefore,
    in local Euler angles $\vartheta$, $\hat{\vartheta}$
    measured along each great circle,
    $T^{1,1}$ can be embedded
    as any hypersurface
    $\vartheta + \hat{\vartheta} = \text{const}$~\cite{Gubser:1998vd}.
    The coset representatives are labelled by the angular
    coordinate along this hypersurface.
    Accordingly,    
    functions on $T^{1,1}$ should depend only on the coset label
    $\vartheta - \hat{\vartheta}$
    and not $\vartheta + \hat{\vartheta}$.}
and carries the metric~\cite{Candelas:1989js}
\begin{equation}
    \label{metric_line}
    \d s_{T^{1,1}}^2 = \frac{1}{9}
        \Bigg(
            \d\psi + \sum_{i=1}^2 \cos{\theta_i} \, \d\phi_i
        \Bigg)^2
        +
        \frac{1}{6} \sum_{i=1}^2
        \Big(
            \d\theta_i^2 + \sin^2\theta_i \, \d \phi_i^2
        \Big) .
\end{equation}
The coordinate ranges are
$\theta_i \in [0, \pi]$,
$\phi_i \in [0, 2\pi]$
and $\psi \in [0, 4\pi]$.
The warp factor $A$ can be computed explicitly
for the unperturbed conifold,
but is
not required for the inflationary analysis.

\begin{figure}
    \centering
    \includegraphics[width=0.7\textwidth]{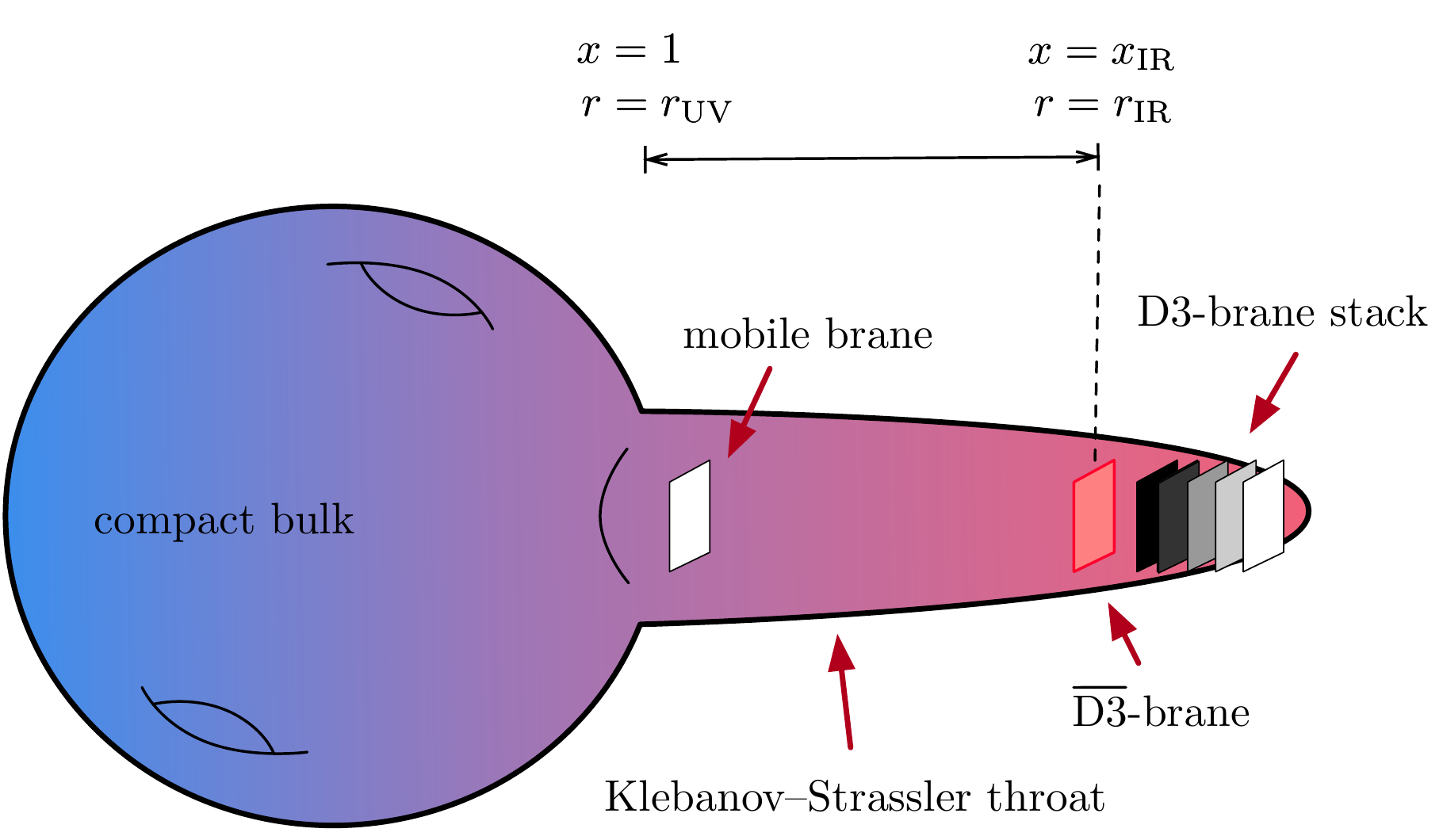}
    \caption[Schematic representation of the
    conifold geometry]
    {\label{fig:conifold}Schematic representation
    of the extradimensional geometry.
    The throat is supported by a stack of
    $\brane{3}$-branes placed at its tip
    and is
    approximately described by the Klebanov--Strassler
    solution. It attaches in the ultraviolet
    ($x=1$, $r = \rUV$) to an unknown
    bulk manifold.
    In the infrared ($r = \rIR$) a
    $\antibrane{3}$-brane
    draws the mobile brane down the throat
    due to their mutual Coulomb attraction.}
\end{figure}

If the branes are exactly parallel,
their displacement is uniquely labelled by
the coordinate $r$.
It
runs from a minimum value $\rIR$,\label{glossary:rIR}
where the geometry is `deformed'
by attaching to a smooth cap
that resolves the singularity at the tip of the cone~\cite{Candelas:1989js},
and
extends in principle
to arbitrarily large $r$.
However,
we usually imagine that the Klebanov--Strassler
solution is cut off
at some large value $\rUV$
and glued to a compact bulk
space.\label{glossary:rUV}%
    \footnote{The six extra dimensions must be compactified,
    unlike the Klebanov--Strassler solution, otherwise
    the effective Planck mass would be $\infty$
    and there would be no dynamical gravity induced in the
    four-dimensional world.}
In the field theory dual,
small values of $r$ correspond to the infrared
and large values correspond to the ultraviolet.
The $\antibrane{3}$-brane is located in the infrared,
where the warp factor minimizes its energy,
and the mobile $\brane{3}$-brane
is drawn from the ultraviolet towards the infrared
end of the cone.
Inflation ends when the mobile brane becomes sufficiently
close to the $\antibrane{3}$ that a tachyonic
instability develops and the brane
pair dissolves into closed
string modes.
In the four-dimensional field theory description this is
a hybrid transition
in which the inflaton is destabilized by a waterfall field.
See Fig.~\ref{fig:conifold}.

\para{Kinetic and potential terms for the brane}
The dynamics of the mobile $\brane{3}$-brane are determined
by the action
\begin{equation}
    \label{eq:nambu-goto}
    S = - T_3 \int \d^4 x \; \sqrt{-\det \gamma_{ab}}
    + T_3 \int \d^4 x \, \sqrt{-g} \; \alpha ,
\end{equation}
where
$x^a$ label coordinates on the brane and
$\gamma_{ab}$ is its induced metric.
There are two contributions.
The first
is the Nambu--Goto action.
This computes
the total energy of the brane, given by its tension
$T_3$ integrated over its worldvolume.
The second is a Chern--Simons term that
couples the brane worldvolume to a four-form
potential $C_4 = \alpha \omega$\label{glossary:C4},
where $\omega = \star(1)$\label{glossary:omega} is the volume form
determined by $g_{ab}$.

We now drop the parallel approximation.
Assuming the brane is embedded in the transverse
dimensions at $X^A = X^A(x^a)$,\label{glossary:phiA}
it follows that
\begin{equation}
    \gamma_{ab} =
    \e{2A} g_{ab}
    +
    \e{-2A} G_{AB} \partial_a X^A \partial_b X^B ,
\end{equation}
where $G_{AB}$ is the metric on the cone, Eq.~\eqref{eq:cone-metric}.
Therefore,
\begin{equation}
    S = -T_3 \int \d^4 x \, \sqrt{-g} \;
    \e{4A}
    \sqrt{\det\Big(
        \delta^a_c
        +
        \e{-4A} G_{AB} g^{ab} \partial_b X^A \partial_c X^B
    \Big)}
    + T_3 \int \d^4 x \, \sqrt{-g} \; \alpha .
\end{equation}

Assuming the brane is moving non-relativistically
it suffices to work only up to quadratic order in derivatives.
The zeroth order term from the determinant
combines with the Chern--Simons
term to generate a potential,\label{glossary:PhiMinus}
\begin{equation}
    \label{eq:nonperturbative-potential}
    V = T_3 \equiv T_3 \Phi_- ,
    \quad \text{where} \quad
    \Phi_- \equiv \e{4A} - \alpha .
\end{equation}
If the Klebanov--Strassler geometry is unperturbed---%
meaning that the infrared $\antibrane{3}$-brane
is absent---%
then $A = A(r)$ and $\alpha = \alpha(r)$ depend only on the radial
coordinate.
Therefore $V = V(r)$ also depends only on $r$,
and the angles $\{ \theta_1, \theta_2, \phi_1, \phi_2, \psi \}$
are flat directions.
However, more careful analysis shows that
in this case $\alpha = \e{4A}$
and the potential vanishes~\cite{Giddings:2001yu}.
Generically, both $A$ and $\alpha$ will depend on $r$ \emph{and} the angles.
This lifts the flat directions.
Notice that the potential depends on the warp
factor,
which produces the
flattening observed in Refs.~\cite{Kachru:2003aw,Kachru:2003sx}.

At second order in derivatives we obtain
\begin{equation}
    \label{eq:field-space-kinetic}
    S = - \frac{T_3}{2}
    \int \d^4 x \, \sqrt{-g} \;
        G_{AB} \partial^a X^A \partial_a X^B ,
\end{equation}
in which the warp factor has cancelled.
Eq.~\eqref{eq:field-space-kinetic}
is the kinetic term for a set of
noncanonical four-dimensional
scalar fields $X^A$
with kinetic mixing matrix $G_{AB}$
inherited from the cone~\eqref{eq:cone-metric}.
Therefore,
if the potential can be chosen suitably,
the $\phi^A$ may
support a phase of slow-roll inflation.
Predictions from inflationary models of this type were
studied by Sasaki \& Stewart~\cite{Sasaki:1995aw}.
The theory was developed up to three-point observables
by a number of authors~\cite{Nakamura:1996da,GrootNibbelink:2001qt,Sun:2005ig,Rigopoulos:2005xx,
Peterson:2011yt,Elliston:2011dr,Kaiser:2012ak,Butchers:2018hds}.

Instead of $r$ we choose to work
in terms of the coordinate $x = r/\rUV$\label{glossary:x}
introduced in Ref.~\cite{Agarwal:2011wm}.
Its range is
$\xIR < x \ll 1$,\label{glossary:xIR}
where $\xIR \equiv \rIR / \rUV$.
The throat attaches
to the compact bulk space
in the region $x \sim 1$.
Further,
if we simplify the brane kinetic term by absorbing the tension
$T_3$ into the metric, we find\label{glossay:GABT11}
\begin{equation}
    \label{eq:practical-conifold-metric}
    G_{AB} \, \d X^A \d X^B \rightarrow
    \rUV^2 T_3 \Big( \d x^2 + x^2 \, \d s^2_{T^{1,1}} \Big)
    \equiv
    \phiUV^2 \Big( \d x^2 + x^2 \, \d s^2_{T^{1,1}} \Big) ,
\end{equation}
where we have defined $\phiUV = \rUV T_3^{1/2}$.\label{glossary:phiUV}
Note that
the fields $\{ x, \theta_1, \theta_2, \phi_1, \phi_2, \psi \}$
appearing in the four-dimensional effective action
all have engineering dimension zero.
To compensate, the metric $G_{AB}$ has engineering dimensions
of $[\phiUV^2] = [\mathrm{M}^2]$.

\para{Field range}
Baumann \& McAllister argued that in a throat carrying
$\brane{3}$-brane charge $Z \gg 1$,\label{glossary:Z}
the field range would be bounded
by $\phiUV < 2 \Mp / Z^{1/2}$~\cite{Baumann:2006cd}.
Following Agarwal et al.~\cite{Agarwal:2011wm}
we generally take $\phiUV = 10^{-1}$
and fix $T_3 = 10^{-2} \Mp^4$.
In~\S\ref{sec:small-studies}
we briefly look at the effect of varying the field-range
bound over the interval
$10^{-1} < \phiUV < 10^{-3}$.

\subsection{The $\brane{3}$-brane potential}
\label{sec:brane-potential}

The remaining task is to enumerate
permitted contributions to
the $\brane{3}$-brane potential, Eq.~\eqref{eq:nonperturbative-potential}.
In the
unperturbed
Klebanov--Strassler geometry
(without the infrared $\antibrane{3}$-brane)
both $A = A(r)$ and $\alpha = \alpha(r)$
can be calculated explicitly~\cite{Giddings:2001yu},
but as explained above this leads to a vanishing potential.
The interpretation is that
gravitational attraction between the
mobile $\brane{3}$-brane and the
$\brane{3}$-brane stack at the tip
is balanced by
repulsion due to their same-sign
charges.
This arrangement cannot support an inflationary epoch.
To generate
a nontrivial potential requires
additional sources,
so that
we no longer expect exact cancellation.

How are we to determine the possible contributions?
The
general formula $V = T_3 \Phi_-$ given
in~\eqref{eq:nonperturbative-potential}
continues to apply,
which reduces the problem to determination of $\Phi_-$.
The supergravity field equations in the throat
can be shown to require
\begin{equation}
    \grad^2 \Phi_- = R_4
    + \frac{\gstring}{96} |\Lambda|^2
    + \e{-4A}|\grad \Phi_-|^2
    + \text{local terms} ,
    \label{eq:Phi-minus-eq}
\end{equation}
where
$\gstring$ is the string coupling,\label{glossary:gstring}
$\grad^2$ is the Laplacian on the conifold~\eqref{eq:cone-metric},
and $R_4$ is the four-dimensional Ricci scalar.\label{glossary:ricci4}
$\Lambda$ is a 3-form flux\label{glossary:Lambda}
that depends on the 3-form field $G_3$\label{glossary:G3} of type IIB
supergravity; for details, see Refs.~\cite{Baumann:2008kq,Baumann:2010sx,Baumann:2014nda}.
The `local terms' represent localized contributions from the 
mobile brane and the
$\antibrane{3}$-brane,
which we now introduce.

\para{Coulomb and mass terms}
First, the local terms generate a
Coulomb attraction between the
mobile brane and the
antibrane,
with potential\label{glossary:VCoulomb}
\begin{equation}
    \label{Coulomb}
    \VCoulomb (x) = D_0 \left( 1 - \frac{27}{64 \pi^2} \frac{D_0}{\phiUV^4} \frac{1}{x^4} \right) .
\end{equation}
The parameter $D_0$ is defined by
$D_0 \equiv 2 T_3 a_0^4$~\cite{Baumann:2008kq},\label{glossary:D0}
where $a_0 \equiv \e{A(\rIR)} \ll 1$.\label{glossary:a0}
It is the smallness of $D_0$,
caused by warping of the conifold,
that makes the potential
sufficiently flat to inflate at modest values of $x$.

Second, the leading effect of
the Ricci term $R_4$ is to generate
the operator
$R_4 \phiUV^2 x^2 / 12$.
During inflation the background geometry is approximately
de Sitter, for which $R_4 = 12H^2$.
If we take the inflationary phase to be supported by the
constant term in~\eqref{Coulomb},
possibly augmented by a second uplift $V_0$\label{glossary:V0},
then
$3 H^2 \Mp^2 \approx V_0 + D_0$.
Here $V_0$ accounts for
constant contributions that do not originate in
the Coulomb interaction,
which
could include distant sources of 
supersymmetry breaking.
In total this yields a mass term for $x$
of the form\label{glossary:Vmass}
\begin{equation}
	\label{MassCurvature}	
	\Vmass = \mu^4 x^2 / 3
	,
\end{equation}
where we have defined~\cite{Agarwal:2011wm}\label{glossary:mu}
\begin{equation}
	\label{eq:mu-def}
    \mu^4 \equiv (V_0 + D_0) \left( \frac{\phiUV}{\Mp} \right)^2 .    
\end{equation}
Note that $\mu$ has mass dimension $[\mathrm{M}]$.

\para{Deformations of the throat}
Third, the throat geometry~\eqref{eq:cone-metric} may be disturbed
because of back-reaction from the passage of the brane.
It may also be deformed
by the suture between the throat and the compact bulk geometry.
Any such disturbances will affect the
dynamics of the brane 
and contribute to its effective potential.

In the vicinity of the suture
an adequate description of $\Phi_-$
will require
boundary conditions that determine how information from
the compact bulk is communicated to the ultraviolet end
of the throat.
This complicated structure
for $\Phi_-$ will generate a large
number of operators
in the effective theory
whose Wilson coefficients depend
sensitively on the ultraviolet data.
In this region there is little hope of
performing a realistic analysis of the model.

On the other hand, in the infrared region $x \ll 1$
we expect that renormalization group running will
suppress most of these operators, leaving only
a handful of the most relevant terms.
In this region
fewer Wilson coefficients must be specified,
making the model significantly simpler to
analyse. In particular, as described in~\S\ref{sec:introduction},
we can parametrize our ignorance
of the ultraviolet boundary data by drawing these unknown
Wilson coefficients from one or more
suitable statistical distributions.

Any deformation of the throat must satisfy~\eqref{eq:Phi-minus-eq}.
We work to leading order in perturbations.
(For details of the approximation scheme being used we
refer to the original literature~\cite{Baumann:2010sx}.)
The local terms and
Ricci scalar generate only the
additive contributions
described above.%
    \footnote{This is not true in general.
    As explained in Ref.~\cite{Baumann:2010sx},
    the effect of the Ricci term is to dress
    each term in the potential with
    higher powers of $x$.
    However, we will truncate
    the brane potential
    before the first of these dressed
    terms appears. See
    the discussion on p.~\pageref{page:truncation}.
    Therefore, for our analysis, it
    suffices to add the Coulomb term and
    mass term to the potential obtained
    from Eq.~\eqref{eq:simple-Phi-minus-eq}.}
The equation to be solved is therefore
\begin{equation}
    \label{eq:simple-Phi-minus-eq}
    \grad_0^2 \Phi_- = \frac{\gstring}{96} |\Lambda|^2 ,
\end{equation}
where $\grad_0^2$ is the unperturbed conifold Laplacian.
Notice that
to solve~\eqref{eq:simple-Phi-minus-eq}
we do not need to know the behaviour of the remaining supergravity fields
except for the dilaton that determines $\gstring$.
Its general solution consists of a particular integral
(or `flux term')
$\PhiPI$
supported by the source term $\gstring |\Lambda|^2/96$
plus a complementary function
(or `homogeneous term')
$\PhiCF$
that satisfies the
homogeneous equation. The complete potential is
therefore\label{glossary:VCF}\label{glossary:VPI}
\begin{equation}
	V = \VCoulomb + \Vmass + \VCF + \VPI ,
\end{equation}
where $\VCF$ and $\VPI$ are the potential terms
generated by $\PhiCF$ and $\PhiPI$, respectively.
We describe their
construction
in~{\S\S}\ref{sec:harmonic-analysis}--\ref{sec:flux-contributions}
below.

\subsection{Harmonic analysis on the conifold}
\label{sec:harmonic-analysis}

\para{The Peter--Weyl theorem}
Both $\VCF$ and $\VPI$ can be analysed
using the methods of harmonic
analysis on Lie groups.
According to the Peter--Weyl theorem,
an orthonormal basis for square-integrable
functions
on a compact Lie group $G$ is furnished by the
matrix coefficients $D^\rep{\mu}_{mm'}$,\label{glossary:WignerD}
summed over all
unitary irreducible representations\label{glossary:replabel}
$\rep{\mu}$~\cite{Peter1927}.
These are defined to
satisfy
\begin{equation}
    D^\rep{\mu}_{mm'}(g) \equiv
    \langle m | \rho(g) | m' \rangle
    ,
\end{equation}
where $|m\rangle$ labels a basis for the representation
$\rep{\mu}$
and $\rho$ is its representation map.

Specifically, for
a square-integrable function $\Phi$ and
$g \in G$,
the Peter--Weyl theorem guarantees
that $\Phi$ can be represented
as the sum
\begin{subequations}
\begin{equation}
    \label{eq:PeterWeyl}
    \Phi(g) = \sum_{\rep{\mu}} \sum_{mm'} c^\rep{\mu}_{mm'} D^\rep{\mu}_{mm'}(g) ,
\end{equation}
where
$c^\rep{\mu}_{mm'}$ are coefficients depending
on $\Phi$.
Hence, the
$D^\rep{\mu}_{mm'}$ function as
harmonics of $G$ in a sense analogous to
Fourier analysis.
Observe that each representation occurs
in~\eqref{eq:PeterWeyl}
with multiplicity equal to its
dimension.
Although we will not need this refinement,
if $\Phi$ transforms in an
irreducible higher-dimensional representation
$\rep{\nu}$ of
$G$, in the sense
\begin{equation}
    \Phi_m^\rep{\nu}(g' \cdot g)
    =
    \sum_{m'} D_{mm'}^\rep{\nu}(g')
    \Phi^\rep{\nu}_{m'}(g) ,
    \label{eq:peter-weyl-higher}
\end{equation}
then the expansion~\eqref{eq:PeterWeyl} is shortened and only the
$\rep{\nu}$ representation is present, with multiplicity one.
For further details
see Salam \& Strathdee~\cite{Salam:1981xd},
who explained the application of~\eqref{eq:PeterWeyl}
to the typical case where $\Phi$ is a supergravity field transforming
in some nontrivial representation of the tangent space $SO(1,3)$ symmetry.

Our interest lies in the case where $\Phi$ is a spacetime
scalar and $G$ is the coset
$T^{1,1} = SU(2) \times SU(2) / U(1)$ described above.
In this situation~\eqref{eq:PeterWeyl} continues to apply, with
\begin{equation}
    \Phi[ L(g) ]
    =
    \sum_\rep{\mu}
    \sum_{mm'}
    c_{mm'}^\rep{\mu}
    D^\rep{\mu}_{mm'}[ L(g) ] ,
    \label{eq:peter-weyl-coset}
\end{equation}
\end{subequations}
where $L(g)$ is the coset representative of $g$.\label{glossary:cosetL}

\para{Application to $T^{1,1}$}
The unitary irreducible representations of $SU(2)$
are labelled
by their spin $\ell$.
The corresponding
matrix coefficients are given by Wigner's
\emph{Darmstellung}
or
$D$-matrix,
\begin{equation}
    D^\ell_{mm'}(\phi, \theta, \vartheta)
    \equiv
    \langle \ell m |
    \e{-\im \phi T_z}
    \e{-\im \theta T_y}
    \e{-\im \vartheta T_z}
    | \ell m' \rangle , 
    \label{eq:WignerDMatrix}
\end{equation}
where
the $T_i$ are generators of $SU(2)$,
$\{ \theta, \phi, \vartheta \}$
are corresponding Euler angles,
and $-\ell \leq m, m' \leq \ell$.
Representations of $SU(2) \times SU(2)$ are
built from the tensor product of a pair
of representations
of spin $\ell_1$, $\ell_2$
associated with the left- and right-hand $SU(2)$ factors.
We distinguish these factors using the labels
$i=1,2$, respectively.
It follows that
the corresponding harmonics are
\begin{equation}
	\Eharmonic^{\ell_1, \ell_2}_{m_1, n_1, m_2, n_2}
	(\theta_1, \phi_1, \vartheta_1; \theta_2, \phi_2, \vartheta_2)
	\equiv
	\mathcal{N}
    D^{\ell_1}_{m_1, n_1}(\phi_1, \theta_1, \vartheta_1)
    D^{\ell_2}_{m_2, n_2}(\phi_2, \theta_2, \vartheta_2)
    \label{eq:su2su2-harmonics}
\end{equation}
where
the quantum numbers $\ell_i$, $m_i$, $n_i$
satisfy the usual constraints for representations 
of $SU(2)$.
The prefactor $\mathcal{N}$ is a normalization to be determined.

In terms of these Euler angles,
$T^{1,1}$ can be embedded in $SU(2) \times SU(2)$
as
a hypersurface $\Sigma$ satisfying
$\vartheta_1 + \vartheta_2 = \text{const}$.
(See the discussion in
footnote~\ref{footnote:T11} on p.~\pageref{footnote:T11},
and the explicit discussion given by
Gubser~\cite{Gubser:1998vd}.)
Eq.~\eqref{eq:peter-weyl-coset}
shows that the harmonics on $T^{1,1}$
follow from~\eqref{eq:su2su2-harmonics}
by restriction to suitable coset representatives,
and
therefore we must project out dependence
on $\vartheta_1 + \vartheta_2$.
The representatives are
labelled by
$\psi = (\vartheta_1 - \vartheta_2)/2$,
where $0 \leq \psi < 4\pi$.
To obtain the correct projection,
note that the
Wigner $D$-matrix can be expressed
in terms of the  `little' $d$-matrix,\label{glossary:Wignerd}
defined by
\begin{equation}
	D^\ell_{mn}(\phi, \theta, \vartheta)
	\equiv
	\e{-\im m \phi} d^\ell_{mn}(\theta) \e{-\im n \vartheta} .	
\end{equation}
Therefore we must choose $n_1 = - n_2$.
Note that $d^\ell_{mn}(\theta)$ is real.

We write $n_1 = - n_2 = R/2$. After making a parity inversion
on the $i=2$ sphere, the metric on $\Sigma$
can be brought to the canonical
form~\eqref{metric_line}.
Using the
transformation rule
$d^\ell_{-m,-n}(\theta) =
d^\ell_{mn}(-\theta)$
we find that
the harmonics\label{glossary:Eharmonic} can be written%
    \footnote{Ceresole et al.\ define a `scalar harmonic
    condition', which in our language can be written
    $m_1 = R/2$, $m_2 = -R/2$~\cite{Ceresole:1999zs,Ceresole:1999ht}.
    Harmonics satisfying this conditions
    depend only on $\Delta \phi = \phi_1 - \phi_2$,
    and not $\phi_1$ or $\phi_2$ separately.
    They are
    `scalar' in the sense that they are uncharged
    under the $U(1)_H$ divisor of $T^{1,1}$.
    Notice, however, that this condition is immaterial
    for the expansion of a typical \emph{spacetime} scalar
    such as~\eqref{eq:PeterWeyl}, which contains representations
    of all dimensions, not just `scalar' representations in the
    sense of Ceresole et al.}
\begin{equation}
	\Eharmonic^{\ell_1, \ell_2}_{m_1, m_2, R/2}
	(\theta_1, \phi_1, \theta_2, \phi_2, \psi)
	=
	\mathcal{N}'
	\exp
	\im
	\bigg(
		\frac{R}{2} \psi
		+
		\sum_i m_i \phi_i
	\bigg)
	d^{\ell_1}_{m_1, R/2}(\theta_1)
	d^{\ell_2}_{m_2, R/2}(\theta_2) ,
	\label{eq:harmonics-little-d}
\end{equation}
where $\mathcal{N}'$ is an adjusted normalization.
To repeat,
the properties of the quantum numbers follow from
the selection rules for representations of $SU(2)$, viz.,
\begin{itemize}[noitemsep]
    \item $\ell_1$ and $\ell_2$ are nonnegative and
    \emph{either} both integers
    \emph{or} both half-integers;
    \item $m_1 \in \{ -\ell_1, \ldots, \ell_1 \}$ and $m_2 \in \{ -\ell_2, \ldots, \ell_2 \}$; and
    \item $R/2 \in \{ -l, \ldots, l \}$ where $l = \min(\ell_1, \ell_2)$.
\end{itemize}

This analysis clearly exhibits the Lie group structure
underlying the harmonics.
For practical calculations, however, we require explicit
formulae for the $\Eharmonic_{LM}$.
Here we borrow
the economical notation
of Baumann et al.~\cite{Baumann:2006th}
in which the harmonics are distinguished by multi-indices
$L = (\ell_1, \ell_2, R)$\label{glossary:L}
and
$M = (m_1, m_2)$.\label{glossary:M}
Specifically
$L$ labels the representation of the harmonic
under the isometry group
$SU(2) \times SU(2) \times U(1)$,
and $M$ runs over the corresponding representation space.

\para{Explicit formulae}
Explicit formulae for the $\Eharmonic_{LM}$
were
obtained by
Gubser
using a direct analysis of their governing
differential equations~\cite{Gubser:1998vd}.
Later, a more extensive discussion was given by
Ceresole et al.~\cite{Ceresole:1999zs,Ceresole:1999ht,Ceresole:1999rq},
who used algebraic methods based on group theory~\cite{Salam:1981xd}.
The details were summarized by Baumann et al.~\cite{Baumann:2006th}.
Expressions for the zero-modes on the conifold
were given in
Ref.~\cite{Baumann:2010sx};
see also Ref.~\cite{Gandhi:2011id}.
The expression~\eqref{eq:harmonics-little-d}
in terms of the little $d$-matrix
was first given in Ref.~\cite{McAllister:2016vzi}.%
    \footnote{Explicit formulae for the $d^\ell_{mn}$
    were given by Wigner~\cite{1959142}.
    Their generating function
    was computed by Schwinger~\cite{osti_4389568}.
	The connexion between harmonics
	on $T^{1,1}$ and Wigner's little $d$-matrix
	was apparently not noticed prior to Ref.~\cite{McAllister:2016vzi}.}
Here we briefly collect these details in a unified notation.

The
$\Eharmonic_{LM}$ are eigenfunctions of the Laplacian
on $T^{1,1}$
with eigenvalue $\Lambda_L$,\label{glossary:LambdaL}
\begin{equation}\label{Eigenvalue_problem}
    \nabla^2_{T^{1,1}} \Eharmonic_{LM}(\Psi)
    =
    -\Lambda_{L} \Eharmonic_{LM}(\Psi) .
\end{equation}
The eigenvalue spectrum depends on a sum of quadratic Casimir
invariants for the representations specified by $L$,
but not the representation-space labels $M$.
It satisfies
\begin{equation}
    \Lambda_{L} = 6
    \Bigg(
        \ell_1(\ell_1+1) + \ell_2(\ell_2+1) - \frac{R^2}{8}
    \Bigg) .
\end{equation}
The necessary nonsingular solutions for
$d^{\ell}_{m,R/2}(\theta)$
are%
	\footnote{These formulae match those quoted in Ref.~\cite{Gubser:1998vd}.
	The four cases given there can be related in pairs using
	an Euler transformation of the hypergeometric function.}
\begin{subequations}
    \begin{equation}
    \begin{split}
        \label{hypa}
		\left(
			\frac{2\ell+1}{2}
		\right)^{1/2}
		d^{\ell}_{m,R/2}(\theta)
		& =
        \JupBare
        \\
        & \equiv
		N^\Upsilon_{LM}
        \left( \sin \theta \right)^{m} 
        \left( \cot \frac{\theta}{2} \right)^{\frac{R}{2}}
        {_2F_1} \left(
            \begin{array}{cc}
                -\ell + m, & 1 + \ell + m \\
                \multicolumn{2}{c}{1 + m - R/2}
            \end{array}
            \; \Big| \; \sin^2 \frac{\theta}{2}
        \right),
    \end{split}
    \end{equation}
    if $m \geq R/2$, and
    \begin{equation}
    \begin{split}
        \label{hypb}
		\left(
			\frac{2\ell+1}{2}
		\right)^{1/2}
		d^\ell_{m,R/2}(\theta)
		& =
        \JomBare
        \\
        & \equiv
		N^\Omega_{LM}
        \left( \sin \theta \right)^{\frac{R}{2}}
        \left( \cot \frac{\theta}{2} \right)^{m}
        {_2F_1} \left(
            \begin{array}{cc}
                -\ell + R/2, & 1 + \ell + R/2 \\
                \multicolumn{2}{c}{1 - m + R/2}
            \end{array}
            \; \Big| \; \sin^2 \frac{\theta}{2}
        \right),
    \end{split}
    \end{equation}
\end{subequations}
if $m < R/2$.\label{glossary:Jup}\label{glossary:Jom}
As explained above,
both solutions are real.
Here, $_2F_1(a, b; c \,|\, z)$
is the Gauss hypergeometric function,
and we have introduced mode functions
$\JBare$
to match the notation of Ref.~\cite{Baumann:2006th}.
It follows from a Sturm--Liouville argument
that the $\JBare$
are orthogonal
for fixed $m$, $R$
in the measure $\sin \theta \, \d \theta$.
Their normalization is fixed
by adjusting
$N^\Upsilon_{LM}$,
$N^\Omega_{LM}$
so that
\begin{equation}
	\int_0^\pi
	\d \theta
	\;
	\sin \theta
	\,
	J_{\ell, m, R}(\theta)
	J_{\ell', m, R}(\theta)
	=
	\delta_{\ell\ell'} .
\end{equation}
If we choose the normalization of the
$\Eharmonic_{LM}$ so that
\begin{equation}
    \Eharmonic_{LM}(\Psi) = \J{1} \, \J{2} \,
    \exp \im \Big(
        m_1 \phi_1 + m_2 \phi_2 + \frac{R}{2} \psi
    \Big),
    \label{eqn:Ylm}
\end{equation}
then the $\Eharmonic_{LM}$ satisfy the
larger orthogonality
condition
\begin{equation}
    \label{eq:harmonic-complete}
    \int \d^5 \Psi \, (-\det G_{T^{1,1}})^{1/2} \;
    \Eharmonic_{LM}(\Psi) \Eharmonic^\ast_{L'M'}(\Psi)
    =
    \delta_{LL'} \delta_{MM'} ,    
\end{equation}
where $\Psi$ stands schematically for the five angles
on $T^{1,1}$,
and $G_{T^{1,1}}$ is its metric.

\para{Zero-modes on the conifold}
Each $T^{1,1}$ harmonic can be promoted to a zero-mode
of the conifold---that is, a solution of the
homogeneous equation
$\grad_0^2 \Phi_- = 0$.
A simple calculation shows that
if $\Eharmonic_{LM}$ is a harmonic on $T^{1,1}$
with eigenvalue $\Lambda_L$, then~\cite{Baumann:2006th}
\begin{equation}
	f_{LM}(r, \Psi)
	= r^{\Delta(L)} \Eharmonic_{LM}(\Psi)
\end{equation}
is a zero-mode of $\grad^2_0$, where\label{glossary:DeltaL}
\begin{equation}
    \label{eq:basic-radial-scaling}
	\Delta(L) \equiv -2 \pm 	\sqrt{4 + \Lambda_L} .
\end{equation}
Therefore the complementary function for $\Phi_-$
can be expressed as a linear combination
of these zero-modes.
It will make a contribution to the brane potential of the form
\begin{equation}\label{Vhomog}
	\VCF(x, \Psi)
    = \mu^4 \sum_{LM}C_{LM} x^{\Delta(L)} \Eharmonic_{LM}(\Psi)
    +
    \text{c.c.} ,
\end{equation}
where the scale $\mu^4$,\label{glossary:CLM}
defined in Eq.~\eqref{eq:mu-def},
has been inserted by hand to account
for the factor of the tension $T_3$
appearing in the dictionary between
$\Phi_-$ and the brane potential $V$.
The $\Clm$
are taken to be unknown
(complex)
Wilson coefficients,
and $\Delta(L)$ determines the radial scaling
of each operator.
In the region
$|x| \ll 1$,
away from the ultraviolet end of the throat,
only a few operators of lowest
scaling dimension will be relevant,
as anticipated 
in the discussion above Eq.~\eqref{eq:simple-Phi-minus-eq}.

Although we would like to keep as many operators as
possible, there are practical limitations.
As we increase the number of terms that are retained,
we incur corresponding costs
from the automated symbolic manipulations
carried out by the {\CppTransport} and {\PyTransport}
platforms, and also in the numerical solution
of the transport equations.
We will see that it is already challenging to solve for the
three-point function in a model of this complexity, so it is
not realistic to attempt to retain operators of very high order.
On the other hand, at a minimum,
we would like to retain operators
that contribute significantly to the effective
cubic couplings.
If these are large
(as suggested by the parametric estimates given
in Ref.~\cite{McAllister:2012am})
they potentially source a large
bispectrum from the
quasi-single-field `QSFI'
mechanism~\cite{Chen:2009we,Chen:2009zp,Baumann:2011su,Gong:2013sma}.
It follows that aggressive truncation of the potential
risks serious misprediction for a key observable of the model.

In Table~\ref{Table:eig_table} we tabulate the
lowest-lying zero modes of $\grad_0^2$ with radial
scaling dimensions that satisfy
$\Delta(L) \leq \DeltaMax = 3.8$.\label{page:truncation}
(This choice was made by
Agarwal et al.\ and Dias et al.~\cite{Agarwal:2011wm,Dias:2012nf}.
McAllister et al.~\cite{McAllister:2016vzi} did not
give their truncation explicitly, but apparently
used the same prescription.)
We use this truncation
in~{\S}\ref{sec:experimental-procedure}
to construct our primary statistical ensemble,
giving sufficient headroom
to capture large QSFI effects.
At this level
there are eleven contributing representations $L = (\ell_1, \ell_2, R)$.
However, it should be remembered that the number of
\emph{modes} is rather larger because
the dimension of these
representations lies between $3$ (for $\ell_1 = 1$, $\ell_2 = 0$ and
vice-versa)
and $9$ (for $\ell_1 = \ell_2 = 1$),
and
as explained in~{\S}\ref{sec:harmonic-analysis}
each representation contributes with multiplicity
equal to its dimension.
There is a unique constant mode with $\ell_1 = \ell_2 = R = 0$
that we omit; it is proportional to the unit operator
and
merely renormalizes the vacuum energy.
Therefore its effect can be absorbed
into $V_0$.
(However, see Table~\ref{Table:seriesIIITable}.)
\ctable[
    label = Table:eig_table,
    caption = {Scalar zero-modes of the conifold Laplacian $\grad_0^2$.
    Tabulated values are the radial scaling dimension $\Delta(L)$;
    $SU(2) \times SU(2)$ representation labels $(\ellI, \ellII)$;
    the $U(1)$ representation label $R$;
    the mode normalization constant for
    $\JBare$;
    and the dimension of the representation.
    We limit inclusion to modes with lowest-lying radial scaling
    dimensions $\Delta(L) \leq 3.8$.
    This gives a total of 73 different modes, all of which occur
    with fixed multiplicities as described in~{\S}\ref{sec:harmonic-analysis}.}
]{LSRRRRR}{}
{
    \toprule
    \multicolumn{1}{C}{\Delta(L)} &
    \multicolumn{1}{c}{decimal scaling} &
    \multicolumn{1}{C}{\ellI} &
    \multicolumn{1}{C}{\ellII} &
    \multicolumn{1}{C}{R} &
    \multicolumn{1}{c}{normalization $\times \ \pi^{3/2}$} &
    \multicolumn{1}{c}{dimension} \NN
    \midrule
    3/2           &  1.5    & 1/2   &  1/2    &  -1  &  3\sqrt{3}/{2}   & 4 \NN
    3/2           &  1.5    & 1/2   &  1/2    &  1   &  3\sqrt{3}/{4}   & 4 \NN
    2             &  2.0    & 1     &  0      &  0   &  9/{4}           & 3 \NN
    2             &  2.0    & 0     &  1      &  0   &  9/{4}           & 3 \NN
    3             &  3.0    & 1     &  1      &  2   &  9\sqrt{3}/{16}  & 9 \NN
    3             &  3.0    & 1     &  1      &  -2  &  9\sqrt{3}/{4}   & 9 \NN
    2\sqrt{7} - 2 &  3.2915 & 1     &  1      &  0   &  9\sqrt{3}/{4}   & 9 \NN
    7/2           &  3.5    & 1/2   &  3/2    &  -1  &  3\sqrt{6}/{4}   & 8 \NN
    7/2           &  3.5    & 1/2   &  3/2    &  1   &  3\sqrt{6}/{2}   & 8 \NN
    7/2           &  3.5    & 3/2   &  1/2    &  -1  &  3\sqrt{6}/{2}   & 8 \NN
    7/2           &  3.5    & 3/2   &  1/2    &  1   &  3\sqrt{6}/{4}   & 8 \NN
    \midrule
                  &         &       &         &      &                  & \multicolumn{1}{r}{\semibold{73}} \NN
    \bottomrule
}

\para{Reality properties of the zero-modes}
Because the `little' $d$-matrices
(or equivalently, the $J$ mode functions)
are real,
complex conjugation simply reverses
the sign of the labels $R$, $m_1$ and $m_2$.
This follows from Eqs.~\eqref{hypa}--\eqref{hypb}
after making an Euler transformation
of the hypergeometric function.
Therefore
modes with $R < 0$
in Table~\ref{Table:eig_table}
are related to those with $R > 0$
by complex conjugation.
A special case of this observation
is that modes with $R = m_1 = m_2 = 0$ are
purely real.

In the interest of clarity, we note that
the sum in~\eqref{Vhomog}
is unrestricted and includes representations with
both signs of $R$.
Moreover, the coefficients $\Clm$
are taken to be independent for each set
of quantum numbers $L$, $M$.
First,
consider a complex mode $\Eharmonic_{LM}$
for which at least one of
the labels $R$, $m_1$ and $m_2$ is nonzero.
For convenience we define
$-L = (\ell_1, \ell_2, -R)$
and $-M = (-m_1, -m_2)$.
The contribution of $\Eharmonic_{LM}$ to $\VCF$
can be written as a sum of $\Eharmonic_{LM}$
and $\Eharmonic_{-L-M}$,
\begin{equation}
    \VCF \supseteq
    \mu^4 x^{\Delta(L)}
    \bigg(
        C_{LM} \Eharmonic_{LM}
        +
        C_{LM}^\ast \Eharmonic_{LM}^\ast
    \bigg)
    =
    \mu^4 x^{\Delta(L)}
    \bigg(
        C_{LM} \Eharmonic_{LM}
        +
        C_{LM}^\ast \Eharmonic_{-L-M}
    \bigg)
    ,
\end{equation}
where the notation $\supseteq$
denotes that $\VCF$ contains the indicated
contribution, together with other contributions
that have not been written.
Meanwhile a similar relation holds for $\Eharmonic_{-L-M}$.
In combination they yield
\begin{equation}
    \VCF \supseteq
    2 \mu^4 x^{\Delta(L)}
    \Re
    \Big(
        D_{LM} \Eharmonic_{LM}
    \Big)
    = 2 \mu^4 x^{\Delta(L)}
    \bigg(
        \Re(D_{LM})
        \Re(\Eharmonic_{LM})
        -
        \Im(D_{LM})
        \Im(\Eharmonic_{LM})
    \bigg)
    ,
    \label{eq:VCFRealPart}
\end{equation}
where $D_{LM} \equiv C_{LM} + C_{-L-M}^\ast$.\label{glossary:DLM}
It follows that we can equivalently restrict the sum
in~\eqref{Vhomog} to $L = (\ell_1, \ell_2, R)$ with $R>0$
provided we adjust the summand
to match~\eqref{eq:VCFRealPart}.
Later we will take the real and imaginary parts of
$C_{LM}$ to be random variables
drawn from some distribution $X$.
It is important to note that the real and imaginary
parts of $D_{LM}$ should then be drawn from
the appropriate distributions for the sum and
difference of two $X$-distributed
random variables, respectively.

Second, consider a real mode with $R = m_1 = m_2 = 0$.
This depends only on (at least one of) $\theta_1$, $\theta_2$,
unless $\ell_1 = \ell_2 = 0$.
In this case it is a constant and is excluded from the sum as
described above.
Therefore the contribution from this mode to
$\VCF$ is
\begin{equation}
    \VCF \supseteq
    2 \mu^4 x^{\Delta(L)} C_{LM} \Eharmonic_{LM}
\end{equation}
where now $C_{LM}$ is real
with numerical value drawn from the distribution $X$.

\subsection{Flux contributions}
\label{sec:flux-contributions}
The final
step is to include the `particular integral'
for~\eqref{eq:simple-Phi-minus-eq}---that is,
the perturbation to $\Phi_-$ sourced by
the square of the 3-form flux $|\Lambda|^2$.
The allowed contributions to $\Lambda$
were enumerated by Baumann et al.~\cite{Baumann:2010sx}
and fall into three distinct `series', distinguished
by the recipe for building the 3-form $\Lambda$
from a seed zero-mode drawn from Table~\ref{Table:eig_table}.
Depending on the details of the recipe, the resulting
fluxes may exhibit an adjusted
radial scaling dimension $\Delta$
or an adjusted $U(1)$ quantum number $R$.
We will not require explicit formulae for the fluxes,
and therefore refer to the literature for details of their
construction.

\para{Flux contributions to $\Phi_-$}
In Tables~\ref{Table:seriesITable}--\ref{Table:seriesIIITable}
we tabulate the required seed representations,
and adjusted radial scaling dimensions,
associated with these flux
series, labelled Series I, II and III in the notation of
Baumann et al.~\cite{Baumann:2010sx}.
We limit attention to representations for which the adjusted
scaling dimension $\Delta$\label{glossary:DeltaNoL} satisfies
$2 < \Delta \leq 5.8$.
At the upper limit,
we will see later that fluxes with $\Delta \leq 5.8$
are sufficient to capture
all
contributions to $\Phi_-$ with radial scaling dimension
$\delta \leq \DeltaMax = 3.8$.
Notice that in this section
we carefully distinguish the adjusted radial scaling dimension
$\Delta$ of the fluxes, and the dimension
$\delta$ of the contribution that is ultimately produced
in the potential.
At the lower limit,
fluxes with $\Delta \leq 2$ do not couple to a
probe $\brane{3}$ brane~\cite{Baumann:2010sx}
and can be discarded.
When summed over all three series there are 197 flux modes.
Clearly the resulting $\brane{3}$-brane potential is
very complicated.
\ctable[
    label = Table:seriesITable,
    caption = {\small Series I fluxes. Tabulated values are the
    adjusted radial
    scaling dimension $\Delta = 1 + \Delta_f$, where $f$ is the scalar seed mode;
    the $SU(2) \times SU(2) \times U(1)$ quantum numbers
    $\ellI$, $\ellII$, $R$ for $f$;
    dimension of the representation;
    and whether the flux is chiral,
    defined to mean $\ell_1 = \ell_2 = R/2$
    for the seed mode.
    We include the 109 modes with radial scaling dimensions
    that satisfy $2 < \Delta \leq 5.8$.},
    width = 0.9\textwidth
]{LSRRRRY}{
    \tnote[a]{The representations with
    $(\ell_1, \ell_2, R) = (1, 0, 0), (0, 1, 0)$ are absent, even though
    they appear in Table~\ref{Table:eig_table}.
    These modes were excluded by
    Baumann et al.~\cite{Baumann:2010sx} without comment,
    apparently because they are projected out of the spectrum by
    the field equation for $G_3$.
    These seed modes \emph{do} appear for the
    Series II and Series III fluxes
    listed in Tables~\ref{Table:seriesIITable}
    and~\ref{Table:seriesIIITable}. In Ref.~\cite{Dias:2012nf}
    they were accidentally excluded from Series II and III.
    In principle this could influence outcomes from the model,
    but see footnote~\ref{foot:DiasClebsch} on p.~\pageref{foot:DiasClebsch}.
    We would like to thank Mafalda Dias for very helpful
    correspondence on these issues.}
}
{
    \toprule
    \multicolumn{1}{C}{\Delta} &
    \multicolumn{1}{c}{decimal scaling} &
    \multicolumn{1}{C}{\ellI} &
    \multicolumn{1}{C}{\ellII} &
    \multicolumn{1}{C}{R} &
    \multicolumn{1}{c}{dimension} &
    type \NN
    \midrule
    5/2             & 2.5     & 1/2     & 1/2     & -1    & 4   & non-chiral    \NN
    5/2             & 2.5     & 1/2     & 1/2     & 1     & 4   & chiral        \NN
    4\tmark[a]      & 4.0     & 1       & 1       & -2    & 9   & non-chiral    \NN
    4               & 4.0     & 1       & 1       & 2     & 9   & chiral        \NN
    \sqrt{28}-1     & 4.2915  & 1       & 1       & 0     & 9   & non-chiral    \NN
    9/2             & 4.5     & 1/2     & 3/2     & -1    & 8   & non-chiral    \NN
    9/2             & 4.5     & 1/2     & 3/2     & 1     & 8   & non-chiral    \NN
    9/2             & 4.5     & 3/2     & 1/2     & -1    & 8   & non-chiral    \NN
    9/2             & 4.5     & 3/2     & 1/2     & 1     & 8   & non-chiral    \NN
    2\sqrt{10}-1    & 5.3246  & 2       & 0       & 0     & 5   & non-chiral    \NN
    2\sqrt{10}-1    & 5.3246  & 0       & 2       & 0     & 5   & non-chiral    \NN
    11/2            & 5.5     & 3/2     & 3/2     & -3    & 16  & non-chiral    \NN
    11/2            & 5.5     & 3/2     & 3/2     & 3     & 16  & chiral        \NN
    \midrule
                    &         &         &         &       & \multicolumn{1}{r}{\semibold{109}} & \NN
    \bottomrule
}
 
\ctable[
    label = Table:seriesIITable,
    caption = {\small Series II fluxes. Tabulated values are the
    adjusted radial
    scaling dimension $\Delta = 2 + \Delta_f$, where $f$ is the scalar seed mode;
    the $SU(2) \times SU(2) \times U(1)$ quantum numbers
    $\ellI$, $\ellII$, $R$ for $f$;
    dimension of the representation;
    and whether the flux is chiral,
    defined to mean $\ell_1 = \ell_2 = R/2$
    for the seed mode.
    We include the 73 modes with radial scaling dimensions
    that satisfy $2 < \Delta \leq 5.8$.},
    width = 0.9\textwidth
]{LSRRRRY}{}
{
    \toprule
    \multicolumn{1}{C}{\Delta} &
    \multicolumn{1}{c}{decimal scaling} &
    \multicolumn{1}{C}{\ellI} &
    \multicolumn{1}{C}{\ellII} &
    \multicolumn{1}{C}{R} &
    \multicolumn{1}{c}{dimension} &
    type \NN
    \midrule
    7/2             & 3.5     & 1/2     & 1/2     & -1    & 4   & non-chiral   \NN
    7/2             & 3.5     & 1/2     & 1/2     & 1     & 4   & chiral       \NN
    4               & 4.0     & 0       & 1       & 0     & 3   & non-chiral   \NN
    4               & 4.0     & 1       & 0       & 0     & 3   & non-chiral   \NN
    5               & 5.0     & 1       & 1       & -2    & 9   & non-chiral   \NN
    5               & 5.0     & 1       & 1       & 2     & 9   & chiral       \NN
    2\sqrt{7}       & 5.2915  & 1       & 1       & 0     & 9   & non-chiral   \NN
    11/2            & 5.5     & 1/2     & 3/2     & -1    & 8   & non-chiral   \NN
    11/2            & 5.5     & 1/2     & 3/2     & 1     & 8   & non-chiral   \NN
    11/2            & 5.5     & 3/2     & 1/2     & -1    & 8   & non-chiral   \NN
    11/2            & 5.5     & 3/2     & 1/2     & 1     & 8   & non-chiral   \NN
    \midrule
                    &         &         &         &       & \multicolumn{1}{r}{\semibold{73}} & \NN
    \bottomrule
}

\ctable[
    label = Table:seriesIIITable,
    caption = {\small Series III fluxes. Tabulated values are the
    adjusted radial
    scaling dimension $\Delta = 3 + \Delta_f$, where $f$ is the scalar seed mode;
    the $SU(2) \times SU(2) \times U(1)$ quantum numbers
    $\ellI$, $\ellII$, $R$ for $f$;
    dimension of the representation;
    and whether the flux is chiral,
    defined to mean $\ell_1 = \ell_2 = R/2$
    for the seed mode.
    We include the 15 modes with radial scaling dimensions
    that satisfy $2 < \Delta \leq 5.8$.},
    width = 0.9\textwidth
]{LSRRRRY}{
    \tnote[a]{Considered as a zero-mode of the scalar Laplacian $\grad_0^2$,
    this mode is a constant.
    It was excluded from Table~\ref{Table:eig_table} because it does not
    contribute to $\VCF$ as explained in the main text.
    It also does not appear in the Series I or II fluxes,
    Tables~\ref{Table:seriesITable}--\ref{Table:seriesIITable},
    because these fluxes are all built from derivatives of the seed $f$.
    However, Series III fluxes include contributions from
    $f$ without differentiation, so this mode can yield a nontrivial flux~\cite{Baumann:2010sx}.}
}
{
    \toprule
    \multicolumn{1}{C}{\Delta} &
    \multicolumn{1}{c}{decimal scaling} &
    \multicolumn{1}{C}{\ellI} &
    \multicolumn{1}{C}{\ellII} &
    \multicolumn{1}{C}{R} &
    \multicolumn{1}{c}{dimension} &
    type \NN
    \midrule
    3\tmark[a]      & 3.0     & 0       & 0       & 0     & 1  & chiral  \NN
    9/2             & 4.5     & 1/2     & 1/2     & 1     & 4  & chiral  \NN
    9/2             & 4.5     & 1/2     & 1/2     & -1    & 4  & non-chiral  \NN
    5               & 5.0     & 0       & 1       & 0     & 3  & non-chiral  \NN
    5               & 5.0     & 1       & 0       & 0     & 3  & non-chiral  \NN
    \midrule
                    &         &         &         &       & \multicolumn{1}{r}{\semibold{15}} & \NN
    \bottomrule
}

The possible contributions to $|\Lambda|^2$ involve
combinations of any two fluxes drawn
from Tables~\ref{Table:seriesITable}--\ref{Table:seriesIIITable},
with the proviso that two chiral modes
can combine only if they belong to the same flux series~\cite{Baumann:2010sx}.
A flux mode is said to be chiral if and only if its
seed scalar mode is chiral in the
sense $\ell_1 = \ell_2 = R/2$~\cite{Baumann:2010sx}.

The contribution to $\Phi_-$ from any pair of fluxes
can be found using the method of Green's functions.
Specifically,
\begin{equation}
	\Phi_-(x) = \frac{\gstring}{96} \int \d^6 y \;
	(-\det G_{AB})^{1/2} \,
	\GreenFunction(x, y) |\Lambda(y)|^2 ,
	\label{eq:FluxesGreenFunction}
\end{equation}
where $G_{AB}$ is the metric~\eqref{eq:cone-metric}
on the conifold
and
$\GreenFunction(x,y)$ is the corresponding Green's function
obtained in Ref.~\cite{Baumann:2006th}.
It has the spectral representation\label{glossary:GreenFunction}
\begin{equation}
    \label{eq:GreenFunction}
	\GreenFunction(x, x') = \sum_{LM}
	\Eharmonic_{LM}(\Psi)
	\Eharmonic_{LM}^\ast(\Psi')
	g_L(r, r') ,
\end{equation}
where $x = (r, \Psi)$, $x' = (r', \Psi')$
are coordinates on the conifold,
and $\Psi$, $\Psi'$ represent the angles on $T^{1,1}$.
The kernel
$g_L(r, r')$ satisfies~\cite{Baumann:2006th}
\begin{equation}
	g_L(r, r') = - \frac{1}{2 \Delta(L) + 1}
	\left\{
		{
			\renewcommand{\arraystretch}{1.3}
			\begin{array}{l@{\hspace{5mm}}l}
				(r')^{-4} (r/r')^{\Delta(L)} & \rIR \lesssim r \leq r' \\
				r^{-4} (r'/r)^{\Delta(L)} & r' \leq r \lesssim \rUV
			\end{array}
		}
	\right.
	.
\end{equation}
As explained above, some of the fluxes listed in
Tables~\ref{Table:seriesITable}--\ref{Table:seriesIIITable}
have modified quantum numbers because they combine
with other ingredients.
For example (now passing to a K\"{a}hler description),
some flux series involve the holomorphic 3-form
$\Omega_{\mathsf{a}\mathsf{b}\mathsf{c}} = q \epsilon_{\mathsf{a}\mathsf{b}\mathsf{c}}$,
where the indices $\mathsf{a}$, $\mathsf{b}$, $\mathsf{c}$
run over complex coordinates on the conifold,
$\epsilon$ is the Levi--Civita tensor, and
$q^\ast q = (-\det G_{AB})^{1/2}$ is the determinant of the K\"{a}hler metric.
However, in $|\Lambda|^2$
these factors cancel 
with contractions involving the inverse metric.
Therefore the angular terms in~\eqref{eq:FluxesGreenFunction}
involve only a product of the seed modes
appearing in these tables.

\para{Radial profile}
The integral in~\eqref{eq:FluxesGreenFunction}
factorizes into an integral over the radius $r$
of the cone
and an integral over the angles on $T^{1,1}$.
First, consider the radial integral.
For the fluxes described in
Tables~\ref{Table:seriesITable}--\ref{Table:seriesIIITable},
radial dependence arises only
from their scaling dimensions.
Accordingly,
given
two fluxes $\Lambda_1$, $\Lambda_2$
and a fixed representation $L$ drawn from the sum in~\eqref{eq:GreenFunction},
the Green's function
produces a radial profile
\begin{equation}
	\text{radial profile}
	\propto
    \int_{\rIR}^{\rUV} \frac{\d r'}{r'} \; g_L(r, r') \Big( \frac{r'}{\rUV} \Big)^{\Delta_1 + \Delta_2} ,
    \label{eq:ProtoRadialPart}
\end{equation}
where the factor $1/r'$ is produced by combining
$(r')^5$ from the Jacobian $(-\det G_{AB})^{1/2}$
and $(r')^{-6}$ from three copies of the inverse
metric needed for the contractions implied by $|\Lambda|^2$.
There is an overall constant of proportionality that we
do not write explicitly.
The result is
\begin{equation}
	\text{radial profile}
    \propto
    \frac{1}{\rUV^4}
	\left(
		\alpha x^{\Delta_1 + \Delta_2 - 4}
		+ \beta x^{\Delta(L)}
		+ \gamma x^{-4 - \Delta(L)} \Big( \frac{\rIR}{\rUV} \Big)^{\Delta(L) + \Delta_1 + \Delta_2}
	\right)	
	,
\end{equation}
where $x = r/\rUV$
as above, and $\alpha$, $\beta$, $\gamma$ are roughly $\Or(1)$ numerical
coefficients.
The term involving the infrared regulator $\rIR$ is small provided
$\rIR / \rUV \ll 1$
and formally vanishes in the limit $\rIR \rightarrow 0$.
Therefore the integral does not accumulate
large contributions from the region $r \sim \rIR$
where we need a precise resolution of the conifold
singularity.
We assume this term is negligible, and it will be dropped in the following
discussion.

The remaining terms source radial profiles $\sim x^{\Delta(L)}$
and $\sim x^{\Delta_1 + \Delta_2 - 4}$.
The $x^{\Delta(L)}$ profile reproduces the radial scaling dimension
associated with the scalar zero-modes of Table~\ref{Table:eig_table}.
This term will modify the coefficients
associated with the $L$-representation
in $\VCF$, Eq.~\eqref{Vhomog}.
Fortunately, this is harmless: our premise is that we
cannot predict these coefficients, which already depend on
ultraviolet data.
Combining two unknown coefficients
merely yields another unknown coefficient.
The other term, scaling like
$x^{\Delta_1 + \Delta_2 - 4}$,
is new.
It will produce
contributions to the potential
involving modes in the $L$-representation,
but with a radial scaling dimension
$\Delta_{12} = \Delta_1 + \Delta_2 - 4$
\emph{different} to $\Delta(L)$.
Note that $\Delta_{12}$ is guaranteed to be positive because
$\Delta > 2$ for all flux modes that participate in the
cross product.
If this dimension falls below the truncation point
then such terms
should be retained.%
    \footnote{In Tables~\ref{Table:seriesITable}--\ref{Table:seriesIIITable}
    we retained terms with $2 < \Delta \leq 5.8$.
    Inspection of the formula
    for $\Delta_{12}$ shows
    that
    to capture contributions to $\Phi_-$ with $\Delta \leq 3.8$
    it is sufficient to consider fluxes with scaling dimension
    in this range.}

\para{Angular terms}
Now consider the angular part of~\eqref{eq:FluxesGreenFunction}.
We have already explained that factors coming from
copies of the
inverse metric in the contraction $|\Lambda|^2$ cancel with
normalization adjustments in the individual fluxes.
Therefore the integrand involves only
the combination
\begin{equation}
    \text{angular part}
    \propto
    \sum_{LM} \Eharmonic_{LM}(\Psi)
    \int \d^5 \Psi' \; (-\det G_{T^{1,1}})^{1/2} \,
    \Eharmonic_{LM}^\ast(\Psi')
    \Eharmonic_{L' M'}(\Psi')
    \Eharmonic_{L'' M''}^\ast(\Psi') .
    \label{eq:ProtoAngularPart}
\end{equation}
The modes
$\Eharmonic_{L' M'}$
and $\Eharmonic_{L'' M''}$
represent the quantum numbers of
the two flux modes contributing to $|\Lambda|^2$.
The two modes with quantum numbers $LM$
contribute to the sum
in the spectral representation of the Green's function,
Eq.~\eqref{eq:GreenFunction}.

The complex conjugation on
$\Eharmonic_{L'' M''}$
can be dropped without loss of generality,
because this merely reverses the labels
$L'' \rightarrow -L''$ and $M'' \rightarrow -M''$.
To build $\Phi_-$
we will sum $(L', M')$ and $(L'', M'')$
over all entries in
Tables~\ref{Table:seriesITable}--\ref{Table:seriesIIITable},
so this reversal is immaterial.
Moreover, since $\Eharmonic$ transforms as a pair
of $SU(2)$ representations,
the tensor product
$\Eharmonic_{L_1 M_1} \Eharmonic_{L_2 M_2}$ can be decomposed
as a direct sum of similar representations.
This follows from the same property of $SU(2)$, expressed
via Clebsch--Gordan coefficients.
Specifically, Wigner's `little' $d$-matrix satisfies
\begin{equation}
    d^{\ell}_{m n}(\beta)
    d^{\ell'}_{m' n'}(\beta)
    =
    \sum_{L = |\ell - \ell'|}^{\ell + \ell'}
    \langle \ell m, \ell' m' | L M \rangle
    \langle \ell n, \ell' n' | L N \rangle
    d^L_{MN}(\beta) ,
\end{equation}
where $M = m + m'$ and $N = n + n'$.
Here,
$\langle \ell m, \ell' m' | L M \rangle$
is an ordinary Clebsch--Gordan coefficient.
The $\JBare$ mode functions used to build the $\Eharmonic_{LM}$
harmonics are related to $d^\ell_{mn}$ via~\eqref{hypa}--\eqref{hypb}.
Therefore,
recalling $L = (\ell_1, \ell_2, R)$, $L' = (\ell_1', \ell_2', R')$,
$M = (m_1, m_2)$ and $M' = (m_1', m_2')$,
we conclude
\begin{equation}
\begin{split}
    \Eharmonic_{L M}(\Psi)
    \Eharmonic_{L' M'}(\Psi)
    =
    \sum_{\suml_1 = |\ell_1 - \ell'_1|}^{\ell_1 + \ell'_1}
    \sum_{\suml_2 = |\ell_2 - \ell'_2|}^{\ell_2 + \ell'_2}
    &
    \sqrt{
        \frac{(2\ell_1+1)(2\ell_1'+1)}{2(2\suml_1 + 1)}
        \frac{(2\ell_2+1)(2\ell_2'+1)}{2(2\suml_2 + 1)}
    }
    \\
    &
    \times
    \langle \ell_1 m_1, \ell_1' m_1' |\suml_1 \summ_1 \rangle
    \langle \ell_1 \frac{R}{2}, \ell_1' \frac{R'}{2} |
        \suml_1 \frac{\sumR}{2} \rangle
    \\
    &
    \times
    \langle \ell_2 m_2, \ell_2' m_2' | \suml_2 \summ_2 \rangle
    \langle \ell_2 \frac{R}{2}, \ell_2' \frac{R'}{2} |
        \suml_2 \frac{\sumR}{2} \rangle
    \Eharmonic_{\sumL \sumM}(\Psi) ,
\end{split}
\label{eq:EharmonicClebschGordan}
\end{equation}
where $\summ_1 = m_1 + m_1'$,\label{glossary:suml}\label{glossary:summ}
$\summ_2 = m_2 + m_2'$,
$\sumR = R + R'$,\label{glossary:sumR}
$\sumL = (\suml_1, \suml_2, \sumR)$\label{glossary:sumL}
and\label{glossary:sumM}
$\sumM = (\summ_1, \summ_2)$.%
    \footnote{\label{foot:DiasClebsch}Ref.~\cite{Dias:2012nf}
    used a direct numerical evaluation
    of~\eqref{eq:ProtoAngularPart} to compute the re-expansion of
    flux cross-products into $\Eharmonic_{LM}$ harmonics,
    but their implementation
    inadvertently neglected the angular
    Jacobian $(-\det G_{T^{1,1}})^{1/2}$.
    This will slightly change numerical values appearing in the
    re-expansion, and it may also change the selection rules
    that couple the quantum numbers $(L, M)$, $(L', M')$
    and $(L'', M'')$.
    Nevertheless,
    based on the numerical evidence to be discussed
    in~\S\ref{sec:small-studies} below
    it seems possible
    that this will not significantly influence the final
    distribution of observables reported by these authors.}

Substitution of~\eqref{eq:EharmonicClebschGordan}
in~\eqref{eq:ProtoAngularPart}
and use of the completeness relation~\eqref{eq:harmonic-complete}
shows that the integral $\d^5 \Psi'$
collapses to $\delta_{L\sumL} \delta_{M\sumM}$.
For fixed $(L', M')$, $(L'', M'')$, the angular part
therefore reproduces the right-hand side of~\eqref{eq:EharmonicClebschGordan}
after suitable relabelling of indices.
To summarize,
consider the cross product between any two fluxes
$\Lambda$, $\Lambda'$
drawn from
Tables~\ref{Table:seriesITable}--\ref{Table:seriesIIITable},
with labels $(L, M)$ and $(L', M')$
and (adjusted) radial scaling dimensions $\Delta$, $\Delta'$.
Unless both fluxes are chiral,
their cross product
makes a contribution to the brane potential, via the particular integral
for $\Phi_-$,
of the form
\begin{equation}
\label{eq:FluxesVPI}
\begin{split}
    \Lambda \times \Lambda'
    \rightarrow
    \frac{g_s}{96}
    \mu^4
    \sum_{\suml_1 = |\ell_1 - \ell'_1|}^{\ell_1 + \ell'_1}
    \sum_{\suml_2 = |\ell_2 - \ell'_2|}^{\ell_2 + \ell'_2}
    &
    \mathscr{A}(L, L', \sumL)
    \alpha x^{\Delta + \Delta' - 4}
    \\
    &
    \times
    \langle \ell_1 m_1, \ell_1' m_1' |\suml_1 \summ_1 \rangle
    \langle \ell_1 \frac{R}{2}, \ell_1' \frac{R'}{2} |
        \suml_1 \frac{\sumR}{2} \rangle
    \\
    &
    \times
    \langle \ell_2 m_2, \ell_2' m_2' | \suml_2 \summ_2 \rangle
    \langle \ell_2 \frac{R}{2}, \ell_2' \frac{R'}{2} |
        \suml_2 \frac{\sumR}{2} \rangle
    \Eharmonic_{\sumL \sumM}(\Psi) .
\end{split}
\end{equation}
As explained above, we have dropped corrections from the infrared
end of the conifold
and from the term scaling like $x^{\Delta(\sumL)}$
which is already included in $\VCF$.

If both fluxes \emph{are} chiral then this expression applies if $\Lambda$ and
$\Lambda'$ are drawn from the same series; if not, their contribution should be
set to zero.
The total potential $\VPI$ should be obtained by summing~\eqref{eq:FluxesVPI}
over all fluxes $\Lambda$, $\Lambda'$.
In this expression
$\mathscr{A}(L, L', \sumL)$\label{glossary:Ascr} is an unknown amplitude
that absorbs the unknown constants of proportionality
in~\eqref{eq:ProtoRadialPart} and~\eqref{eq:ProtoAngularPart}.
It depends on the recipes used to construct $\Lambda$ and $\Lambda'$
from their seed zero-modes,
and also the amplitude with which these `building block' fluxes appear in
the supergravity solution for $G_3$.
It also absorbs the normalization factor
that appears under the
square-root in Eq.~\eqref{eq:EharmonicClebschGordan}.

The final result is very complicated,%
    \footnote{In Refs.~\cite{Agarwal:2011wm,Baumann:2014nda}
    the combined potential from \emph{both}
    the $\Phi_-$ complementary function and particular
    integral was represented in the form of
    Eq.~\eqref{Vhomog}; see Eq.~(5) of Ref.~\cite{Agarwal:2011wm}
    and Eq.~(5.57) of Ref.~\cite{Baumann:2014nda}.
    Although this method of presentation was no doubt intended to
    suppress needless complexity, we believe
    that the more precise form of~\eqref{eq:FluxesVPI}
    is helpful.
    In particular, in~\eqref{eq:FluxesVPI}
    it is clear that the radial scaling dimension for each 
    term cannot be computed just from knowledge of the
    representation $L$ for the harmonic with which it appears,
    as~\eqref{eq:basic-radial-scaling} and~\eqref{Vhomog}
    would imply.
    Terms generated from the cross-product between fluxes
    may occur with a radial scaling dimension
    $\Delta + \Delta' - 4$ that has no simple relation to the
    harmonic $\Eharmonic_{\sumL\sumM}$ with which they are
    partnered in the potential.}%
    \textsuperscript{,}%
    \footnote{In Agarwal et al.~\cite{Agarwal:2011wm}, a list of lowest
    scaling dimensions was given in Eq.~(6).
    However, the value $\sqrt{28}-3$ in this list
    should not appear.
    It is absent from the similar list given in Ref.~\cite{Baumann:2009qx}.
    We would like to thank Nishant Agarwal for confirmation of
    this observation.}
and depends on constants
such as $\mathscr{A}(L, L', \sumL)$
that we cannot predict.
To use it in a practical analysis one must make a number of largely
arbitrary choices.
Unfortunately,
prior analyses of this model have generally not documented
the choices made to convert
Eq.~\eqref{eq:FluxesVPI}
to a practical expression for the potential.
In~\S\ref{sec:small-studies} we discuss numerical experiments
in which the flux-sourced contributions described in this
section are dropped,
and find that this significantly affects
the mass spectrum.
The impact on observables is more modest
but not negligible,
and therefore a precise specification of $\VPI$
is necessary.

Our choices are as follows.
First, we do not attempt to model the numerical coefficient
$\gstring \alpha \mathscr{A} / 96$ that normalizes Eq.~\eqref{eq:FluxesVPI}.
We collect these numerical factors into a single
statistical Wilson coefficient
$\mathscr{C}_{\sumL \sumM}$\label{glossary:fluxCLM}
whose real and imaginary parts are assumed to be
drawn from the same distributions
that characterize the coefficients $C_{LM}$
in Eq.~\eqref{Vhomog};
see Table~\ref{Table:ParameterGlossary}.
Second, we \emph{do} track the numerical value of the Clebsch--Gordan
factors.
If
the same mode
$x^{\Delta + \Delta' - 4} \Eharmonic_{\sumL \sumM}$
is produced from more than one cross-product of the
fluxes in Tables~\ref{Table:seriesITable}--\ref{Table:seriesIIITable},
we add their numerical
coefficients coherently to produce a single numerical prefactor.
An alternative would be to treat each occurrence as an independent
random variable, rather than add the amplitudes coherently.
Yet another choice would be to model the Clebsch--Gordan coefficients
as a Kronecker-$\delta$, equal to zero if the Clebsch--Gordan factors give zero
and unity otherwise.
Our procedure is intended to model, at least approximately,
cases where the Clebsch--Gordan coefficients are unusually large or small,
without causing a proliferation of parameters that unnecessarily
enlarge the sample space.
In practice the Clebsch--Gordan values typically do not vary significantly for the
range of quantum numbers we are using and are almost always
$\Or(1)$.

Notwithstanding the foregoing discussion,
the analysis discussed in~\S\ref{sec:small-studies}
does offer hope that
the precise
procedure used to model the amplitude of individual contributions
to the potential will not radically alter
the final distribution of observables.
A detailed understanding will require further numerical
work that is beyond the scope of this paper.

\para{Summary: the complete brane potential}
To summarize, the total potential for the $\brane{3}$-brane
consists of:
\begin{itemize}[noitemsep]
    \item the Coulomb term~\eqref{Coulomb},
    \item the mass term~\eqref{MassCurvature}
    generated by coupling to the four-dimensional
    Ricci scalar,
    \item the homogeneous terms~\eqref{Vhomog}
    generated by the complementary function for $\Phi_-$,
    \item the terms generated by~\eqref{eq:FluxesVPI},
    with amplitudes modelled as described above,
    from the particular integral for $\Phi_-$.
\end{itemize}
\section{Experimental procedure}
\label{sec:experimental-procedure}
In~\S\ref{sec:software-stack}
we describe
our software stack
and sampling strategy,
and our procedure for collecting observables.
In~\S\ref{sec:adiabatic-limit}
we explain our definition of an adiabatic limit,
based on studying eigenvalues of the mass matrix.
Finally, in~\S\ref{sec:code-comparison} we compare the performance of
our pipelines
and the resulting distributions,
and show that (with some caveats) these
demonstrate good agreement.

\subsection{Software stack and sampling strategy}
\label{sec:software-stack}
We employ
two separate pipelines
to
harden our analysis against numerical and implementation errors.
One pipeline is based on the {\CppTransport}
platform~\cite{Seery:2016lko,Butchers:2018hds}.
This is a {\Cpp} framework
for computation of inflationary observables,
up to and including those derived from the
three-point function,
based on an implementation of the `transport'
method~\cite{Mulryne:2009kh,Mulryne:2010rp,Dias:2015rca,Dias:2016rjq}.
The second pipeline is based on {\PyTransport}~\cite{Mulryne:2016mzv,Ronayne:2017qzn}.
This is an independent Python implementation of the same
transport system,
but making different numerical choices and using a different
numerical integrator.
Neither pipeline uses the slow-roll approximation,
except to set initial conditions for each correlation function.

Although {\CppTransport} and {\PyTransport} are related, they are
not equivalent:
their implementation details differ, including the
exact set of equations that are solved
and the underlying computer algebra system used to perform
symbolic computations.%
  \footnote{The systems used are {\GiNaC}~\cite{Bauer:2001ig} for
  {\CppTransport} and {\SymPy} for {\PyTransport}.}
Therefore comparison between
these pipelines is not empty.
Differences in their output
can be regarded as an indication
of the `implementation error' from our inability to
perform perfectly accurate computations.

At the base of the software stack we use a shared
Python script that builds
versions of Tables~\ref{Table:eig_table}
and \ref{Table:seriesITable}--\ref{Table:seriesIIITable}
and
combines them according to the rules
of
Eqs.~\eqref{Vhomog} and~\eqref{eq:FluxesVPI}
to obtain their contribution
to the $\brane{3}$-brane potential.
We restrict attention to operators with radial
scaling dimension $\delta \leq \DeltaMax = 3.8$, as explained
in~\S\S\ref{sec:harmonic-analysis}--\ref{sec:flux-contributions}
above.
The script writes out {\CppTransport} and {\PyTransport} model files
containing canonical forms for the potential, its first three derivatives, the components of the
field-space metric and its inverse, and the components of the
field-space Riemann tensor $\tensor{R}{^A_B_C_D}$.
By sharing expressions for these quantities we ensure that both pipelines
perform their calculations using the \emph{same} parametrization,
so that subsequent analyses
compare like to like.
After this stage, symbolic manipulations carried out by the pipelines
are independent.

\subsubsection[{\CppTransport} pipeline]{{\CppTransport}
\label{sec:cpptransport-pipeline}
pipeline\footnote{An early version of the {\CppTransport}
pipeline was written by Sean Butchers, whom we thank
for assistance in preparing this section.}}
The {\CppTransport} translator converts the model file
into a custom {\Cosmosis} module~\cite{Zuntz:2014csq}.
It expects an input datablock containing the
{\numparams}
parameters of the inflationary
model,%
  \footnote{For details of the data flow through a 
  {\Cosmosis} pipeline, see Ref.~\cite{Zuntz:2014csq}.}
and uses {\CppTransport}'s internal solver
to obtain values for the corresponding background evolution
and $n$-point functions. These are written into the outgoing datablock
for use by later stages of the pipeline.

This implementation is used to generate
our primary trajectory catalogue
and compute observables for its members.
There are two steps.
First, as explained in~\S\ref{sec:introduction},
we generate a catalogue of inflationary solutions
by sampling over 450,000,000 trajectories.
Our methodology is essentially that proposed
by Easther et al.~\cite{Easther:2013rva}.
We use the {\apriori}
sampler
(part of the default {\Cosmosis} package)
to repeatedly draw realizations of the
parameters listed in Table~\ref{Table:ParameterGlossary}.
Each trajectory is evolved from
fixed initial conditions
$x = 0.9$
and $\theta_1 = \theta_2 = \phi_1 = \phi_2 = \psi = 1$,
with the field velocities set to zero.
The observables do not depend significantly
on these choices, as we explain 
in~\S\ref{sec:small-studies} below.
The calculation terminates when
either: (1) inflation
exits gracefully as $\epsilon$ smoothly approaches unity,
or (2) the $\brane{3}$/$\antibrane{3}$ pair
dissolve in a hybrid transition,
taken to occur when their separation
is smaller than $\Delta x = 0.02$.
The Lagrangian parameters and Wilson coefficients
for the subset of roughly {\precataloguesize} trajectories that inflate
for $N > 60$ e-folds constitute the required catalogue.
A candidate trajectory is rejected if any of the following conditions
apply:
\begin{itemize}[noitemsep]
  \item it is not initially inflating,\label{page:tajectory_reject_criteria}
  \item it does not reach the hybrid transition that describes
  $\brane{3}$/$\antibrane{3}$ annihilation
  while inflation is still ongoing,
  or within a cutoff of 10,000 e-folds,
  \item the brane is ejected from the ultraviolet end of the throat,
  \item the potential becomes negative at any point in the evolution,
  \item numerical overflow, underflow or an integration error occurs.%
    \footnote{During integration, {\CppTransport} automatically
    tests for the following error conditions:
    (a) $H^2$ becoming negative;
    (b) $\epsilon \equiv - \dot{H}/H^2$ becoming negative;
    (c) $\epsilon$ becoming greater than $3$;
    (d) $V$ becoming negative;
    (e) any component of a correlation
    function becoming $\infty$ or \texttt{NaN}.}
\end{itemize}

\para{Catalogue of observables}
Second, the completed catalogue is processed
to determine inflationary observables for each trajectory.
The calculation is broken into reusable components
that are assembled as a second {\Cosmosis} pipeline,
controlled programmatically rather than coupled to a sampler.
For each entry in the catalogue the pipeline performs
the following steps:

\stephead{1}{Two-Point Function}
  \stepsidehead{Power spectra}
  It was explained in~\S\ref{sec:introduction}
  that the power spectrum in the $\brane{3}$/$\antibrane{3}$
  model is often not scale invariant (or even monotonic),
  and therefore summary observables
  measured at a single scale are frequently a poor predictor
  of the goodness-of-fit to observation.
  Nevertheless, they have some uses.
  They allows us to compare with previous
  analyses, and they are still a convenient way to organize
  our catalogue of trajectories.

  For these reasons
  we collect
  summary power spectrum observables
  $\As$, $\At$, and $r$,
  evaluated
  at $\kpiv = 0.002 \, \Mpc^{-1}$.
  Here, $\As \equiv \dimP_\zeta(\kpiv)$ and
  $\At \equiv \dimP_h(\kpiv)$ measure
  (respectively)
  the amplitude
  of the dimensionless
  power spectra for the uniform-density gauge
  curvature perturbation $\zeta$,
  and for tensor modes.
  They are defined in terms of equal-time
  correlation functions,
  \begin{subequations}
  \begin{equation}
      \label{eq:Pzeta-def}
      \langle \zeta(\vect{k}) \zeta(\vect{k}') \rangle
      = (2\pi)^3 \delta(\vect{k} + \vect{k}')
      P_\zeta(k)
      = (2\pi)^3 \delta(\vect{k} + \vect{k}')
      \frac{2\pi^2}{k^3} \dimP_\zeta(k) ,
  \end{equation}
  \begin{equation}
      \label{eq:Ptensor-def}
      \langle h_{s}(\vect{k}) h_{s'}(\vect{k}') \rangle
      = (2\pi)^3 \delta(\vect{k} + \vect{k}')
      \delta_{ss'}
      \frac{\pi^2}{2k^3} \dimP_h(k) ,
  \end{equation}
  \end{subequations}
  where $h_s(\vect{k})$ is a tensor perturbation
  of polarization $s$ ($s = +, \times$) in a normalization where
  the polarization matrices $\vect{e}^s$ satisfy
  $\tr( \vect{e}^s \cdot \vect{e}^{s'} ) = 2 \delta^{ss'}$.
  With this definition
  the tensor spectrum is conventionally normalized
  and the tensor-to-scalar ratio
  satisfies $r \equiv \At/\As$.

  \stepsidehead{Spectral indices}
  We evaluate the $\zeta$ spectral index $\ns$
  at $k=\kpiv$,
  \begin{equation}
    \ns - 1 \equiv \left. \frac{\d \ln \dimP_\zeta}{\d \ln k} \right|_{k=\kpiv} ,
    \label{eq:spectral-indices-defs}
  \end{equation}
  Because of the flatness of the tensor spectrum,
  there are trajectories for which
  it is not straightforward
  to collect a reliable numerical estimate
  of the spectral index
  $\nt \equiv \d \ln \dimP_h / \d \ln k$.
  For a detailed discussion of this and
  other computational details
  see~\S\ref{sec:computational-issues}
  below.

  \stepsidehead{Matching equation}
  To relate physical scales to a horizon-exit time
  we use the matching
  equation~\cite{Liddle:2003as,Adshead:2008vn,Adshead:2010mc},%
      \footnote{See Eq.~(20) of Ref.~\cite{Adshead:2010mc},
      from which
      we have dropped the slow-roll approximation.
      Note that there is a minor
      typo in the version of this equation
      that appears in Ref.~\cite{Adshead:2010mc};
      the correct numerical constant
      appearing in it
      should be 55.98, not 55.75.
      The numerical constant quoted in~\eqref{eq:matching-equation}
      includes this correction.
      We thank Peter Adshead for helpful correspondence.}
  \begin{equation}
    N(k) = 59.57 - \ln \frac{k}{\kpiv}
    + \ln
    \left(
        \frac{H_k}{10^{16} \, \GeV}
        \frac{\Mp^{1/2}}{\Hend^{1/2}}
    \right)
    ,
    \label{eq:matching-equation}
  \end{equation}
  where $N(k)$ is the horizon exit time of the physical mode $k$,
  measured in e-folds from the end of inflation.
  The corresponding Hubble rates are $H_k$ and $\Hend$,
  respectively.
  We assume that reheating completes instantaneously,
  and that decay products from break-up of the
  scalar condensates
  thermalize into radiation.

\stephead{2}{Three-Point Function}
  \stepsidehead{Equilateral and folded configurations}
  We measure
  the amplitude of three-point correlations
  for two indicative $\langle \zeta \zeta \zeta \rangle$
  bispectrum configurations.
  In the Fergusson--Shellard
  parametrization,%
    \footnote{The momenta $\vect{k}_1$, $\vect{k}_2$, $\vect{k}_3$
    that participate in a three-point function
    such as $\langle \zeta(\vect{k}_1) \zeta(\vect{k}_2) \zeta(\vect{k}_3)
    \rangle$ satisfy the `triangle' condition
    $\vect{k}_1 + \vect{k}_2 + \vect{k}_3$ as a consequence
    of statistical translation invariance.
    This makes the correlator a function only of $k_1$, $k_2$, $k_3$.
    In the Fergusson \& Shellard parametrization we set
    $k_t = k_1 + k_2 + k_3$ to be the perimeter of the momentum triangle.
    Then~\cite{Fergusson:2006pr},
    \begin{align*}
      k_1 & = \frac{k_t}{4}(1 + \alpha + \beta) , \\
      k_2 & = \frac{k_t}{4}(1 - \alpha + \beta) , \\
      k_3 & = \frac{k_t}{2}(1 - \beta) .
    \end{align*}
    An equivalent parametrization had earlier been introduced
    by Rigopoulos, Shellard \& van Tent~\cite{Rigopoulos:2004ba}.}
  these
  are:
  (1) an equilateral configuration
  $\{ k_t = 3\kpiv, \alpha = 0, \beta = 1/3 \}$;
  and (2) a folded configuration
  $\{ k_t = 3\kpiv, \alpha = 0, \beta = 0.005 \}$.
  We report the correlation as a measurement of
  the reduced bispectrum
  $\fNL(k_1, k_2, k_3)$,
  defined by
  \begin{equation}
    \label{eq:reduced-bsp-def}
    \fNL(k_1, k_2, k_3) =
    \frac{6}{5}
    \frac{B_\zeta(k_1, k_2, k_3)}{P_\zeta(k_1) P_\zeta(k_2) + \text{cyclic}} ,
  \end{equation}
  where `$+$ cyclic' implies that the preceding term is to be summed over
  cyclic permutations of the momenta $k_1$, $k_2$, $k_3$.
  The $\zeta$ bispectrum $B_\zeta$ satisfies
  \begin{equation}
    \langle \zeta(k_1) \zeta(k_2) \zeta(k_3) \rangle
    =
    (2\pi)^3 \delta(\vect{k}_1 + \vect{k}_2 + \vect{k}_3)
    B_\zeta(k_1, k_2, k_3)
  \end{equation}
  and the correlator is computed at equal times.
  When evaluated on our representative equilateral
  and folded configurations
  we denote the reduced bispectrum by $\fNLeq$ and $\fNLfold$,
  respectively.

  \stepsidehead{Squeezed configurations}
  As explained in~\S\ref{sec:introduction}
  and Appendix~\ref{sec:run_time},
  it is too time-consuming to compute three-point correlations
  on a squeezed configuration for the entire primary catalogue.
  Instead, we sample the squeezed configurations
  $\{ k_t = 3\kpiv, \alpha = 0, \beta = 0.9 \}$
  and
  $\{ k_t = 3\kpiv, \alpha = 0, \beta = 0.95 \}$
  on a separate catalogue of trajectories
  to determine how their amplitudes correlate with the
  equilateral and folded configurations.

\stephead{3}{Adiabatic Limit}
  The mass spectrum is computed
  from the
  eigenvalues of the mass matrix
  ${M^A}_B$~\cite{Sasaki:1995aw,Nakamura:1996da,Dias:2015rca},
  \begin{equation}
      {M^A}_B
      = \grad^A \grad_B V
      - \tensor{R}{_J^A_B_K} \dot{X}^J \dot{X}^K
      - \frac{3 + \epsilon}{\Mp^2} \dot{X}^A \dot{X}_B
      + \frac{\dot{X}^A \ddot{X}_B + \dot{X}_B \ddot{X}^A}{H \Mp^2} ,
      \label{eq:mass-matrix}
  \end{equation}
  where the $X^A$ are the scalar fields
  $\{ x, \theta_1, \theta_2, \phi_1, \phi_2, \psi \}$
  appearing
  in
  Eqs.~\eqref{eq:field-space-kinetic}--\eqref{eq:practical-conifold-metric}.
  Indices on $\dot{X}^A$ and
  $\ddot{X}^A \equiv \dot{X}^B \grad_B \dot{X}^A$
  are raised and lowered using the conifold metric $G_{AB}$
  normalized as in Eq.~\eqref{eq:practical-conifold-metric},
  $\grad_A$ is the covariant derivative compatible
  with $G_{AB}$,
  and $\tensor{R}{_A_B_C_D}$ is the Riemann tensor
  constructed from $\grad_A$.
  We sample these eigenvalues at
  $N=55$, $N=2.5$, $N=1$ and $N=0$
  e-folds before the end of inflation.
  In~\S\ref{sec:adiabatic-limit}
  below we explain how these are used
  to detect the onset of an adiabatic limit.

\stephead{4}{Likelihood Function}
  Finally,
  where possible
  we compute the CMB likelihood
  for this trajectory using
  the
  Planck2015 likelihood code~\cite{Aghanim:2015xee}.%
    \footnote{In fact, this is not done for all trajectories.
    In some cases we have $N \geq 60$ e-folds of inflation from
    the initial conditions, but too few total e-folds to allow $4.5$ e-folds of
    subhorizon evolution for
    the largest scale
    $k = 10^{-6} \, \Mpc^{-1}$
    needed for the {\CLASS} computation of the $C_\ell$.
    In such cases there is a choice between rejecting the trajectory
    or foregoing the Planck likelihood.
    We choose the latter.}
  The $\zeta$ power spectrum is sampled
  at 100 logarithmically-spaced wavenumbers
  in the range $10^{-6} \, \Mpc^{-1} \leq k \leq 50 \, \Mpc^{-1}$.
  This sample is passed to
  {\CLASS} via the {\Cosmosis} pipeline
  and used to compute the CMB angular power spectra
  $C^{TT}_{\ell}$,
  $C^{TE}_{\ell}$
  and $C^{EE}_{\ell}$.
  We do not vary the parameters of the post-inflationary
  cosmology, which are fixed to their Planck2015 best-fit
  values~\cite{Ade:2015xua}.
  The bundled {\Cosmosis} \texttt{planck}
  module is used
  to calculate the likelihood for all these
  $C_\ell$.
  If desired,  
  any other likelihood could be substituted
  in this step.

\subsubsection{Computational issues}
\label{sec:computational-issues}
\para{Ultra slow-roll inflation}
In the $\brane{3}$/$\antibrane{3}$ model it is known that
extended epochs of inflation are typically
realized near an inflexion point in the
potential~\cite{Baumann:2007np,Agarwal:2011wm,Dias:2012nf}.
Therefore we must allow for the possibility that some
inflationary trajectories enter a phase of ultra slow-roll
dynamics~\cite{Kinney:2005vj,Martin:2012pe,Namjoo:2012aa,Mooij:2015yka,Romano:2016gop}.
Such phases are
characterized by: (1) $V' \approx 0$,
(2) a small and rapidly decaying
value of $\epsilon \equiv -\dot{H}/H^2$,
and (3) $\eta \approx -6$.
Previously,
the possibility of an ultra slow-roll phase in this model does not
appear to have been considered.

At the level of the background,
both {\CppTransport} and {\PyTransport} implement the full scalar
field dynamics and therefore capture all ultra slow-roll effects.
However, initial conditions for each correlation function are
estimated using analytic expressions that assume slow-roll dynamics~\cite{Dias:2016rjq}.
This is harmless if the dynamics are close to slow-roll while the momenta
characterizing an individual correlation function are exiting the horizon,
even if a transition to ultra slow-roll inflation occurs later.
If slow-roll does not apply
the procedure is still mostly harmless for the two-point function,
because the slow-roll result $\dimP_{\delta\phi} \sim H^2$
continues to apply during ultra slow-roll inflation~\cite{Kinney:2005vj}.
Therefore the initial condition will be significantly
inaccurate only for
modes that exit
during a transition between
slow-roll and ultra slow-roll phases.

The prospects for the three-point function are
less straightforward
because the slow-roll initial condition is corrected by terms of order
$\eta$~\cite{Chen:2013aj,Chen:2013eea,Cai:2017bxr},
which is large during ultra slow-roll.
It follows that
numerical three-point functions
computed by
{\CppTransport} and {\PyTransport} must be
treated with caution
for scales exiting during an ultra slow-roll phase.
Nevertheless,
if the calculation starts sufficiently far before horizon exit
and the $\Or(\eta)$ displacements do not take the initial condition out of the
basin of attraction of the true solution,
we may still expect
the results to be valid.
We return to this question in~\S\ref{sec:results-background}.

\begin{figure}
  \centering
  \includegraphics{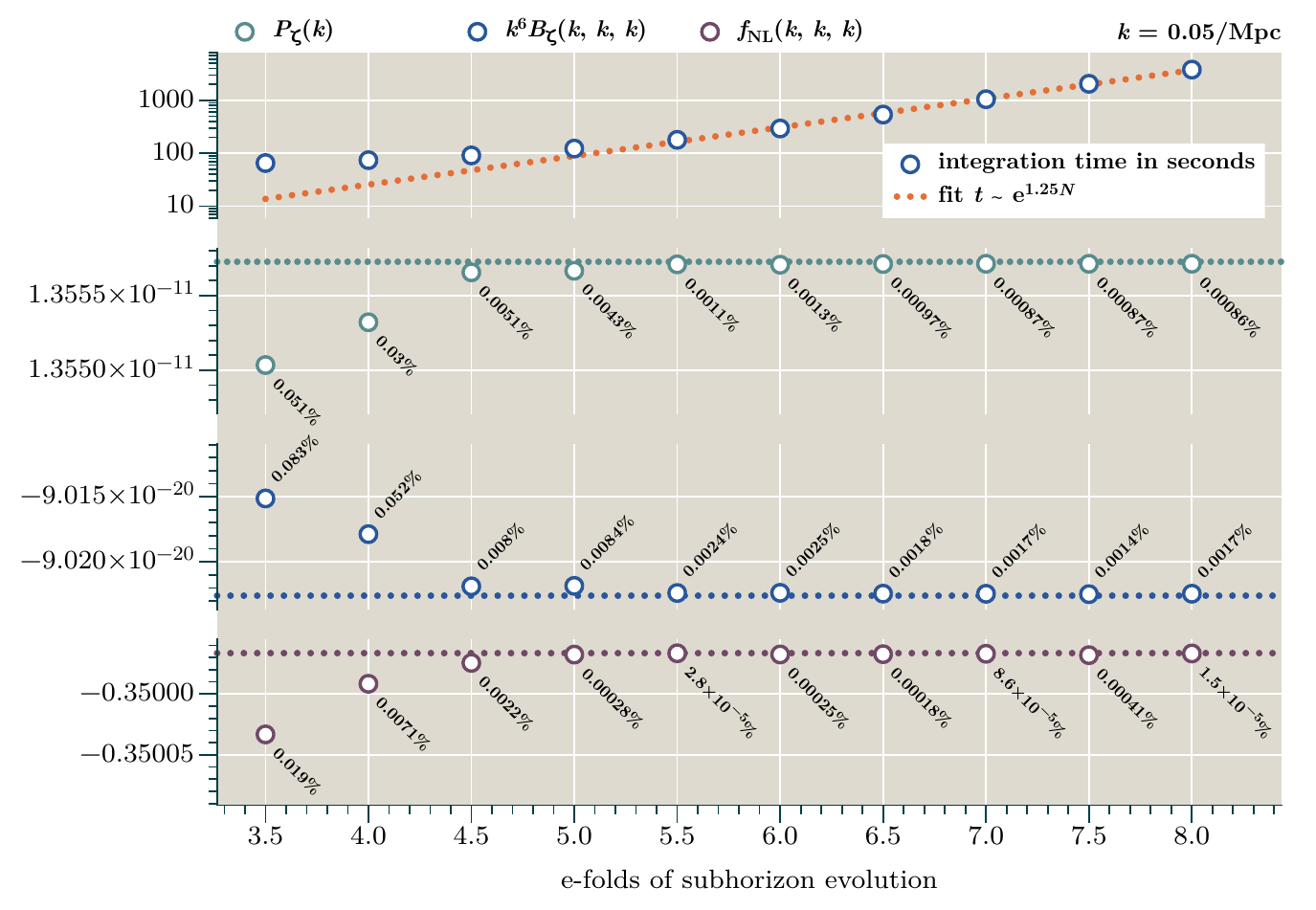}
  \caption[Convergence test with increasing subhorizon e-folds]
  {\label{fig:subhorizon-convergence}Dependence on
  number $\Nsubh$ of e-folds of subhorizon evolution
  for the representative trajectory {\#}32327. \semibold{Top panel}:
  integration time in seconds. The dotted orange line
  shows a fit to the exponential dependence
  $t \sim \e{1.25N}$, which is valid for $\Nsubh \gtrsim 5.5$.
  Increasing the number of subhorizon e-folds is very
  expensive, but the following panels show there are
  diminishing returns
  for $\Nsubh \gtrsim 4.5$.
  \semibold{Second panel}:
  Convergence of $\dimP_\zeta$ evaluated
  at $k = 0.05 \ \Mpc^{-1}$. The asymptote is extracted
  by fitting a function of the form
  $a - b\e{-c N}$ and measuring $a$.
  The labels show the percentage deviation from this
  asymptotic value.
  \semibold{Third panel}:
  Same as second panel,
  but for $k^6 B_\zeta(k, k, k)$ measured on
  an equilateral configuration of side $k = 0.05 \, \Mpc^{-1}$.
  \semibold{Bottom panel}:
  Same as second panel,
  but for $\fNL(k,k,k)$ measured on the same equilateral
  configuration.}
\end{figure}

\para{Power spectrum amplitudes}
The amplitudes
$\As$ and $\At$, and hence $r$,
are computed directly.
In a transport implementation,
numerical accuracy is usually determined
by the number of e-folds of subhorizon evolution;
see Ref.~\cite{Dias:2016rjq} for a detailed discussion.
We use $4.5$ e-folds, which
(subject to the caveats below for the tensor
power spectrum) we find to be a reasonable compromise
between accuracy and integration time.
We have performed a small number of spot-checks to
test convergence with increasing
subhorizon e-folds, but
these do not show significant improvement:
see Fig.~\ref{fig:subhorizon-convergence}.
Accuracy also depends on the choice of stepper.
We find that
Runge--Kutta (Dormand--Prince and Fehlberg)
and Adams--Bashforth--Moulton
methods sometimes exhibit instabilities,
especially in three-point amplitudes,
although in our
tests these did not propagate to $\zeta$ observables.
The Bulirsch--Stoer variable order
method produces fast, high-precision solutions
without significant instabilities.
However, we do not find that the choice of stepper
has a significant effect on our final
distributions.
For the construction of our primary catalogue we
use the Dormand--Prince
$4^{\text{th}}$/$5^{\text{th}}$-order method.

\para{Numerical precision}
The integrations needed for the $\brane{3}$/$\antibrane{3}$
model are complex and involve a large number of intermediate
steps.
If significant cancellations occur,
there is a risk of accumulating inaccuracies from roundoff error.
To test whether extended precision is needed
we have compared a subsample of $1,000$ trajectories
from our
main catalogue using
\texttt{double} and \texttt{long double}
precision.%
  \footnote{The meaning of \texttt{long double} is
  implementation-dependent, but on our production
  Linux platform with the GCC compiler
  this is an 80-bit extended precision format.
  For comparison, \texttt{double} is a 64-bit format.}
In nearly all cases we find only sub-percent shifts in $\ns$.
However, for $\fNLeq$
we find
$\sim 9\%$ of trajectories exhibit shifts larger than $1\%$,
and $\sim 2\%$ of trajectories exhibit shifts larger than $5\%$.
A handful of trajectories shift by $10\%$ or more.
Therefore, while the enhanced precision is not required in
most cases, it
apparently \emph{is} needed to accurately predict
three-point observables for certain trajectories---%
but we will see in~\S\ref{sec:code-comparison}
that these differences do not seem to be reflected in
the distribution over an entire catalogue.
Nevertheless, we prefer to use the extended precision
calculation out of an abundance of caution.

{\CppTransport} supports arbitrary precision arithmetic
(although with a significant performance penalty),
so although we have not done so it would be possible to perform
the calculation with even higher precision than
\texttt{long double} to verify that it has
properly converged.
Similar benefits from use of \texttt{long double} precision
are known to occur in models of ultra slow-roll inflation,
which has features that are similar to the
$\brane{3}$/$\antibrane{3}$ model.

\para{Spectral indices}
Computation of spectral indices is more
challenging, especially for the tensor
power spectrum which is extremely flat.
Several numerical strategies are available.
When applied to the $\zeta$ spectral index
$\ns$ these methods all
yield consistent results,
but none are entirely satisfactory for the
tensor spectral index $\nt$.

To be concrete, we
collect the fields $X^A$ and the corresponding momenta
$\pi^A \equiv \d X^A / \d N$
into a single phase-space coordinate
$\pcoord{X}^{\pidx{A}} = (X^A, \pi^B)$
and
define the
two-point function for $\delta\pcoord{X}^{\pidx{A}}$
to satisfy
\begin{equation}
  \langle
    \delta\pcoord{X}^{\pidx{A}}(\vect{k}_1)
    \delta\pcoord{X}^{\pidx{B}}(\vect{k}_2)
  \rangle
  = (2\pi)^3
  \delta(\vect{k}_1 + \vect{k}_2)
  \tensor{\Sigma}{^{\pidx{A}}^{\pidx{B}}}(k) ,
  \label{eq:scalar-2pf}
\end{equation}
where $k = |\vect{k}_1| = |\vect{k}_2|$.
The $\zeta$ power spectrum $P_\zeta$
can be written
$P_\zeta(k) = \tensor{N}{_{\pidx{A}}}
\tensor{N}{_{\pidx{B}}}
\tensor{\Sigma}{^{\pidx{A}}^{\pidx{B}}}(k)$,
where explicit expressions for the coefficients
$\tensor{N}{_{\pidx{A}}}$ are
known~\cite{Sasaki:1995aw,Mulryne:2009kh,Dias:2014msa,Dias:2015rca,Dias:2016rjq}.
They become independent of $k$ on superhorizon scales.
The first option is to write a transport equation for the
`spectral matrix'
$\tensor{n}{^{\pidx{A}}^{\pidx{B}}}
\equiv
\d \tensor{\Sigma}{^{\pidx{A}}^{\pidx{B}}} / \d \ln k$,
which can be used to compute
$\ns$~\cite{Dias:2012nf,Dias:2015rca,Dias:2016rjq},
\begin{equation}
  \label{eq:spectral-index-spectral-matrix}
  \ns - 1 = 3 +
  \frac{\tensor{N}{_{\pidx{A}}} \tensor{N}{_{\pidx{B}}}
    \tensor{n}{^{\pidx{A}}^{\pidx{B}}}}
  {\tensor{N}{_{\pidx{C}}} \tensor{N}{_{\pidx{D}}}
    \tensor{\Sigma}{^{\pidx{C}}^{\pidx{D}}}} .
\end{equation}
We briefly review this approach in
Appendix~\ref{sec:transport_spectral}.
It is conceptually clean, but as explained
in Ref.~\cite{Dias:2015rca}
it can happen that
we require more e-folds of subhorizon evolution
to obtain good numerical results
for $\tensor{n}{^{\pidx{A}}^{\pidx{B}}}$ than for
$\tensor{\Sigma}{^{\pidx{A}}^{\pidx{B}}}$.
This is partly because to compute $\ns-1$ we
effectively
subtract the leading
term from the right-hand side
of~\eqref{eq:spectral-index-spectral-matrix}.

A similar expression applies for $\nt$,
although the calculation is simpler because no gauge
transformation is required.
For both $\ns$ and $\nt$
we attempt to accelerate convergence by using
initial conditions that include
subleading terms in both the slow-roll expansion and
$k/(aH)$.
We find that
Eq.~\eqref{eq:spectral-index-spectral-matrix}
gives results for $\ns$ that agree with other methods,
but its counterpart for $\nt$
does not always
yield good results even with a large number
of subhorizon e-folds.

A second option is to fit
a function of the form $\dimP = A_\star (k/\kpiv)^{n}$
to sampled
values of the power spectrum $\dimP$ near the pivot
scale $\kpiv$, and extract the spectral
index from the fit for $n$.
This approach has the advantage that 
it requires only knowledge of the power spectrum
and not the spectral matrix.
The disadvantage is that the fit can be thrown off
by small inaccuracies in the computed amplitude,
perhaps caused by noise or other
numerical artefacts.
If the spectrum has non-negligible tilt these
do not usually affect the measured spectral index.
However, for the $\brane{3}$/$\antibrane{3}$ model,
based on analytic estimates
we expect roughly
$10^{-14} \lesssim \nt \lesssim 10^{-2}$.
Because $\nt$ is so small,
fluctuations
in excess of
$\delta \ln \dimP_h \sim \nt \, \delta \ln k$
can be present between samples
with $k$-spacing $\delta \ln k$
(even with the high-precision Bulirsch--Stoer stepper).

Specifically,
for some trajectories
we find the tensor power spectrum
to be contaminated by
oscillations
of very small amplitude.
These can spoil automated measurement
of $\nt$.
The source of these oscillations is not clear,
but they are almost certainly not physical.
Their amplitude typically decreases when
we allow more subhorizon e-folds.
On some trajectories this is sufficient
to extinguish the oscillations,
but on others their amplitude appears
nearly stable.
Based on this, we speculate that
they are possibly a discretization artefact.
If so, the same effect
(or a closely related one)
may be responsible for the
poor outcomes from the $\tensor{n}{^{\pidx{A}}^{ \pidx{B}}}$
transport equation,
which include positive values for $\nt$
at some values of $k$.
This is incompatible with the strong energy
condition, which implies that $H$ should
decrease.
We see similar results from attempts to
fit for $\nt$, which can
yield positive values by
catching the rising edge of
an oscillation.
This effect can be mitigated by
binning
the power spectrum before performing
the fit, but it is difficult to do this
in an automated way
without risking errors from
over-smoothing.
The significant challenge
entailed in obtaining an accurate
estimate of $\nt$ was already noted
by McAllister et al.~\cite{McAllister:2012am}.

In practice,
the sampling pipeline
fits a quadratic polynomial
to 15 logarithmically-spaced power spectrum samples
between $k = 1.986 \times 10^{-3} \, \Mpc^{-1}$
and $k = 2.014 \times 10^{-3} \, \Mpc^{-1}$.
We have performed spot-checks on roughly
$1,200$ trajectories to compare the
$\zeta$
spectral index computed this way
with Eq.~\eqref{eq:spectral-index-spectral-matrix}.
With
$4.5$ e-folds of subhorizon evolution
we find these are typically consistent within
$1\%$.
For a handful of trajectories, we make a further
confirmation
that these estimates
are \emph{also} consistent with fits performed `by hand'
using a different range of $k$.
Unfortunately,
for $\nt$
we find that these fitting procedures
typically disagree.
We conclude that computing the tensor spectral index
using either method is not acceptable for the
$\brane{3}$/$\antibrane{3}$ model.

A third option is to use an analytic approximation
for the tensor power spectrum to
estimate $\nt \approx -2 \epsilon$,
which requires only knowledge of the background.
This approach has the advantage that it does not
depend on the accuracy with which we can compute the
power spectrum amplitude.
Conversely, it is blind to information
provided by the transport calculation
that is not included in the analytic approximation.
We find negligible correlation
between measurements
using our numerical procedure and those
obtained from $\nt \approx -2\epsilon$,
but this is to be expected in light of the
foregoing discussion.

In conclusion,
using the estimate
$\nt \approx -2\epsilon$
is apparently the least unsatisfactory
option.
In~\S\ref{sec:Results}
the values of $\nt$ we quote are derived using this method
with $\epsilon$ sampled at 60 e-folds before the
end of inflation.
Based on `by hand' fits
to the smoothed tensor
power spectrum, we believe the resulting
values of $\nt$ are accurate within a factor of $2$
(but sometimes much better).
For now, it is prudent to treat our $\nt$
estimates with caution.

\subsubsection{{\PyTransport} pipeline}
\label{sec:pytransport-pipeline}
The second pipeline is based on {\PyTransport}.
Its main purpose is to perform a number of smaller
($\sim${\minicataloguesize} trajectory)
complementary studies that explore the dependence of observables
on discrete choices made in~\S\ref{sec:construct-potential}.
We also use it to test for consistency with the principal catalogue.
Symbolic computation of the potential
and curvature tensors is shared with the {\CppTransport}
pipeline, but otherwise there is no code re-use.

There are some important differences.
The {\PyTransport} pipeline collects less fine-grained
metadata about the trajectory.
Observables are mostly computed as explained above,
except that to give discrepancies an opportunity to manifest
we apply a different fitting prescription for the spectral
index.
This is based on fitting a spline to five sampled
power spectrum values
for $k$-modes with horizon exit values
spaced $0.3$ e-folds apart.
We use 4.5 e-folds of subhorizon evolution, as above,
but {\PyTransport} does not offer
an option to change the stepper or value type
and
therefore we use the built-in
Runge--Kutta 4/5$^{\text{th}}$-order solver
and \texttt{double} precision arithmetic.
{\PyTransport} does not currently implement calculation
of the tensor power spectrum, so we do not sample $r$.
The pipeline is not controlled by {\Cosmosis},
but uses a custom sampling layer
that
draws parameter combinations from the
priors listed in Table~\ref{Table:ParameterGlossary}
until a prescribed number of trajectories supporting
$N > 60$ e-folds of inflation have been sampled.
However,
the criteria for rejecting trajectories are the same as those
given on p.~\pageref{page:tajectory_reject_criteria}.
The {\PyTransport} pipeline does not compute
$C_\ell^{TT}$,
$C_\ell^{TE}$,
$C_\ell^{EE}$,
or the CMB likelihood function.

Finally, we apply a different prior for $Q$.
Specifically, we choose
$Q \sim \beta'(4.16, 494)$, where the `beta-prime'
(or `inverted beta')
distribution
$\beta'(\alpha, \beta)$
is
characterized by shape parameters
$\alpha$, $\beta$
and
has the probability distribution function
$\Prob(x) = x^{\alpha-1}(1+x)^{-\alpha-\beta}/\BetaFunction(\alpha, \beta)$,
where $\BetaFunction(x,y) = \Gamma(x)\Gamma(y)/\Gamma(x+y)$
is the Euler $\beta$-function.
This choice is not motivated by physics, but rather sampling
efficiency. We will see below that values of $Q$ that frequently support
$N > 60$ e-folds of inflation are
tightly clustered. It is this phenomenon that underlies the choice
$Q \sim U(0,0.04)$ made in Table~\ref{Table:ParameterGlossary},
which enhances sampling efficiency but has the drawback that 
it excludes the region $Q > 0.04$ completely.
To assist in exploring this region
we use the opportunity
provided by
the {\PyTransport} pipeline to
introduce a prior that cannot be implemented
using {\Cosmosis}.
The beta-prime distribution
samples the region of parameter space that is preferential
for obtaining $N > 60$ e-folds of inflation, while still exhibiting broad
tails that allow less-likely regions to be explored.
We find that the posterior distribution for $Q$
is completely consistent with the
posterior produced from the
more restrictive prior used by {\CppTransport}.

In Table~\ref{Table:SmallStudies}
we list the different ensembles
to be analysed using the {\PyTransport} pipeline.
In each case the priors
match those given in Table~\ref{Table:ParameterGlossary}
except for the stated variations.
Each sample comprises roughly $18 \times 10^{3}$
trajectories,
except for the $\alpha=0$ study which uses a smaller number
of trajectories
($\sim 5 \times 10^3$).
\ctable[
    label = Table:SmallStudies,
    caption = {\small `Small' studies performed using the {\PyTransport} pipeline},
    doinside = {\StandardTable\renewcommand{\arraystretch}{1.2}\rowcolors{2}{gray!20}{white}},
    width = 0.9\textwidth
]{XR}{
\tnote[a]{Notice that with our conventions, $\phiUV$ also appears in the
potential via the definition of the dimensionless radial
coordinate $x$; cf. Eqs.~\eqref{Coulomb} and~\eqref{eq:mu-def}.
Therefore variation of $\phiUV$
does not \emph{only} vary the size of the throat,
but also adjusts the scale of some terms in the potential.}
\tnote[b]{Agarwal et al.\ studied the potential with
the same truncations used here, that is
$\DeltaMax \in \{ 3.8, 3, 2 \}$~\cite{Agarwal:2011wm}.
Dias et al.\ studied
the cases $\DeltaMax \in \{ 3.8, 3 \}$~\cite{Dias:2012nf}.}
\tnote[c]{This prior corresponds to switching off terms in the potential
sourced by the flux product $\gstring |\Lambda|^2 / 96$.\
Only the Coulomb term, mass term and
the sum of scalar zero-modes~\eqref{Vhomog} are retained.}
\tnote[d]{Setting $\alpha=0$ allows us to compare
with the prior analyses reported
by Agarwal et al.~\cite{Agarwal:2011wm},
Dias et al.~\cite{Dias:2012nf},
and McAllister et al.~\cite{McAllister:2012am}.
This study is smaller than the others and comprises
$\sim 5 \times 10^3$ trajectories.}
}
{
    \toprule
    \semibold{study} & \multicolumn{1}{c}{\semibold{prior}} \NN
    \midrule
    initial conditions &
        x \sim N(0.9, 0.02);
        \theta_i \sim U(0,\pi);
        \phi_i \sim U(0,2\pi);
        \psi \sim U(0,4\pi) \NN
    size of throat\tmark[a] &
        \phiUV \sim N(0.1, 0.02) \NN
    sensitivity to truncation\tmark[b] &
        \DeltaMax \in \{ 3.8, 3, 2 \} \NN
    homogeneous potential\tmark[c] &
        \mathscr{C}_{\sumL, \sumM} = 0 \NN
    drop $V_0$ uplift\tmark[d] & \alpha=0 \NN
    \bottomrule
}

\subsection{The adiabatic limit}
\label{sec:adiabatic-limit}
To determine whether the observables
we collect are related to quantities observable in the
CMB or large-scale structure,
we must understand whether the dynamics become
adiabatic before the end of inflation.
Where this happens the perturbations are typically conserved
through the subsequent evolution, provided they remain on
superhorizon scales.
On the other hand, if the evolution does not become adiabatic
then the value of each observable may evolve
during and after reheating
until all isocurvature modes are
exhausted~\cite{Wands:2000dp,Rigopoulos:2003ak,Weinberg:2003sw,
Weinberg:2004kr,Weinberg:2004kf,
Lyth:2004gb,Meyers:2013gua,Hotinli:2017vhx}.
Here, `adiabatic' has its usual cosmological meaning that there is
effectively a single trajectory followed by each
patch of spacetime smoothed on some superhorizon scale.
The difference between neighbouring patches can only be a time offset
$\delta t$ along this trajectory, from which all other
perturbations can be derived.

To diagnose the emergence of an adiabatic trajectory we
inspect the eigenvalues of the mass matrix~\eqref{eq:mass-matrix}
collected in \textsc{Step 3} of~\S\ref{sec:cpptransport-pipeline}.
Note that the relevant mass matrix
is not merely the covariant Hessian
$\grad^A \grad_B V$
that would describe the mass matrix for the scalars alone,
but includes mixing with scalar modes of the metric.
To obtain the correct mass spectrum it is critical to account
for this mixing~\cite{Mukhanov:1985rz,Sasaki:1986hm}.

On an inflationary trajectory there will typically be one
massless or tachyonic mode that is the would-be
conserved Goldstone mode
$\zeta \sim \delta \phi_{\text{ad}} / (\sqrt{2\epsilon} \Mp)$
associated with broken
time translation invariance
along the adiabatic direction~\cite{Dias:2014msa}.
Eigenvectors in the subspace orthogonal to this adiabatic
direction span the available isocurvature modes,
and their corresponding eigenvalues
determine their growth or decay.
Fluctuations in a direction with eigenvalue $m^2$
typically evolve like
$s(N) = s_0 \e{-\eta (N-N_0)}$,
where $s_0$ is the amplitude at
a fiducial time $N=N_0$
and $\eta = m^2 / (3H^2)$.
Therefore fluctuations decay rapidly
in any `heavy' direction where the eigenvalue
satisfies $m^2 \gtrsim 3H^2$.

Exponential suppression implies
that isocurvature modes rapidly become
small but are never completely extinguished,
so there is no unique criterion to determine when a
trajectory has become `sufficiently' adiabatic.
We choose to sample the mass spectrum at $N=0$, $N=1$ and $N=2.5$
e-folds before the end of inflation.
The trajectory is declared to be adiabatic if
the following conditions apply at all three sample points:
(1) one eigenvalue of the mass-matrix is tachyonic,
and (2)
$N-1$ eigenvalues are heavy
in the sense $m^2/(3 H^2) > 1$~\cite{GarciaBellido:1995qq,
GarciaBellido:1996qt,Elliston:2011dr,Seery:2012vj}.%
  \footnote{This is a sharper criterion than the
  one proposed in Refs.~\cite{Dias:2012nf,Seery:2012vj},
  in which it was suggested by analogy with the formation
  of caustics that adiabaticity could be associated with
  regions where the dilation $\theta$ of a narrowly collimated
  bundle of trajectories becomes large.
  For a flow of inflationary trajectories
  the dilation is approximately given by a normalized
  sum of eigenvalues
  $\theta \approx \sum_i m_i^2/H^2$.
  Therefore
  the criterion $\theta \gg 1$ is necessary but not sufficient
  to yield an adiabatic limit in our sense.
  For example, it can happen that $\theta \gg 1$
  but more than one eigenvalue remains light.
  In this case the trajectories converge onto a sheet
  rather than degenerating to a single adiabatic
  trajectory.}
This implies a minimum suppression of $\e{-2.5} \approx 8 \times 10^{-2}$
in each isocurvature direction,
but usually substantially more.

\subsection{Agreement between pipelines}
\label{sec:code-comparison}
We now turn to the question of compatibility
between the two pipelines,
which enables us to assess the integrity of our
numerical computations.

\para{Trajectory-level agreement}
First, we have performed a number of spot-checks
to verify that the pipelines yield compatible
results given the same input data.
With matching initial conditions and
parameter values,
we typically find agreement to better than $0.1\%$ for $\ns$
and better than $0.5\%$ for the three-point correlation amplitudes.
We should regard these
as a lower limit
on the implementation error for individual trajectories
in the catalogue.

\begin{figure}
  \centering
  \includegraphics{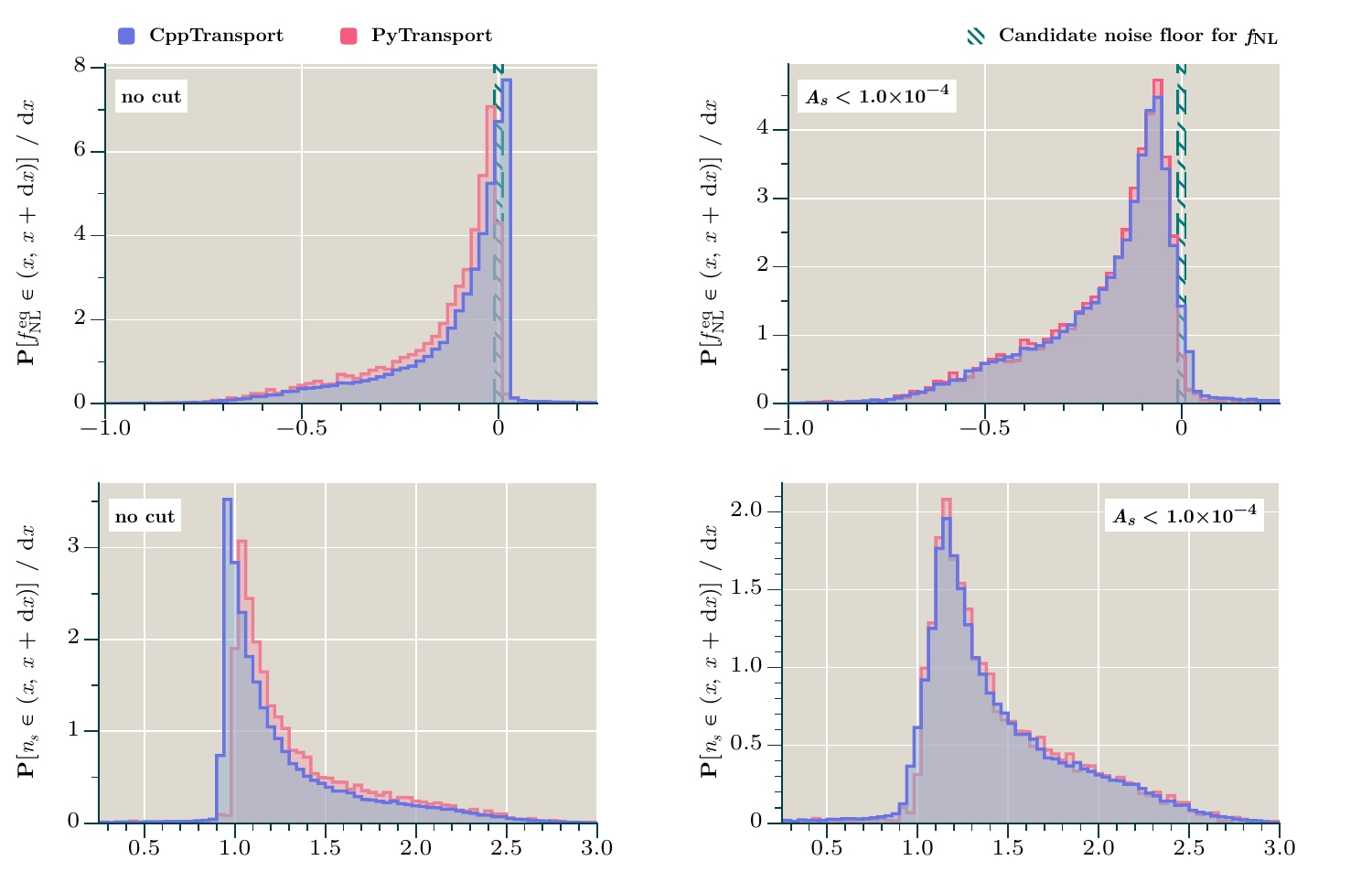}
  \caption[{\CppTransport}/{\PyTransport} comparison:
  Distribution of $\fNLeq$ and $\ns$]
  {\label{fig:code-compare-4up}Comparison of distributions
  for $\fNLeq$ and $\ns$
  derived from the {\CppTransport} (blue)
  and {\PyTransport} (red)
  pipelines.
  Before cutting on $\As$, the {\CppTransport} distributions
  contain {\precatalogueexact} trajectories
  and the {\PyTransport} distributions contain
  {\compareprecatalogueexact} trajectories.
  \semibold{Left column}:
  no cuts applied.
  The distributions are qualitatively similar
  but disagree in detail.
  \semibold{Right column}:
  applying the cut $\As < \Ascut$
  brings the distributions into agreement.
  The green hatched region
  is common to Figs.~\ref{fig:code-compare-4up},
  \ref{fig:discrepant-trajectories}
  and~\ref{fig:fNL-results}.
  Its interpretation is described in the second
  bullet point on p.~\pageref{bullet:noise}.}
\end{figure}

\para{Catalogue-level agreement}
Second,
to test agreement at the level of the catalogue as a whole,
we construct a `small' {\PyTransport} catalogue
using the same
$\phiUV$, $\DeltaMax$,
initial conditions and parameter priors used to construct
our primary catalogue.
In the left-hand column of Fig.~\ref{fig:code-compare-4up}
we show the resulting
distributions
of $\fNLeq$ and $\ns$
for {\CppTransport} (blue) and {\PyTransport} (red).
They are qualitatively similar but different in detail.
For both observables
the most obvious
difference is the change in
amplitude and location of the peak.
In the right-hand column
we show the same distributions with the cut
$\As > \Ascut$.
The amplitude and location of each peak, and the
structure of the tails, now show excellent agreement.

A similar effect can be achieved by
cutting out trajectories for which
$|\fNLeq| \lesssim \fNLeqcut$.
This removes the region around the peak
of the distribution for $\fNLeq$,
and (although not obvious
from Fig.~\ref{fig:code-compare-4up})
the resulting transfer of statistical weight
into the tails brings the distributions in agreement.
The nontrivial outcome
(also for the cut on $\As$) is that a \emph{single}
cut brings multiple distributions into alignment.
The underlying reason,
to be demonstrated in~\S\ref{sec:twopf-observables},
is that $\As$, $\ns$ and $\fNLeq$
are all highly correlated in this model.
However,
choosing to
cut on $\As$ removes marginally fewer
trajectories.

\begin{figure}
  \centering
  \includegraphics{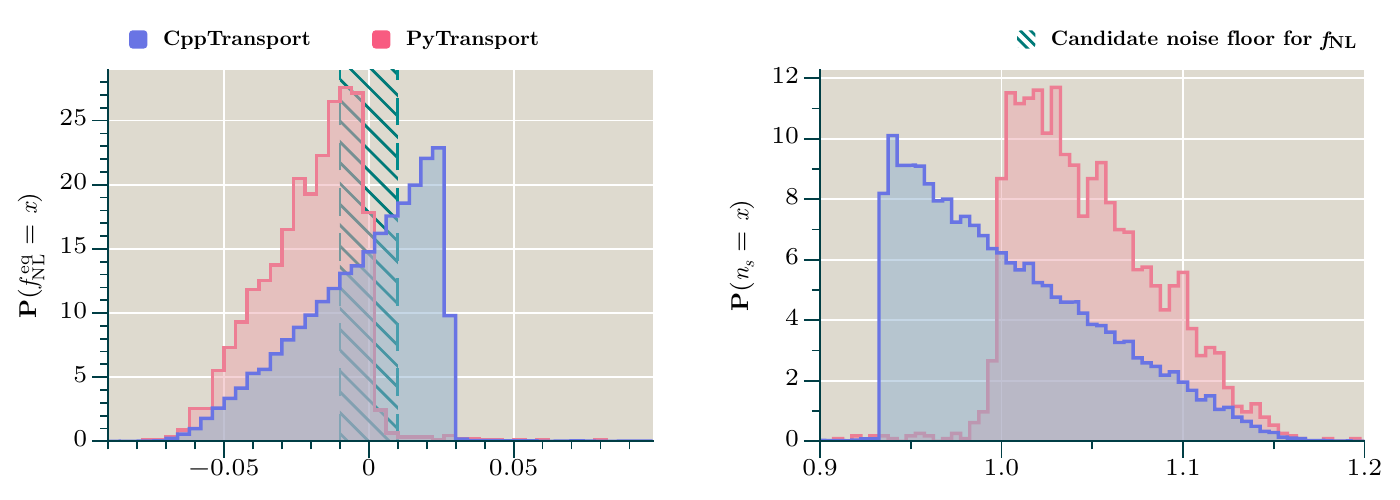}
  \caption[{\CppTransport}/{\PyTransport} comparison:
  Distribution of $\fNLeq$ and $\ns$ in the
  cut region $\As > 10^{-4}$]
  {\label{fig:discrepant-trajectories}Distribution
  of $\fNLeq$ and $\ns$ in the cut region
  $\As > \Ascut$.
  {\CppTransport} produces a smooth distribution
  over this range.
  The apparent falloff near
  the boundary value $\fNLeq = - 2.8 \times 10^{-2}$
  is a binning artefact,
  whereas the falloff near
  $\fNLeq = + 2.8 \times 10^{-2}$
  is the steep falloff to the right of the peak
  visible in Fig.~\ref{fig:code-compare-4up}.
  In comparison, the peak/cutoff
  values for the {\PyTransport}
  distribution occur at
  $\fNLeq > 0$
  and $\ns < 1$.
  The green hatched region
  is common to Figs.~\ref{fig:code-compare-4up},
  \ref{fig:discrepant-trajectories}
  and~\ref{fig:fNL-results}.
  Its interpretation is described in the second
  bullet point on p.~\pageref{bullet:noise}.}
\end{figure}

The origin of this discrepancy is not
completely clear. In the discussion below
we enumerate a number of possibilities that we
believe are \emph{not} the cause.
In Fig.~\ref{fig:discrepant-trajectories}
we plot the distribution of $\fNLeq$ and $\ns$
in the cut region,
which clearly exhibits the difference in structure
of the peak.
The $\fNLeq$ and $\ns$ distributions are both
characterized by a sharp peak and a one-sided tail.
The similarity in shape of the distribution is due
to the strong correlation between $\fNLeq$ and $\ns$.
The `missing' tail is so sparsely populated that there
are barely any samples beyond the peak,
which therefore serves as a cutoff.

We have considered a number of possible explanations
for this discrepancy.
\begin{itemize}
  \item First, it is not caused by
  disagreement between the pipelines
  for trajectories that populate the region
  $\fNLeq \gtrsim 0$, $\ns \lesssim 1.0$
  where the {\PyTransport} pipeline produces almost
  no statistical weight.
  Comparison of
  output from both pipelines shows excellent agreement
  for trajectories producing observables in this region.

  \item \label{bullet:noise} Second, one could imagine that
  small values of $\fNLeq$ are simply unreliable
  because they are dominated by noise.
  This explanation has the drawback that it would not
  naturally explain the disagreement in $\ns$.
  However, it is a possible interpretation
  for the distribution of $\ShiftEqFold \equiv \fNLeq - \fNLfold$
  in the right-hand panels of Fig.~\ref{fig:fNL-results}
  (see below).
  In a single-field model $|\ShiftEqFold|$
  should be proportional to $\nt$~\cite{Maldacena:2002vr}
  and therefore
  negligible on most trajectories,
  whereas Fig.~\ref{fig:fNL-results} shows that
  it is typically of order $10^{-2}$.
  This might happen if each $\fNL$
  were contaminated by noise at this level.
  The different behaviour of {\CppTransport}
  and {\PyTransport} could be ascribed to
  the differing ODE solvers.
  The green hatched regions in
  Figs.~\ref{fig:code-compare-4up},
  \ref{fig:discrepant-trajectories}
  and~\ref{fig:fNL-results}
  indicate the region that should be excluded
  in this interpretation.
  Coincidentally it is roughly the same region that must be
  excluded to bring the distributions reported by each
  pipeline into agreement.

  However, this interpretation does not seem viable.
  It is clear from Fig.~\ref{fig:discrepant-trajectories}
  that trajectories
  for which $\fNLeq$ falls within the green hatched
  region form part of a smooth distribution that extends
  to much larger values of $|\fNLeq|$.
  In the noise interpretation we would be obliged to assume
  that this structure can somehow be ascribed to
  properties of the noise---and that noise
  contaminates
  values of $\fNLeq$ for which Fig.~\ref{fig:fNL-results}
  gives no reason to believe it is significant.
  In our judgement this does not appear probable.

  \item Third, the discrepant region is associated
  with large values of $\As$---indeed, much larger
  than the observationally-allowed window.
  At sufficiently large $\As$ the tree-level approximation
  will break down, making
  all predictions unreliable~\cite{Dias:2015rca,Dias:2016rjq}.
  It is \emph{possible}
  (if perhaps unlikely) that $\As \approx 10^{-4}$ is already
  large enough for the leading loop correction
  to become important, especially if the power spectrum runs
  to large values on small scales.

  While this suggests we should already be skeptical
  of observables computed from trajectories yielding
  large $\As$, there seems no reason for
  {\CppTransport} and {\PyTransport} to fail in different
  ways if the numerical integration remains under control.
\end{itemize}

We therefore reject each of these proposed
explanations.
Instead, we apparently must conclude
that
the most likely explanation
is the performance of the samplers---that is,
a systematic
difference in the way the pipelines draw parameter
combinations leading to trajectories that populate this
region.
However, we have not managed to
identify an error in either pipeline that would
cause such a difference.
Therefore, to be conservative,
we impose the cut $\As > \Ascut$ when discussing observable
distributions in~\S\ref{sec:Results}.
In this region there is excellent agreement between
the pipelines, and we have good reason to believe that
the reported distributions are robust.

In any case,
as explained above, this cut (or a similar one) is likely to be
required to exclude trajectories for which
the tree-level approximation is inadequate.
Further work is required to improve our understanding
of all these issues.
\section{Results}
\label{sec:Results}
We now present key outputs from our sampling
procedure.
Agarwal et al.~\cite{Agarwal:2011wm}
previously discussed the relationship
between $\ns$ and $r$, but
the reliability of their
predictions was unclear in
regions of parameter space where the single-field
approximation was insufficient.
A subsequent analysis performed by Dias et al.~\cite{Dias:2012nf}
yielded
comparable results, but also certain differences of detail.
Unfortunately, although their
computation was more accurate, their catalogue of 564 inflating
trajectories was much smaller.
McAllister,
Renaux-Petel \& Xu~\cite{McAllister:2012am}
considered a significantly larger catalogue,
but their primary interest was in the frequency
of occurrence of multiple-field effects and they
did not report distributions for observables.
Relationships involving
the observables $\nt$, $\fNLeq$, $\fNLfold$
and $\fNLsq$
have not yet been studied.

In this section our intentions are twofold.
First, we use our primary catalogue of
{\cataloguesize} trajectories to characterize
correlations among the observables
$\As$, $\ns$, $\At$, $\nt$, $r$, $\fNLeq$ and $\fNLfold$.
This enables us to compare
(up to certain ambiguities)
with the results of
previous analyses.
However, despite their convenience, we emphasize that
these observables often have limited utility.
A full likelihood analysis is often needed to
determine the goodness-of-fit for each trajectory.
Second,
the small catalogues listed in
Table~\ref{Table:SmallStudies}
are used to study changes to
these distributions when we vary discrete features
of the model, such as the truncation of the potential
or our choice of initial conditions.

Except for
Figs.~\ref{fig:small-homog-massdist}--\ref{fig:small-homog-observables},
all distributions reported in this section
respect the cut $\As < \Ascut$
discussed in~\S\ref{sec:code-comparison}.

\subsection{Background evolution and mass spectrum}
\label{sec:results-background}
\begin{figure}
    \centering
    \includegraphics{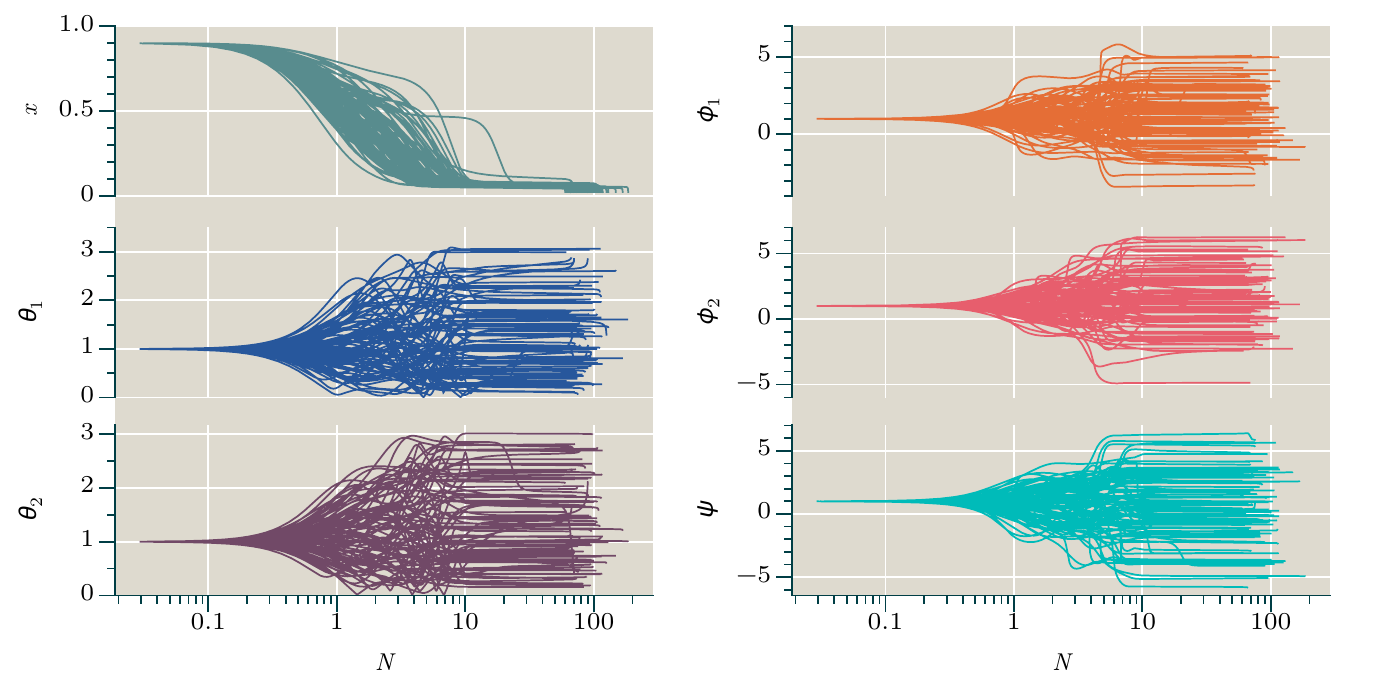}
    \caption[Background evolution for 100 sample trajectories]
    {\label{fig:results-background-ensemble}Background evolution for
    a subsample of 100 realizations. The horizontal scale shows
    e-folding number $N$ measured from the initial time.}
\end{figure}

\para{Field evolution}
In Fig.~\ref{fig:results-background-ensemble}
we plot the evolution of the background fields for a subset of
100 trajectories as a function of
e-folding number $N$
measured from $N=0$ at the initial data.
Typical trajectories show very similar evolution
for the radial position $x = r/\rUV$,
characterized by the onset of rapid motion after a few e-folds
followed by an extended loitering period
as the inflationary potential flattens
at small values of $x$.
Very similar behaviour was described by
Dias et al.~\cite{Dias:2012nf}.
The angular fields show more variability,
but in most cases their values become constant after $\sim 10$
e-folds.
This is
an indication that
trajectories frequently
evolve to an adiabatic
limit.

To express this quantitatively
we
apply the criteria for adiabaticity
given in~\S\ref{sec:adiabatic-limit},
according to which a trajectory is adiabatic if
each heavy isocurvature mode satisfies
$m^2 / H^2 > 3$.
Before applying any cuts,
we find that $64\%$ of trajectories
become adiabatic by the end of inflation.
For trajectories that respect the cut $\As < \Ascut$
the corresponding figure is $62\%$.
If the adiabaticity condition is relaxed
to $m^2 / H^2 > 1$ for the heavy
eigenstates, the fraction of
adiabatic trajectories increases
to $\sim 95\%$.
These proportions
are consistent between our pipelines
and appear roughly
consistent with the conclusions
of previous
studies~\cite{Agarwal:2011wm,Dias:2012nf,McAllister:2012am}.

\begin{figure}
    \centering
    \includegraphics{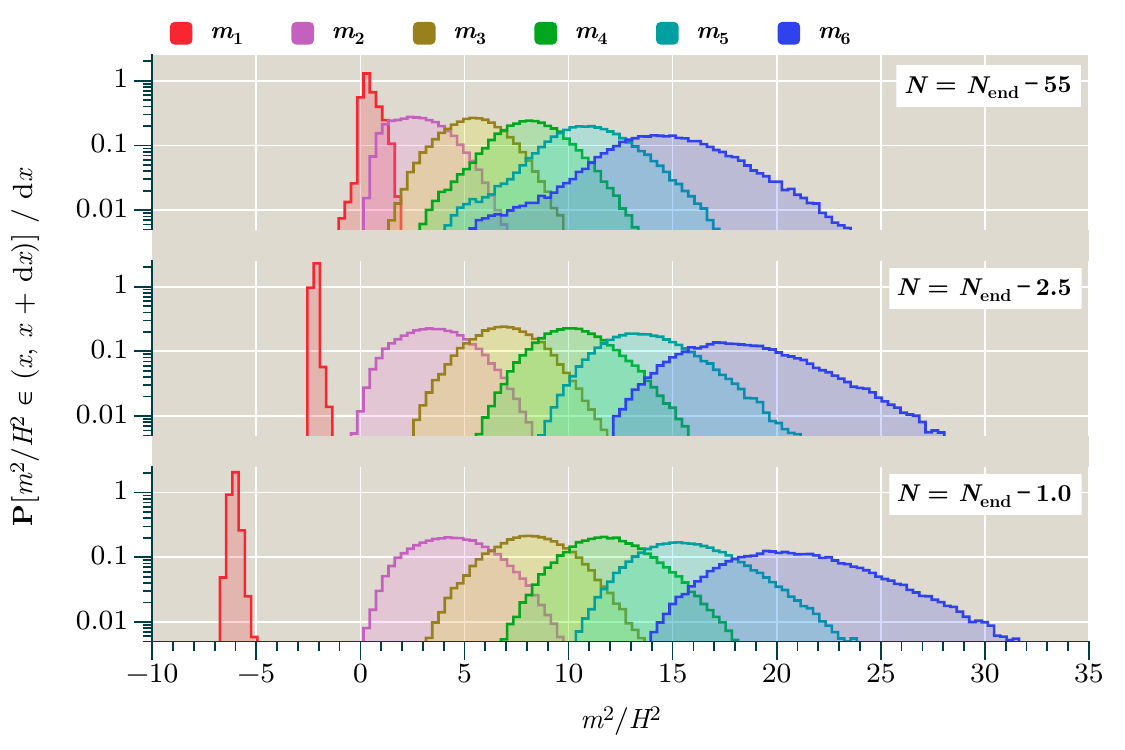}
    \caption[Evolution of mass spectrum in the primary catalogue]
    {\label{fig:results-mass-spectrum}Evolution of the mass spectrum
    over the period of observable inflation.
    Each panel contains six histograms showing the distribution of $m^2/H^2$
    for the ordered eigenvalues $m^2$ associated with the mass matrix.
    \semibold{Top}: 55 e-folds before the end of inflation.
    \semibold{Middle}: 2.5 e-folds before the end of inflation.
    \semibold{Bottom}: 1 e-fold before the end of inflation.}
\end{figure}

\para{Mass spectrum}
In Fig.~\ref{fig:results-mass-spectrum} we show the evolution of the mass
spectrum over the period of observable inflation. In each panel we overlay
histograms for the ordered eigenvalues of the mass matrix
${M^A}_B$ given in Eq.~\eqref{eq:mass-matrix}.
The top panel shows the mass spectrum 55 e-folds before the end of inflation,
which can be regarded roughly as the time of horizon exit
for modes contributing to the CMB.
The spectrum is relatively closely packed and evenly spaced.
The lightest mode is most sharply defined and extends to tachyonic values.
The middle and bottom panels show the spectrum at 2.5 e-folds
and 1 e-fold
before the end of inflation, respectively.
In the middle panel,
the spectrum is broader and the heavier modes have shifted to
slightly higher masses.
The lightest mode has become increasingly tachyonic.
Similar behaviour was reported in Ref.~\cite{McAllister:2012am}.
Between the middle and bottom panels
the distribution of higher-lying heavy modes is stable, but the
lightest mode moves even further towards tachyonic values
and develops a sizeable gap relative to the rest of the spectrum.
This behaviour is expected. On an adiabatic trajectory,
$\zeta = \delta \phi / \sqrt{2\epsilon}$
is conserved~\cite{Rigopoulos:2003ak,Lyth:2004gb},
and
$\epsilon$ is typically growing near the end of
inflation if there is a graceful exit.
Therefore $\delta\phi$ must also grow, requiring
the adiabatic direction in field-space to be a tachyon.
Before applying cuts
we find that all trajectories exhibit at least one tachyonic
mode
at 1 e-fold before the end of inflation,
but only $0.2\%$ exhibit a second tachyon.%
    \footnote{\label{footnote:tachyons}One
    might have some reservations regarding
    the emergence of multiple tachyonic
    states
    with large values of $|m^2/H^2|$
    given that our pipeline is based on tree-level
    codes.
    In single-field inflation,
    or multiple-field inflation near an adiabatic limit,
    this is harmless
    because $\zeta$ is exactly massless
    and therefore stable
    (even at loop-level)
    even though $\delta \phi$ is a
    tachyon~\cite{Senatore:2009cf,Assassi:2012et}.
    The situation with multiple tachyons is less clear.
    In this paper we continue to assume that
    a tree-level calculation gives
    an honest representation of the phenomenology,
    but we note
    that the issue does not yet appear to have been
    adequately explored in the literature.}
There are no trajectories exhibiting three tachyons.
The relative occurrence of two tachyonic modes
is essentially the same for trajectories that respect
the cut $\As < \Ascut$.

Our mass spectra are in qualitative but not quantitative agreement with
Dias et al.~\cite{Dias:2012nf}
and
McAllister et al.~\cite{McAllister:2012am}.
These references reported mass spectra near horizon exit for
a mode contributing to the CMB.
Dias et al.\ did not discuss the spectrum at later times,
whereas McAllister et al.\ found
only mild evolution between
horizon exit of CMB modes and the end of inflation.
In comparison with Dias et al.\
the shape of our
mass distribution at horizon-exit shows good
agreement, but the detailed numerical values of the masses are different.
This possibly points to a difference in
treatment of the `particular integral' modes~\eqref{eq:FluxesVPI}
sourced by bulk fluxes,
which contribute significantly to the masses of the heavy
eigenstates
(see~\S\ref{sec:small-studies}).

In comparison with McAllister et al.\
the numerical values of the masses are similar, but
the shape of the distribution is different.
We reproduce their conclusion that
contributions to the
mass matrix from mixing with the metric are generally
smaller than contributions from the Hessian
$\grad^A \grad_B V$ when CMB scales are leaving the horizon,
although the mixing contributions
increase in importance towards the end of inflation
and are eventually necessary to keep $\zeta$ conserved
on an adiabatic trajectory.
McAllister et al.\ observed
a mild tachyonic drift of the lightest eigenstate,
but at late times the effect is more significant in our realizations.
This may be attributable to gravitational mixing.
Further, they found that typically the masses $m_i^2$ of the $i=3,4$
and $i=5,6$ isocurvature states were degenerate
at the level of individual realizations.
We do not observe this degeneracy,
even if we approximate the mass matrix by the Hessian.
This apparently points to
an underlying difference in the construction of our potentials,
perhaps again caused by a differing treatment of the
flux-sourced contributions.

\begin{figure}
    \centering
    \includegraphics{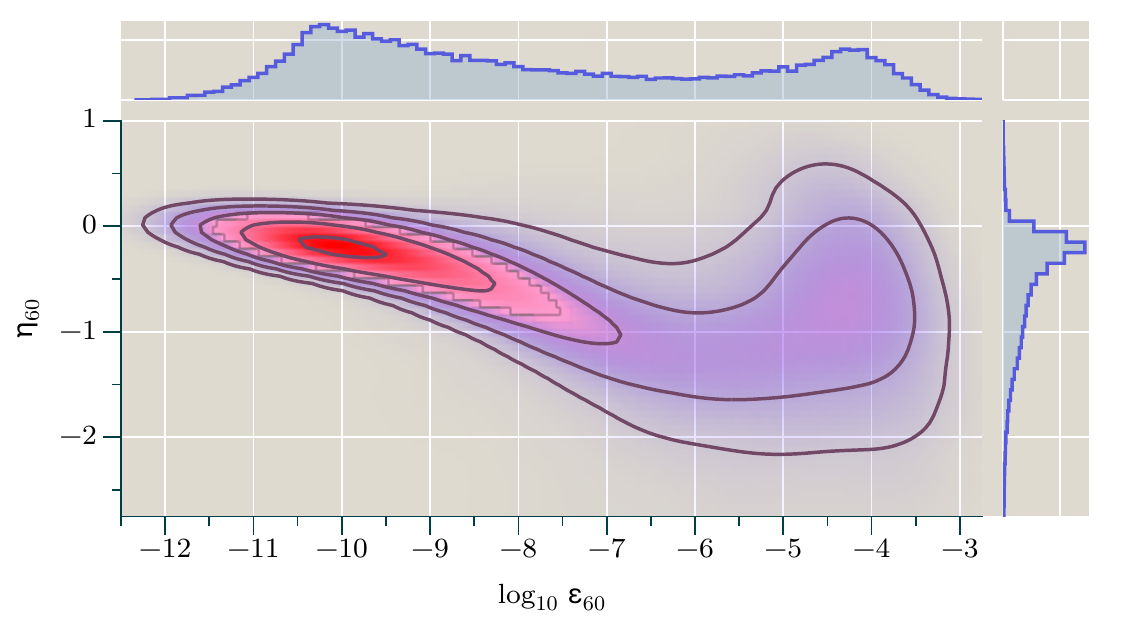}
    \includegraphics{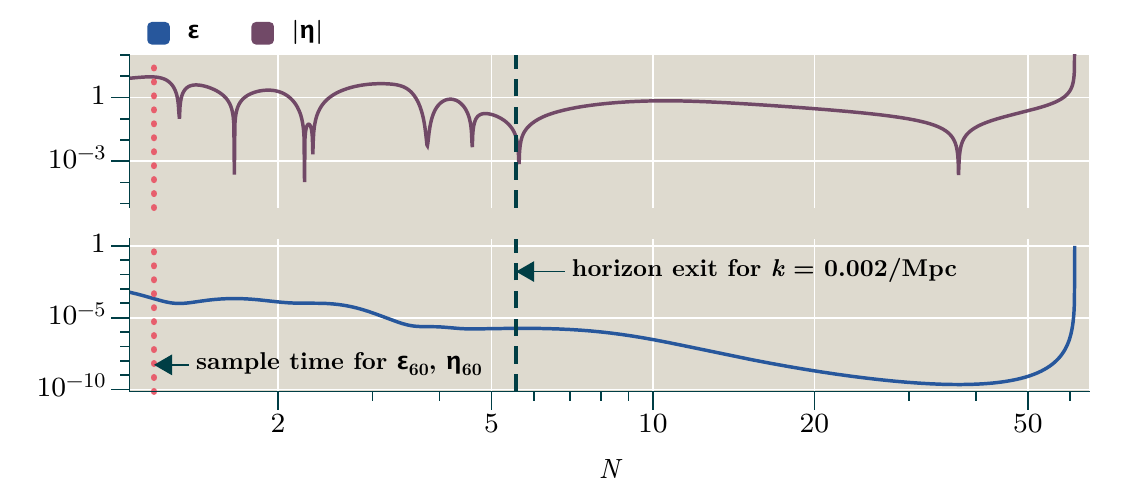}
    \caption[Distribution of slow-roll parameters
    at 60 e-folds]
    {\label{fig:usr-results}\semibold{Top}:
    Distribution of the slow-roll parameters
    $\epsilon$ and $\eta$ at 60 e-folds before the end of inflation,
    and their correlation.
    \semibold{Bottom}: representative time evolution of the
    $\epsilon$ and $\eta$ parameters.
    The vertical green dashed line shows the horizon exit
    time for $\kpiv = 0.002 \, \Mpc^{-1}$.
    The vertical orange dotted line shows the
    time at which we sample $\epsilon_{60}$, $\eta_{60}$.
    The principal features appearing in the plot are
    typical, including the slow evolution of $\epsilon$
    to very small values before a rapid increase
    as inflation ends.
    On this trajectory $|\eta|$ is briefly $\sim 5$
    for a period of roughly one e-fold
    just before horizon exit.
    In general,
    although excursions to large positive and negative values are present,
    they are transient.}
\end{figure}

\para{Slow-roll parameters}
In the upper plot of Fig.~\ref{fig:usr-results} we plot the distribution of
the slow-roll parameters
$\epsilon \equiv -\dot{H}/H^2$
and $\eta \equiv \d \ln \epsilon / \d N$,
and their mutual correlation.
These parameters are measured at $N=60$ e-folds prior to the
end of inflation, which we denote by the subscript `$60$'.
The $\epsilon_{60}$ distribution is bimodal.
It would be interesting to understand whether this is related
to the effect described by Frazer~\cite{Frazer:2013zoa}
in which peaks in the distribution function of some observable $O$
are related to critical points in the map
$O = O(\theta)$ giving $O$ as a function of the field-space
coordinates $\theta$ on a suitable initial hypersurface.

The main weight of the distribution is centred near $\epsilon_{60} \sim 10^{-10}$,
which is just a little larger than the typical value $\epsilon \sim 10^{-12}$
reported by McAllister et al.~\cite{McAllister:2012am},
although their evaluation time was not specified.
The secondary peak is near $\epsilon_{60} \sim 10^{-4}$.
In comparison, McAllister et al.\ reported only $7\%$ of samples
yielded $\epsilon > 10^{-8}$ and no samples with
$\epsilon > 10^{-6}$.
In our full catalogue we find $\sim 50\%$ of samples
yield $\epsilon_{60} > 10^{-8}$
and $\sim 30\%$ yield $\epsilon_{60} > 10^{-6}$.
We find no cases where $\epsilon_{60} > 10^{-3}$.
The conclusion is apparently that typical values
of $\epsilon$ in our catalogue are a few orders of
magnitude larger than those reported
by Agarwal et al.\ and McAllister et al.
The distribution of $\epsilon$ is also broader.
This is perhaps related to the inclusion of $\alpha$
in our sampling procedure,
which effectively adjusts $H$
while leaving gradients of the potential
unchanged.

\para{Ultra slow-roll inflation}
Fig.~\ref{fig:usr-results}
demonstrates that,
while $\epsilon_{60}$ is always very small, $\eta_{60}$
has excursions to large positive and negative values,
although rarely as large as $\eta_{60} \approx 6$.
This suggests that full-blown ultra slow-roll is unlikely
to occur, although there may be periods during which
$\epsilon$ is being suppressed---albeit less dramatically.
Inspection of a subsample of trajectories exhibiting large $|\eta_{60}|$
suggests this is the case.
In the lower plot of Fig.~\ref{fig:usr-results}
we show the time evolution of $\eta$
for a trajectory belonging to this subsample.
Excursions to modestly large $|\eta|$ \emph{are} present,
although on this trajectory
they occur before horizon exit of $\kpiv$.
These excursions are associated with periods
during which $\epsilon$ decays
in a way similar to the ultra slow-roll phenomenology,
but less extreme because $|\eta|$ is not as large.

We have not encountered any trajectories for which
the behaviour of $\eta$
clearly
supports a diagnosis of full-blown ultra slow-roll.
This does not exclude the possibility that,
for
some trajectories
in our catalogue, the initial conditions for observables
might be affected by transiently large $|\eta|$.
A full analysis of these effects, if they occur,
is beyond the scope of this paper.
Here we only note that
in both the full catalogue and the subsample
satisfying WMAP7 constraints on $\As$ at $\WMAPsigma$,
no more than $\sim 5\%$ of trajectories
exhibit $|\eta_{60}| > 2$.%
    \footnote{We use the WMAP7 limits rather than
    more recent Planck values to simplify comparison
    with earlier analyses that used WMAP data.}
Assuming this fraction is representative of the
proportion of trajectories that could be affected,
ultra slow-roll-like effects appear unlikely to
distort the final distribution of observables.

\para{$Q$ parameter and Wilson coefficients}
We now
consider the posterior distribution of
$Q$ and $\alpha$,
after applying the cut $\As < \Ascut$.
(For definitions, see the bottom two lines of
Table~\ref{Table:ParameterGlossary}.)
Both parameters affect the relative scale of terms in the potential,
and therefore influence the likelihood of finding a `delicate'
region of field space where the potential is sufficiently flat to inflate.
\begin{figure}
    \centering
    \includegraphics{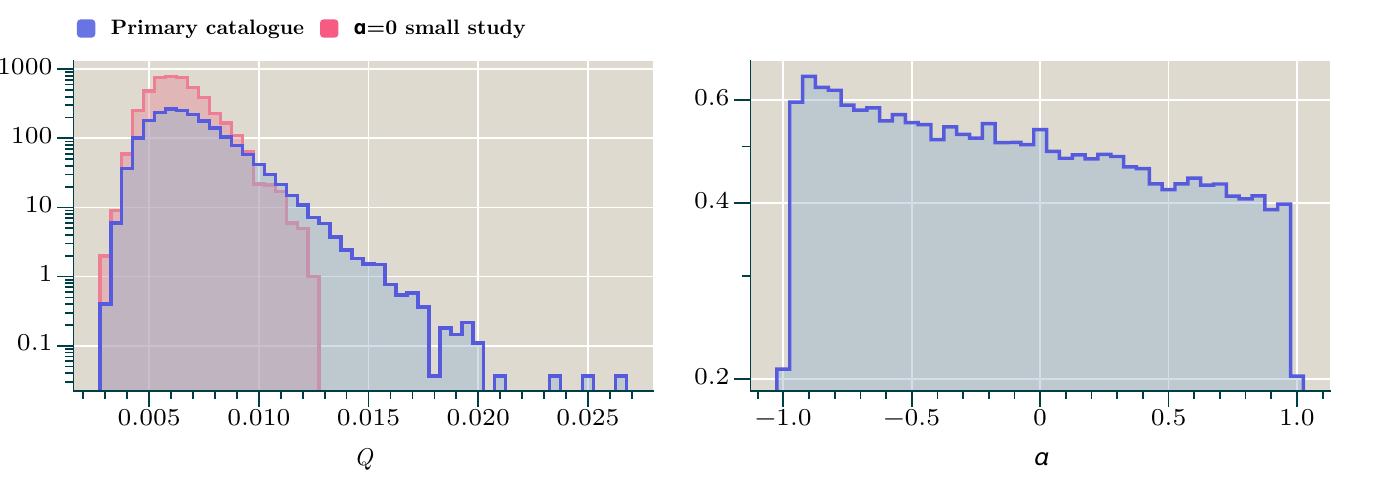}
    \caption[Posterior distribution of $Q$ and $\alpha$]
    {\label{fig:results-Q-alpha}Posterior distributions.
    \semibold{Left}: posterior for $Q$ (blue).
    The corresponding distribution for the $\alpha=0$
    study (red; see Table~\ref{Table:SmallStudies})
    is shown for comparison.
    \semibold{Right}: posterior for $\alpha$.}
\end{figure}

The left panel of Fig.~\ref{fig:results-Q-alpha} shows that the probability
of obtaining an extended epoch of inflation depends strongly on $Q$,
with successful realizations clustering tightly around
$Q \sim 0.006$.
This differs from the value $Q \sim 0.04$ reported by
Agarwal et al.~\cite{Agarwal:2011wm}.
However, as explained in that reference,
the narrow range of $Q$ for which prolonged inflation can be realized
reflects the need to carefully balance Coulomb attraction with
repulsion from the bulk contributions for typical values of the
Wilson coefficients $C_{LM}$.
Repulsion from the bulk terms scales with $Q$, and the precise
point of balance depends on how many terms are retained.
The numerical value of $Q$ therefore has no physical significance.
However, the discrepancy supports
our suggestion of a systematic
difference between typical trajectories in our catalogue
and those of Refs.~\cite{Agarwal:2011wm,Dias:2012nf}.
It is not yet clear whether the difference in $Q$ is caused
by the same difference responsible for the difference in
typical values of $\epsilon$.

Meanwhile, the $\alpha$ distribution is very roughly
flat
on both sides of $\alpha = 0$,
with a small bias to values near
the endpoint $\alpha = -1$
where the constant uplift
$V_0 + D_0$
to the
vacuum energy disappears.
This is a selection effect caused by the cut
$\As < \Ascut$.
We have already seen that
increasing $\alpha$ adjusts
$H$ while leaving gradients of the potential
unchanged, causing $\epsilon$ to decrease.
The net result is that $\dimP_\zeta$ must increase.
Imposition of
an upper limit on
$\dimP_\zeta$ will therefore depopulate the high-$\alpha$
part of the distribution.

When comparing with the results of Ref.~\cite{Agarwal:2011wm}
and Ref.~\cite{Dias:2012nf} it should be remembered that these
references set $\alpha=0$.%
    \footnote{Although this is not said explicitly in either
    reference, we understand it to be the case.
    We thank Nishant Agarwal for helpful correspondence
    on this issue.}
The $\alpha=0$ study
discussed in Table~\ref{Table:SmallStudies}
and~\S\ref{sec:small-studies}
(also plotted in Fig.~\ref{fig:results-Q-alpha})
confirms that the posterior $Q$ distribution
changes when we drop $\alpha$ as a sampling parameter.

\subsection{Two- and three-point observables}
\label{sec:correlated-observables}
We are now in a position to examine the correlation between
the summary statistics
$\As$, $\At$, $\ns$, $\nt$,
$r$, $\fNLeq$, $\fNLfold$
and $\fNLsq$.
To be clear, we recall that these
are defined in
Eqs.~\eqref{eq:Pzeta-def}--\eqref{eq:Ptensor-def},
\eqref{eq:spectral-indices-defs},
and~\eqref{eq:reduced-bsp-def}.
Only the relationship between the scalar spectral index
$\ns$ and the tensor-to-scalar ratio $r$
has previously been studied~\cite{Agarwal:2011wm,Dias:2012nf}.

\subsubsection{Two-point observables}
\label{sec:twopf-observables}
\begin{figure}
    \centering
    \includegraphics[scale=1.0]{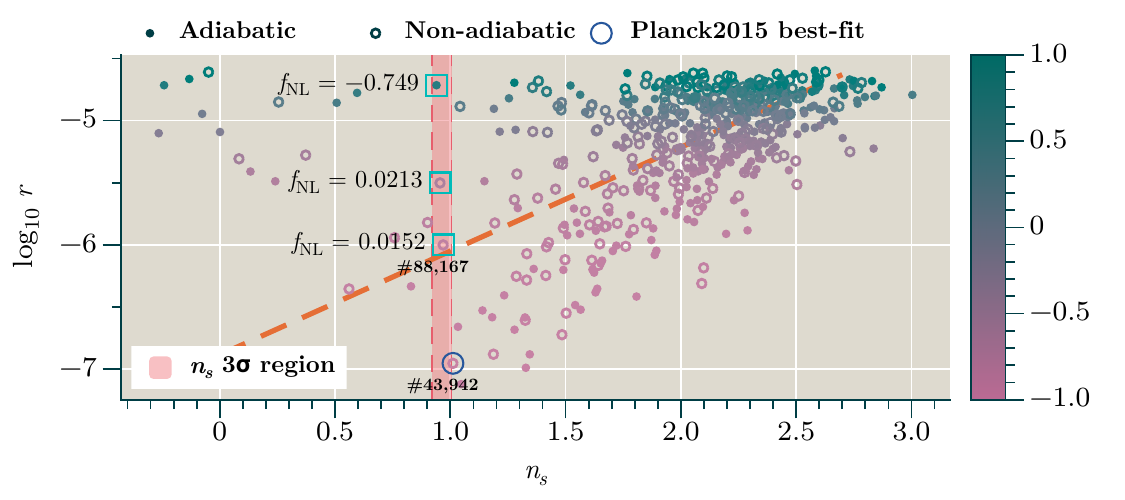}
    \caption[$\ns$ vs. $r$ for primary catalogue]
    {\label{fig:r-ns-detail}Correlation between $r$ and $\ns$
    for the 528 trajectories in our principal catalogue
    that satisfy the WMAP7 constraint on $\As$ at $\WMAPsigma$.
    The points are colour-coded by their value of $\alpha$.
    The dashed orange line shows the approximate
    fit $\log_{10} r \approx -6.78 + 0.785 \ns$,
    computed for the region $0.5 \leq \ns \leq 2.5$.
    The best-fit trajectory (\TrajBestfit) according to the Planck2015
    likelihood is highlighted (enclosed in an open circle).
    We also highlight trajectories whose values of $\As$ and $\ns$
    fall within the WMAP7 bound at $\WMAPsigma$ 
    (enclosed in open squares).
    Trajectories {\TrajBestfit} and {\TrajNofit} are discussed
    in the text, and are also marked in
    Figs.~\ref{fig:consistency-bound} and~\ref{fig:fNL-results}.
    Power spectra for Trajectory {\TrajBestfit}
    are given in Fig.~\ref{fig:traj43942},
    and spectra for Trajectory {\TrajNofit}
    are given in Fig.~\ref{fig:traj88167}.}
\end{figure}

\para{Failure of scale invariance and monotonicity}
In Fig.~\ref{fig:r-ns-detail}
we plot the relationship between $\ns$ and $r$
for the subsample
in which $\As$ is compatible with
the $\WMAPsigma$ WMAP7 constraint
$\As = (2.43 \pm 0.33) \times 10^{-9}$~\cite{Larson:2010gs}.
This distribution enables a comparison
with previous analyses.

We highlight
the WMAP7 $\WMAPsigma$ region
$\ns = 0.963 \pm 0.014$
in red~\cite{Larson:2010gs,Komatsu:2010fb}.
It is populated only by a handful of trajectories,
subject to the caveats mentioned below.
This is somewhat surprising.
In our other catalogues,
the proportion of trajectories that fall within the
WMAP7-allowed region is relatively larger,
perhaps by a factor of $\Or(10)$.
Presumably, this
happens because the allowed region corresponds to such
a small fraction of the model's large parameter space
that our sampling is not entirely representative
even with {\cataloguesize} trajectories.
However, in this paper our aim is not to optimize the
fit to current datasets, but to explore the statistical
distribution of observables for typical values of the
parameters appearing in the Lagrangian.
We expect the sampling in these typical regions to be
more accurately representative.

In Fig.~\ref{fig:r-ns-detail}
the trajectories that fall in the allowed region
are highlighted by enclosing cyan squares.
One of these is adiabatic; the other two are non-adiabatic.
The adiabatic trajectory has
an unusually
large amplitude of three-point correlations,
$\fNLeq = -0.749$, to be discussed
in~\S\ref{sec:threepf-observables} below.
One might expect the best-fit trajectory
to be one of these three.
However, according to the
Planck2015 $TT$+$TE$+$EE$ likelihood,
the best-fit is
Trajectory {\TrajBestfit}
with log-likelihood
$\ln \lik \sim -600$.
It produces a spectrum
for which $\ns$ measured at $k = 0.002 \, \Mpc^{-1}$
is marginally blue
and does not fall in the WMAP7 $\WMAPsigma$-region at all.
This trajectory is highlighted by an enclosing dark blue circle.
Note that the general trend with $\alpha$ is
opposite to the catalogue as a whole, in
which increasing $\alpha$ is correlated with
\emph{decreasing} $\epsilon \sim r/8$
as explained above.
In Fig.~\ref{fig:r-ns-detail}
the smallest $r$ are correlated with
the smallest
$\alpha$.

\begin{figure}
    \centering
    \includegraphics[valign=t]{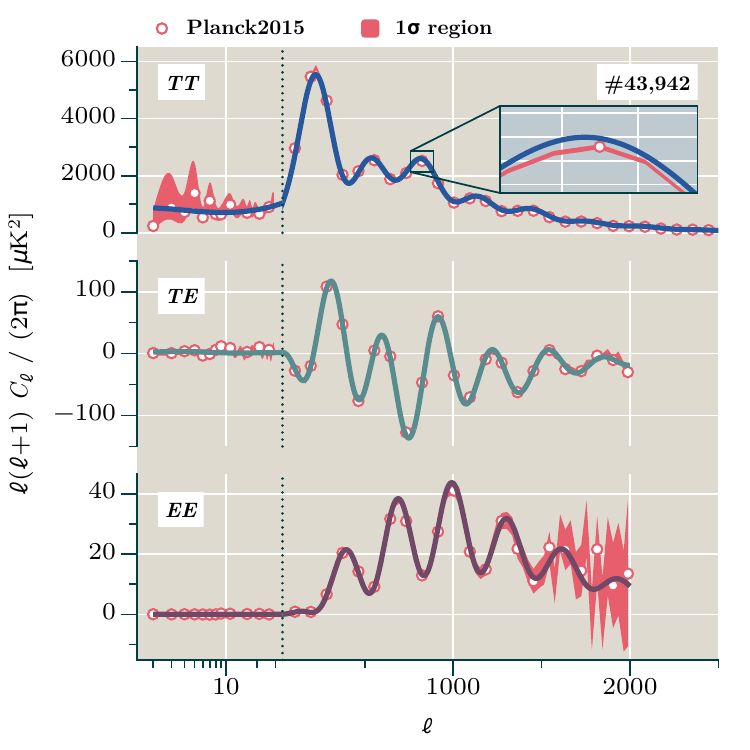}
    \includegraphics[valign=t]{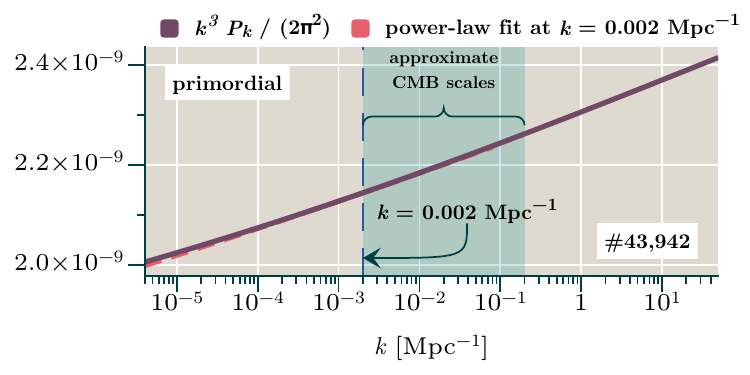}
    \caption[Power spectra for best-fit trajectory {\TrajBestfit}]
    {\label{fig:traj43942}Power spectra for
    our best fit trajectory
    {\TrajBestfit}.
    \semibold{Left}: angular spectra
    $C_\ell^{TT}$,
    $C_\ell^{TE}$ and $C_\ell^{EE}$
    for the temperature and $E$-mode
    fluctuations, and their cross-correlation.
    \semibold{Right}: primordial
    dimensionless power spectrum
    $\dimP_\zeta(k) = k^3 P(k) / (2\pi^2)$.
    The power-law fit at $k = 0.002 \, \Mpc^{-1}$ is marked.
    Note that the primordial spectrum has very little running, and
    is accurately given by the power-law fit
    over a wide range
    $10^{-6} \, \Mpc^{-1} \lesssim k \lesssim 50 \, \Mpc^{-1}$.
    The green shaded region highlights the scales
    $0.002 \, \Mpc^{-1} \, \lesssim k \lesssim 0.2 \, \Mpc^{-1}$
    that (approximately) contribute significantly
    to the angular power spectrum for $\ell \lesssim 2000$.}
\end{figure}

\begin{figure}
    \centering
    \includegraphics[valign=t]{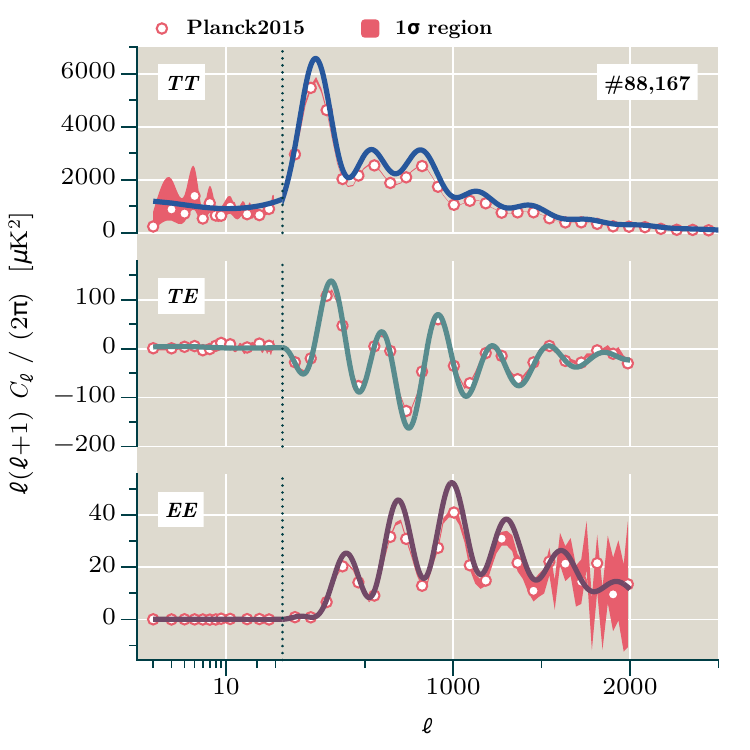}
    \includegraphics[valign=t]{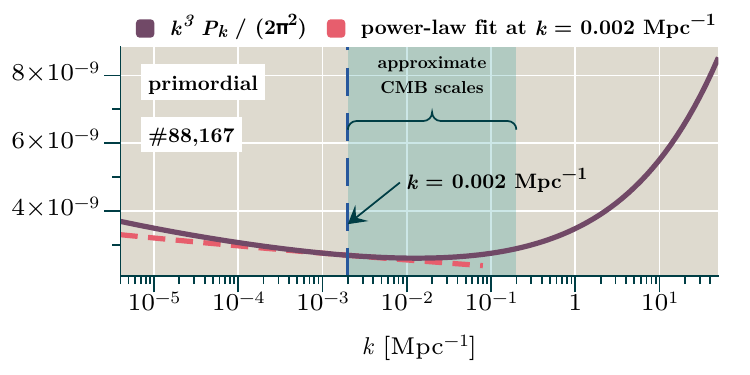}
    \caption[Power spectra for trajectory {\TrajNofit}]
    {\label{fig:traj88167}Power spectra for
    Trajectory {\TrajNofit}.
    The left and right panels
    match Fig.~\ref{fig:traj43942}.}
\end{figure}

\begin{figure}
    \centering
    \includegraphics{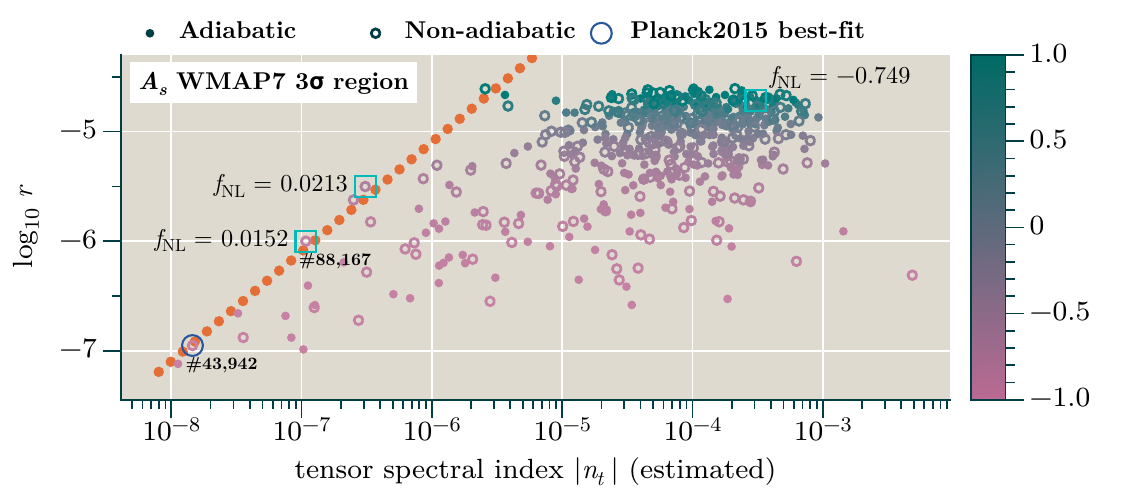}
    \caption[$r$ vs. $\nt$ for primary catalogue]
    {\label{fig:consistency-bound}Tensor-to-scalar ratio $r$ against
    $\nt$ for trajectories satisfying the WMAP7 constraint on $\As$
    at $\WMAPsigma$. The orange dotted line shows the consistency `bound' $r = 8|\nt|$,
    and the points are colour-coded by their value of $\alpha$.
    The best-fit trajectory is highlighted with a large open
    circle, as in Fig.~\ref{fig:r-ns-detail}.}
\end{figure}

In Fig.~\ref{fig:traj43942}
we plot the primordial power spectrum
for the best-fit trajectory,
together with
the corresponding angular spectra $C_\ell^{TT}$,
$C_\ell^{TE}$, $C_\ell^{EE}$,
and the  Planck2015 allowed region at
$1\sigma$~\cite{Aghanim:2015xee}.
In Fig.~\ref{fig:traj88167}
we plot the same quantities for
Trajectory {\TrajNofit},
which is one of the non-adiabatic trajectories
that falls in the WMAP7 $\WMAPsigma$
region for $\As$ and $\ns$.
For ease of comparison
these trajectories are labelled in
Figs.~\ref{fig:r-ns-detail}, \ref{fig:consistency-bound}
and~\ref{fig:fNL-results}.

The situation is clear.
Fig.~\ref{fig:traj88167}
demonstrates that
Trajectory {\TrajNofit} produces
a scale dependent, non-monotonic power spectrum.
Its form is similar to
a portion of
the characteristic non-monotonic
shape emphasized by Dias et al.~\cite{Dias:2012nf}.
Strictly, the angular spectra $C_\ell^{XY}$ computed
from this primordial spectrum are unreliable, because
$\dimP_\zeta(k)$
will be modified by quenching of isocurvature modes before 
it is communicated to the CMB.
However, for the present discussion this is not of primary
concern.
What \emph{is} important is that
the apparent near scale invariance
suggested by measurement of $\ns$
at $k = 0.002 \, \Mpc^{-1}$ is evidently fictitious;
in fact, $\ns$ varies significantly over the observable
range.
Our estimates for the $C_\ell$ show that this trajectory
significantly
overpredicts the correlation amplitude for $TT$ and $EE$.

Conversely, the best-fit trajectory
{\TrajBestfit}
produces a nearly featureless power-law spectrum
over the entire observable range of $k$;
only a very small running is visible for $k \lesssim 10^{-4}
\, \Mpc^{-1}$.
From the inset zoom panel in $TT$
it can be seen that although the overall fit is good,
the trajectory very slightly overpredicts the amplitude
near the third peak, but (not shown) underpredicts near
the first peak---%
as might be expected for a blue primordial spectrum.
It is probable that further trajectories can be found 
that yield an even better fit.
Dias et al.\ already observed that
nonmonotonic $\dimP_\zeta(k)$ occur relatively frequently
in this model due to the characteristic behaviour of
$\epsilon$ as trajectories roll
towards, through and away from the inflexion point.
We discuss this in more detail on p.~\pageref{para:redblue}
below.
To correctly assess the goodness-of-fit for these examples
we cannot rely on summary statistics such as
$\As$ and $\ns$, but instead require a realistic likelihood
calculation.

Note that because $H$ is usually close to constant
while observable scales are leaving the horizon, the tensor
power spectrum is commonly featureless.
Therefore $r$ will inherit scale dependence from $\dimP_\zeta(k)$,
and---if it were not so small---%
predictions for the observability of gravitational waves
would also require careful treatment.

\para{$\ns$--$r$ correlation}
Fig.~\ref{fig:r-ns-detail}
should be compared
with Fig.~9 of Agarwal et al.~\cite{Agarwal:2011wm}
and Fig.~17 of Dias et al.\ (arXiv version 2)~\cite{Dias:2012nf}.
The same qualitative features are visible in all these plots, but there
are quantitative differences.
In Fig.~\ref{fig:r-ns-detail}
the general trend is for larger values
of $\ns$ to be correlated with
larger values of $r$,
with no clear separation between adiabatic and
non-adiabatic trajectories.
The samples fill out a wedge-shaped region,
producing considerable scatter
for very red values of $\ns$.
Most trajectories cluster in the opposite limit
near $\ns \approx 2$,
where the spectrum is very blue.
These trajectories have an unfavourable
CMB likelihood.

In Agarwal et al.\ the observed values of $\ns$ cover
$0.93 \lesssim \ns \lesssim 1.10$
and the corresponding values of $r$
cover
$-14 \lesssim \log_{10} r \lesssim -11$.
Our values of $\ns$ are typically rather more blue,
which could be attributed to the single-field
approximation used by Agarwal et al.
In Dias et al.~\cite{Dias:2012nf}
the corresponding range of
$r$ is not clear because of the choice of axes,
but is plausibly
$-9 \lesssim \log_{10} r \lesssim -7$;
the authors of Ref.~\cite{Dias:2012nf} did not comment on
the discrepancy with Ref.~\cite{Agarwal:2011wm}.
Our range of $r$ is different again but closer to Dias et al.,
spanning roughly
$-4.6 \lesssim \log_{10} r \lesssim -7.0$.
The reason for this substantial variation in $r$ between
different analyses is not clear.
Trajectories with
small $\epsilon$ in the range
$-12 \lesssim \log_{10} \epsilon \lesssim -10$
\emph{are} present in our catalogue, which would correspond to
values of $\log_{10} r$ much closer to those of
Agarwal et al. However,
they are associated
with $\As$ that are outside the observable window.
The $\alpha=0$ study discussed in~\S\ref{sec:small-studies}
suggests that the distribution of
observables is not significantly affected
by our inclusion of $\alpha \neq 0$ during sampling,
so the discrepancy is
apparently caused a structural
difference in the potential
or a difference in sampling methodology.

One possible explanation is variation of the energy scale at which
the brane `loiters' near the inflexion point in its potential.
The inflexion point is itself a consequence of delicate
cancellations between the attractive
Coulomb force and repulsive bulk contributions,
and (as has already been said)
the exact location of the balance point depends on
the treatment of the bulk terms.
Small changes to the cancellations that produce the inflexion
point could perhaps change the corresponding value of $V$
significantly.

Second, we sample parameters at the fixed scale $\kpiv = 0.002 \, \Mpc^{-1}$,
yielding a range of horizon-exit times
corresponding roughly to $55 \lesssim \Npiv \lesssim 57$, where
$\Npiv$ measures horizon exit of the scale $\kpiv$ in terms of e-folds prior to
the end of inflation.
Agarwal et al.\
reported values at the fixed horizon exit time
$\Npiv = 60$~\cite{Agarwal:2011wm} , and
Dias et al.\ used $\Npiv = 55$~\cite{Dias:2012nf} .
For slow-roll inflation such small shifts in the evaluation time often lead
only to small shifts in observables, but in this model
the character of the trajectories can change depending
on the relative position of the initial conditions and
the inflexion point.
(For example, compare the power spectrum
for Trajectory {\TrajNofit} given in Fig.~\ref{fig:traj88167}.)
A systematic difference in
the evaluation point
could perhaps
modify statistical properties of the observables.

Finally, Agarwal et al.\ used a simple single-field approximation to compute
$\As$ which will produce unreliable estimates where multiple-field effects
are significant.
We have already noted that
the same applies to the spectral index, especially if the power spectrum
is not close to scale invariance.
This may explain the different range of $\ns$ observed
between our catalogues.
To decide which of these possible causes, if any,
contribute significantly
to the differences in the $\ns$--$r$
relation would require a forensic analysis
of each implementation, which is beyond the scope of this paper.
We comment further on these issues in~\S\ref{sec:discussion}.

Moving beyond the difference in normalization,
the form of the $\ns$--$r$
relationship is qualitatively similar
in all analyses.
In Agarwal et al.\ the correlation is
tightest and the relationship is approximately linear.
In Dias et al.\ and our Fig.~\ref{fig:r-ns-detail}
the correlation is compatible with linearity, but there is
considerable scatter and it is not clear that other
functional forms are excluded.
Assuming linearity, however, the slope of the relationship differs
between analyses with our Fig.~\ref{fig:r-ns-detail}
being shallowest and Dias et al.\ being steepest.

\para{Red and blue spectral indices}
The majority of our trajectories yield blue spectral indices,
although Figs.~\ref{fig:traj43942}
and~\ref{fig:traj88167} show that any numerical value
for $\ns$ should be interpreted with care.
The same effect was reported by
Agarwal et el.~\cite{Agarwal:2011wm},
Dias et al.~\cite{Dias:2012qy}
and McAllister et al.~\cite{McAllister:2012am},
who all observed that when $\epsilon \ll 1$
(which is the case in the vicinity of the inflexion point),
the spectral index inherits its sign from $V''$,
where $'$ denotes a derivative in the adiabatic direction
in field space.
Therefore, modes leaving the horizon before the
inflexion point are blue-tilted,
whereas those leaving the horizon after the inflexion point
are red-tilted. This is one cause of
non-monotonicity in the power spectrum.
\label{para:redblue}

Agarwal et al.,
Dias et al.\ and McAllister et al.\ noted that this
effect limited the number of trajectories producing
red $\ns$, because it is more difficult to obtain
sufficient e-folds of inflation after crossing the inflexion
point.
Our sample contains populations of trajectories with
red and blue values of $\ns$ that conform to these expectations,
which is a primary driver for the preponderance of blue
values $\ns > 1$ in Fig.~\ref{fig:r-ns-detail}.
We also find a relatively larger population of
trajectories yielding red $\ns$
where horizon exit occurs \emph{prior} to crossing the
inflexion point,
but \emph{before} approaching an adiabatic limit.
This invalidates expectations based on the sign of $V''$.
Further, these trajectories typically pass through a sequence
of critical points
where slow-roll may not be a good approximation.
This population of red-$\ns$ trajectories
does not appear to have been identified
in previous analyses.
However,
their spectra clearly cannot be monotonic,
so it is not yet clear whether their detailed properties can
be observationally acceptable.

\para{Consistency equation}
In single field models the tensor-to-scalar ratio
and tensor spectral index are related to leading
order in slow-roll by the
`consistency relation' $r = -8 \nt$~\cite{Copeland:1993zn,Copeland:1993jj}.
In multiple-field models this is weakened to an inequality
$r \leq -8 \nt$, also valid only to leading order in slow-roll
and assuming that all modes contributing to $\zeta$ are massless%
    \footnote{The massless condition was not discussed in Ref.~\cite{Sasaki:1995aw}
    and has frequently not been stated in the literature, but it is required.
    In the analysis of Ref.~\cite{Sasaki:1995aw} it appears in the assumption
    that the scalar two-point function is proportional to the
    kinetic mixing matrix.
    This need not be true if the fields have a non-negligible mass matrix,
    as may be the case in the $\brane{3}$/$\antibrane{3}$ model.
    For a similar discussion from a different perspective, see~{\S}4.3.4
    of Ref.~\cite{McAllister:2012am}.}%
~\cite{Sasaki:1995aw}.
This follows from the Cauchy--Schwarz inequality applied
to the projection
$\zeta = \tensor{N}{_{\pidx{A}}} \delta \tensor{\pcoord{X}}{^{\pidx{A}}}$
from field fluctuations
$\delta \tensor{\pcoord{X}}{^{\pidx{A}}} = (\delta X^A, \delta \pi^B)$
onto $\zeta$.
In Fig.~\ref{fig:consistency-bound} we plot
$r$ against our estimated $\nt$
for trajectories satisfying the WMAP7 bound on $\As$ at $\WMAPsigma$.
The `consistency bound' is represented by the orange dotted line,
and is respected by a clear majority of trajectories.
A small number of trajectories exhibit marginal transgressions.
The most likely
explanation is that these are effectively single-field models
that \emph{should} lie exactly on the bound, but our
procedure for estimating $\nt$ has produced a result that is
slightly too small.

In the full catalogue we observe a population of trajectories
that exhibit more significant (but still not dramatic)
transgression of the `consistency bound'.
There are several possible explanations of this effect, including
misprediction of $\nt$ as suggested above. However, it is also possible
that these are trajectories for which the massless approximation fails
and the fields become \emph{anticorrelated} at horizon exit,
leading to a reduction in the final power spectrum amplitude
below what would be predicted based on the adiabatic mode alone.
If this reduction is sufficiently dramatic it could
cause a violation of the massless consistency bound.

This anticorrelation
was observed by McAllister et al., who described it as
`destructive interference'~\cite{McAllister:2012am};
see the discussion in~{\S}4.3.4 of this reference.
Our results apparently reproduce their observations.
After performing `by hand' fits to the
tensor power spectrum in order to obtain
the most accurate possible measurements of $\nt$,
we find that residual violations of the consistency
bound are no more than a few percent.
It is not yet clear whether this is
a genuine effect caused by violation of the massless
condition,
or is simply an unmodelled error in fitting
to the tensor power spectrum.
To produce a convincing demonstration of violation
of the consistency bound, or otherwise,
would require an investment
in higher-quality numerical computations of the
tensor spectral index.
It would be interesting to return to this question in future work.

\subsubsection{Three-point correlations}
\label{sec:threepf-observables}

\para{Equilateral and folded configurations}
Next consider the amplitude of three-point correlations, measured by
$\fNLeq$ and $\fNLfold$,
which
represent the major new results
presented in this paper.
(Recall that we sample these quantities at a fixed
scale $k_t = 3\kpiv = 3 \times 0.002 \, \Mpc^{-1}$.
For details of our observables, see~\S\ref{sec:cpptransport-pipeline}.)
In general $\fNLeq$ and $\fNLfold$
need have no simple relation,
but in a single-field model they are connected by Maldacena's
formula%
    \footnote{Compare Eq.~(4.9) of Maldacena, arXiv version~5~\cite{Maldacena:2002vr}.
    Note there is a sign flip of $\fNL$ between this equation
    and Eq.~\eqref{eq:maldacena-relation},
    because Ref.~\cite{Maldacena:2002vr} defined $\fNL$
    with the opposite sign convention.}
\begin{equation}
    \label{eq:maldacena-relation}
    \fNL(k_1, k_2, k_3)
    =
    - \frac{5}{12} (\ns-1)
    + g(k_1, k_2, k_3) \nt ,
\end{equation}
where $g(k_1, k_2, k_3)$ is a calculable function
of the momenta $k_i$~\cite{Maldacena:2002vr}, but does not
depend on slow-roll parameters or other model-dependent data.
Hence, in a single-field model, the `residual'
$\ShiftEqFold \equiv \fNLeq - \fNLfold$
is proportional to $\nt$ multiplied by a fixed factor
depending only on the momentum configuration.

\begin{figure}
    \centering
    \includegraphics[scale=1.0]{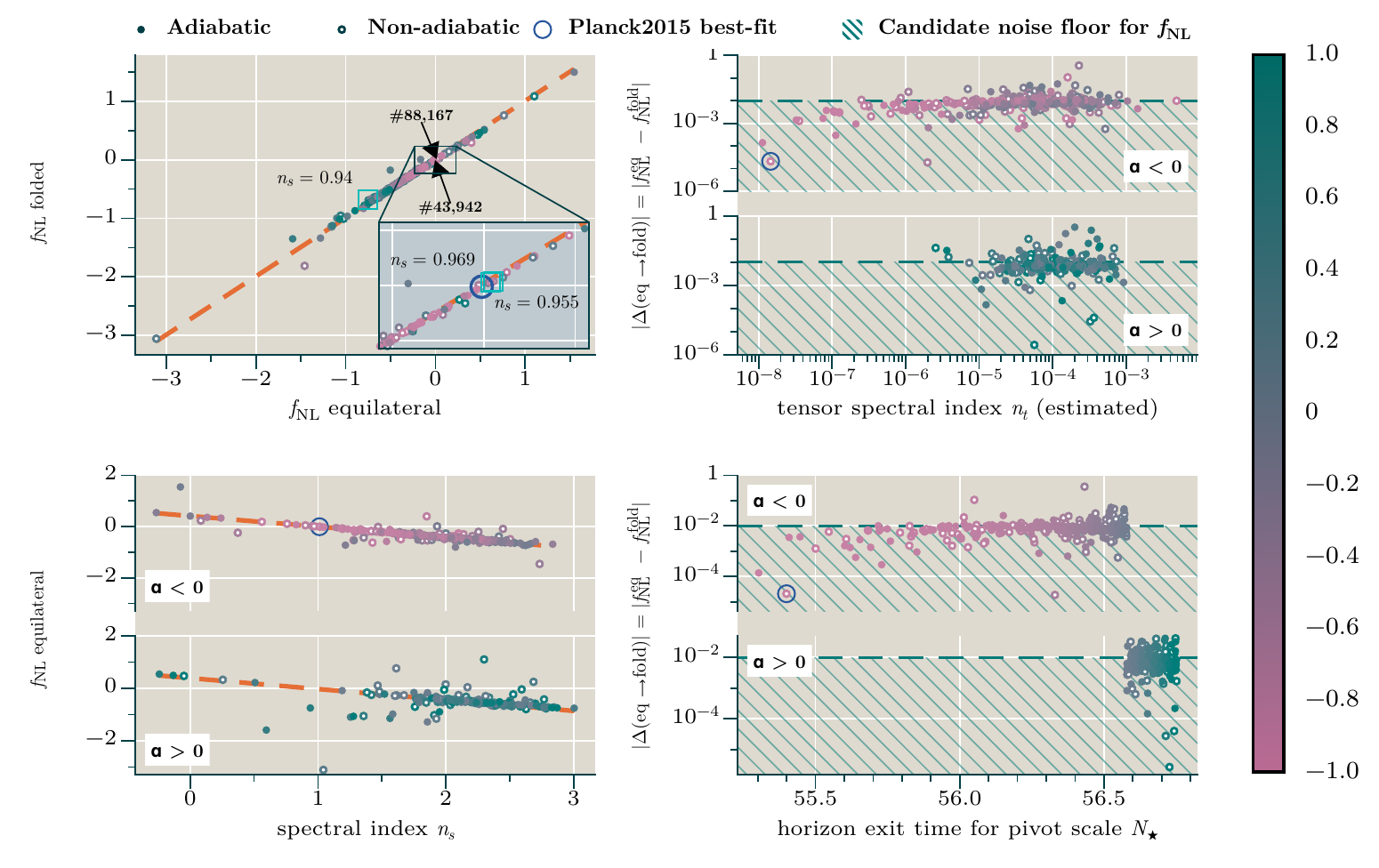}
    \caption[$\fNLeq$ vs. $\fNLfold$ for primary catalogue]
    {\label{fig:fNL-results}Relation between three-point
    correlation amplitudes
    in equilateral and folded configurations
    with $k_t = 3 \kpiv = 3 \times 0.002 \, \Mpc^{-1}$.
    (For details of the three-point configurations we sample,
    see~\S\ref{sec:cpptransport-pipeline}.)
    Plotted points correspond to trajectories
    that
    satisfy the $\WMAPsigma$ 
    WMAP7 constraint on $\As$.
    Except in the top-left panel, the populations with
    $\alpha > 0$ and $\alpha < 0$ are plotted separately
    to aid comparison.
    \semibold{Top left}: $\fNLfold$ against $\fNLeq$.
    The dashed orange line has the functional form $\fNLfold = a + \fNLeq$,
    where $a = 8.3 \times 10^{-3}$ is fit to the measured values
    with correlation coefficient $R > 0.995$.
    \semibold{Bottom left}: $\fNLeq$ against $\ns$.
    The dashed orange line has functional form
    $\fNLeq = b - (5/12)(\ns-1)$ as in Eq.~\eqref{eq:maldacena-relation}.
    For $\alpha > 0$ we find $b = -0.0270$ and for $\alpha < 0$
    we find $b = -0.00286$.
    \semibold{Top right}: the `residual'
    $|\ShiftEqFold| \equiv |\fNLeq - \fNLfold|$
    against $\nt$ (estimated from $\epsilon_{60}$ as described
    in~\S\ref{sec:computational-issues}).
    \semibold{Bottom right}: $|\ShiftEqFold|$
    against $\Npiv$, where $\Npiv$ measures the horizon exit time
    (in e-folds before the end of inflation) of the pivot scale
    $\kpiv = 0.002 \, \Mpc^{-1}$.
    The interpretation of the green hatched region
    was discussed in~\S\ref{sec:code-comparison}.
    \semibold{Colour bar}: in each panel, points are colour-coded by
    their value of $\alpha$.
    In all four panels
    the best-fit trajectory is highlighted
    as in
    Figs.~\ref{fig:r-ns-detail}--\ref{fig:consistency-bound}.}
\end{figure}

In Fig.~\ref{fig:fNL-results} we plot
$\fNLeq$ and $\fNLfold$
for trajectories that
satisfy the WMAP7 constraint on $\As$ within $\WMAPsigma$.
As in Figs.~\ref{fig:r-ns-detail} and~\ref{fig:consistency-bound}
the points are colour-coded by their value of $\alpha$,
with values near $\alpha = +1$ coded green and
values near $\alpha = -1$ coded purple.
Previous analyses have assumed $\alpha=0$.
We highlight the Planck2015 $TT$+$TE$+$EE$ best-fit trajectory,
described above,
which has small amplitudes
$\fNLeq \approx -4.73 \times 10^{-3}$
and
$\fNLfold \approx -4.75 \times 10^{-3}$.
We will see below that these values
are typical when both $\As$ and $\ns$ fall in the
observationally-allowed window.

The top-left panel
of Fig.~\ref{fig:fNL-results}
shows tight correlation between $\fNLfold$ and $\fNLeq$,
as would be predicted by~\eqref{eq:maldacena-relation}.
In single-field cases the small scatter is due to the smallness of $\nt$.
However, even in cases where multiple-field effects are important,
this panel shows that $\fNLeq$ and $\fNLfold$ remain highly
correlated.
This correlation does not depend significantly on $\alpha$
and is a consequence of the bispectrum `shape' usually being close to
local~\cite{Babich:2004gb},
which makes $\fNL(k_1, k_2, k_3)$ nearly independent of configuration.
However, the scatter is larger over the full catalogue, showing that
three-point correlations do not always have an exactly local shape
when $\As$ is not in the observational range.

The bottom-left panel shows $\fNLeq$ as a function of $\ns$.
For ease of comparison we break out the populations $\alpha > 0$
(bottom plot) and $\alpha < 0$ (top plot) separately.
In both cases
a group of effectively single-field trajectories is visible
that accurately match the dashed orange lines,
each of
which is chosen to have the slope $-5/12$
predicted by~\eqref{eq:maldacena-relation}.
This is especially clear for small values of $\ns$, where almost all
points cluster close to the line.
For larger values of $\ns$ the scatter becomes more significant,
especially for $\alpha > 0$,
and may
indicate that multiple-field effects are relevant in this region.
Even here, however, the $-(5/12)(\ns-1)$ dependence
predicted by~\eqref{eq:maldacena-relation}
is strikingly well reproduced.
Clearly, this dependence is the principal
driver for large values of $|\fNL|$.
A notable exception is the single adiabatic trajectory
in the WMAP7 $\WMAPsigma$ regions for both $\As$
and $\ns$ (see Fig.~\ref{fig:r-ns-detail}).
This has $\ns - 1 \approx -0.06$ but
a large amplitude
$\fNLeq = -0.749$, derived from
an abrupt change of angular minimum
to be discussed in~\S\ref{sec:abrupt-change-minima}.
However, the plot shows that
examples of this kind are relatively rare.

In the top-right panel we plot the absolute value
of the `residual' $|\ShiftEqFold|$
as a function of our estimated $\nt$.
To be clear, we repeat that
these are computed using the analytic
approximation $\nt = - 2 \epsilon_{60}$.
According to
Eq.~\eqref{eq:maldacena-relation},
$|\ShiftEqFold|$
should be proportional to $\nt$,
which
on logarithmic axes
would
correspond to a line with unit slope.
This is \emph{not} what we find;
instead, $|\ShiftEqFold| \approx 10^{-2}$
is roughly constant,
independent of $\nt$.
Here there is a clear segmentation of the populations $\alpha \gtrless 0$.
The $\alpha > 0$ population has more scatter
and is restricted to $\nt \gtrsim \times 10^{-6}$.
The $\alpha < 0$ population extends
(with decreasing density) to $\nt \sim 10^{-8}$
with roughly constant amplitude.

The possibility that
this behaviour is caused by
contamination from numerical noise was
rejected in~\S\ref{sec:code-comparison}.
Instead, we must apparently attribute it
to multiple-field effects.
The size of the effect is comparable to the
running from equilateral to squeezed configurations
to be discussed in~\S\ref{sec:squeezed-configurations}
below,
which is consistent with a multiple-field origin.

Finally, the bottom-right panel shows $|\ShiftEqFold|$ as a function
of $\Npiv$, the horizon-exit time for $\kpiv = 0.002 \, \Mpc^{-1}$.
The structure is very similar to the top-right panel,
but the segmentation is even clearer
with each population confined to nearly exclusive regions.
The $\alpha > 0$ population is restricted
to $56.6 \lesssim \Npiv \lesssim 56.8$,
whereas the $\alpha < 0$ population
is restricted to the wider range
$55.3 \lesssim \Npiv \lesssim 56.6$.

\begin{figure}
    \centering
    \includegraphics[scale=1.0]{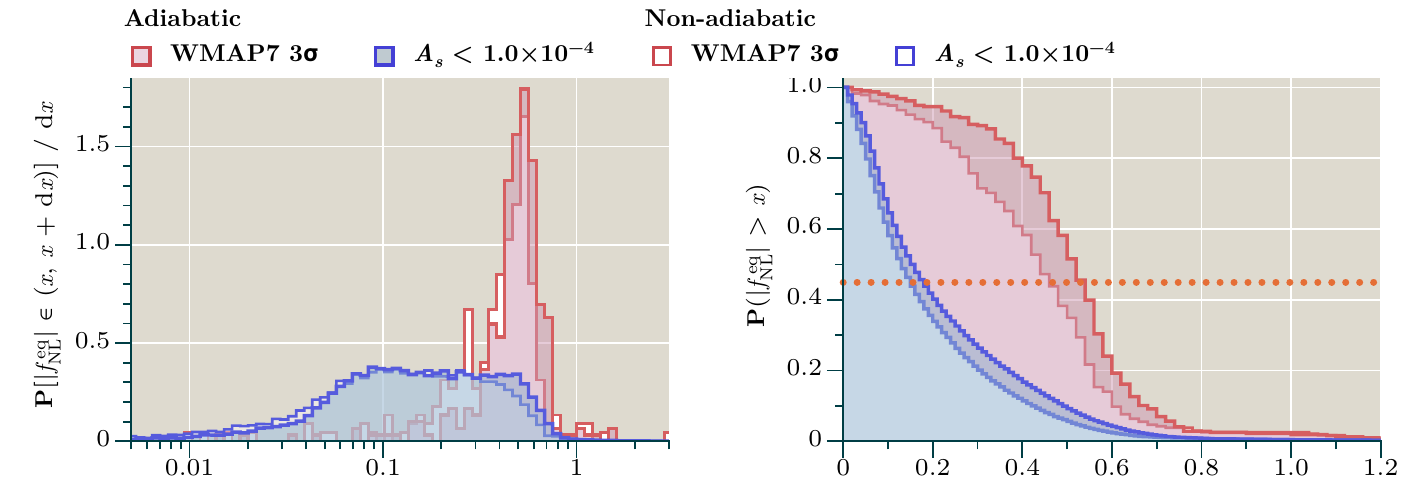}
    \caption[Distribution functions for $|\fNLeq|$]
    {\label{fig:fNLeq-CDF}Distribution functions for
    $|\fNLeq|$.
    \semibold{Left}: probability density function
    $\Prob\big[ |\fNLeq| \in (x, x + \d x) \big] / \d x$
    for different cuts on the catalogue.
    The blue histogram applies the minimal cut $\As < \Ascut$.
    The red histogram represents the subsample of trajectories
    satisfying the WMAP7 constraint on $\As$ at $\WMAPsigma$.
    \semibold{Right}:
    Tail distribution
    $\Prob(|\fNLeq| > x)$,
    measured using different cuts on $\As$.
    The orange horizontal dotted line marks the point where
    $\Prob(|\fNLeq| > x) = 0.45$,
    which for the WMAP7 cut corresponds roughly to $|\fNLeq| > 0.5$.}
\end{figure}

\para{Amplitude distribution for $|\fNL|$}
A key question is the typical
amplitude of $\fNL(k_1, k_2, k_3)$ on
observationally accessible configurations,
because this determines whether non-Gaussian effects are detectable.
In Fig.~\ref{fig:fNLeq-CDF}
we plot distribution functions for $|\fNLeq|$.
(We do not give separate distributions for
$\fNLfold$
because Fig.~\ref{fig:fNL-results}
shows it to be highly correlated with $\fNLeq$.)

Fig.~\ref{fig:r-ns-detail}
shows that many trajectories yield
$\ns \approx 2$,
so the estimate $\fNL \approx - (5/12)(\ns - 1)$
suggests we should expect a concentration
near $\fNL \approx -0.8$.
It should be borne in mind that
bispectra satisfying this estimate
will be strongly scale dependent whenever the spectrum,
and likewise $\ns$, are strongly scale dependent.

This expectation is approximately borne out
the detailed distribution
for $|\fNLeq|$ given in the
left panel of Fig.~\ref{fig:fNLeq-CDF}.
For the minimal cut $\As < \Ascut$ (blue),
the distribution is rather flat for
$0.1 \lesssim |\fNLeq| \lesssim 0.6$.
There is a gently decaying tail to smaller absolute
values, and more abrupt decay for larger values.
There is almost no weight in the distribution
for $|\fNLeq| > 0.8$,
as can be seen by comparison with the right panel
showing the tail distribution.

For the subsample of trajectories that satisfy
the WMAP7 constraint on $\As$ at $\WMAPsigma$ (red),
the distribution is clustered
in the region $0.5 \lesssim |\fNLeq| \lesssim 0.6$,
as suggested by the distribution for $\ns$.
The tail to low absolute values $|\fNLeq| \lesssim 0.1$
is heavily depopulated
and
there are only a handful of samples
for which $|\fNLeq| \gtrsim 1$.
In the right-hand panel, the orange horizontal
dotted line marks the point where $\Prob(|\fNLeq| > x) = 0.45$,
chosen because
for the WMAP7 sample it corresponds roughly to
$|\fNLeq| > 0.5$.
In the full catalogue
$|\fNLeq|$ exceeds $0.5$ much less frequently,
in no more than $8\%$ of cases.
This reflects an approximate trend within our catalogue
(already mentioned in~\S\ref{sec:code-comparison}),
that larger values of $\As$ are correlated with smaller
values of $\fNL$.

Unfortunately these large
$|\fNL|$ can not be regarded as an observable signature
in our own universe,
because they derive from values of $\ns-1$ of order unity.
For trajectories that satisfy observable
constraints on \emph{both} $\As$ and $\ns$,
it appears that $|\fNL|$ will typically be small---as is the
case for our best-fit trajectory.

\subsubsection{Squeezed configurations}
\label{sec:squeezed-configurations}
Finally, we turn to squeezed configurations.
In Appendix~\ref{sec:run_time}
we show that modestly
squeezed isosceles configurations with $\beta = 0.9$ and
$\beta = 0.95$
(corresponding to squeezings $k_3 / k_t = 0.05$ and
$0.025$, respectively)
require integration times in the range
$1,000 \, \seconds$ to $2,000 \, \seconds$.
Taking the CMB to receive contributions from
approximately
$0.005 \, \Mpc^{-1}$
to $0.2 \, \Mpc^{-1}$,
the maximum observable squeezing is roughly
$k_3 / k_t \approx 0.0125$
or $\beta \approx 0.975$.
Therefore these estimates can be taken as a reasonably
reliable lower limit
on the compute time required for predictions in the
observational range.
This would increase processing time
for the entire catalogue by a factor
perhaps in the range $3$ to $5$,
which is
prohibitive given the already sizeable
computational demands for obtaining observables.

\begin{figure}
    \centering
    \includegraphics{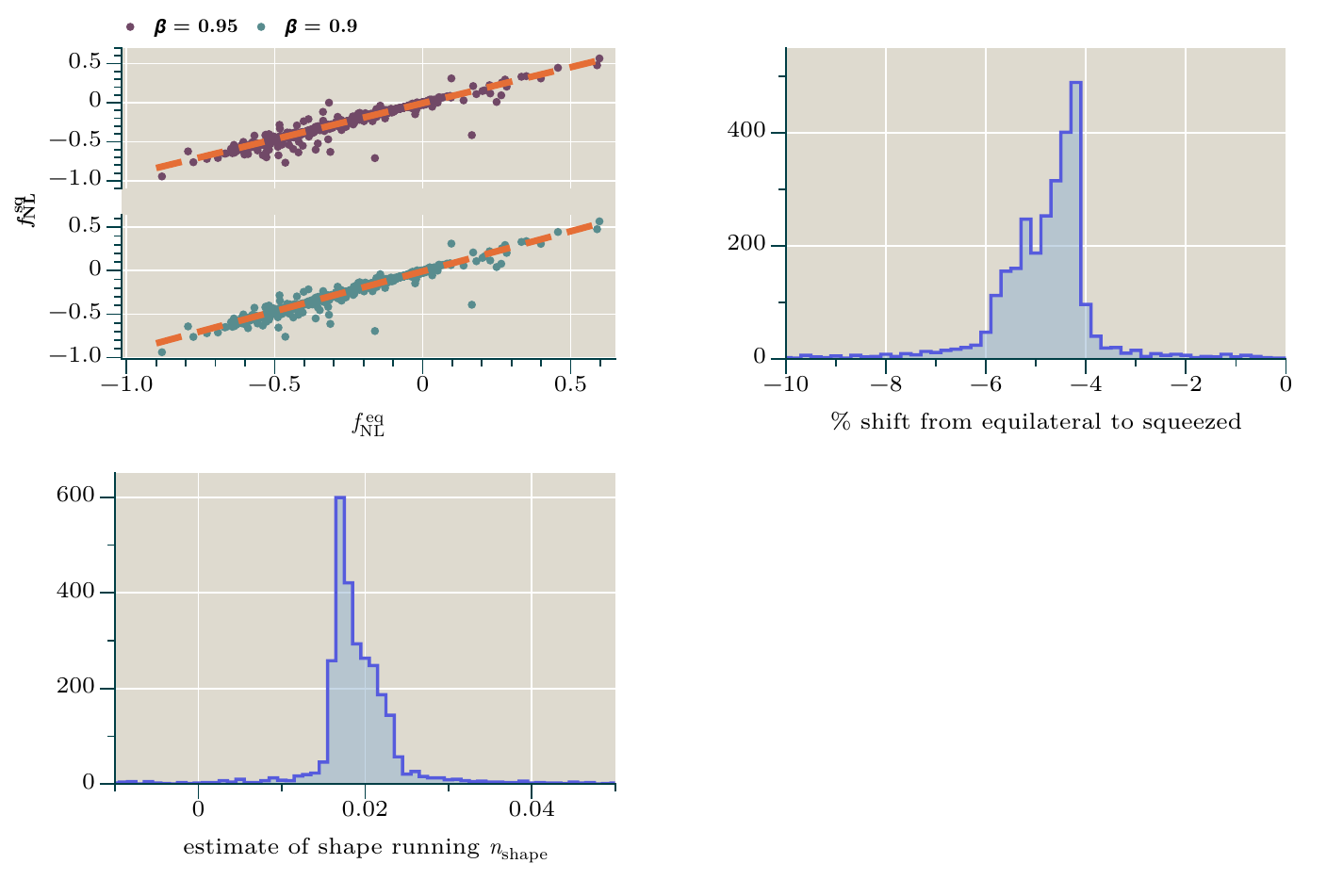}
    \caption[Analysis of $\fNLsq$ and shape response]
    {\label{fig:fNL-squeezed}Analysis of the squeezed
    amplitude $\fNLsq$.
    We impose our standard cut $\As < \Ascut$.
    \semibold{Top left}: $\fNLsq$
    against $\fNLeq$.
    Green points correspond to $\beta = 0.9$ and squeezing
    $k_3/k_t = 0.005$. Brown points correspond to
    $\beta = 0.95$ and squeezing
    $k_3/k_t = 0.0025$.
    The dashed line has approximate form
    $\fNLsq = -0.006 + 0.92 \fNLeq$, which is common
    for both squeezings.
    \semibold{Bottom left}:
    Distribution of the estimated `shape running'
    parameter $\nshape$.
    \semibold{Top right}:
    Distribution of percentage shift between
    $\fNLeq$ and $\fNLsq$ for $\beta = 0.95$}  
\end{figure}

\para{Gelaton and QSFI effects}
Instead, we study the relationship between $\fNLsq$
and $\{ \fNLeq$, $\fNLfold \}$
on a smaller sample of
roughly {\squeezedcataloguesize}
trajectories.
In the top-left panel of Fig.~\ref{fig:fNL-squeezed} we plot
the computed value of $\fNLsq$ against $\fNLeq$
for trajectories that satisfy the WMAP7 bound
on $\As$ at $\WMAPsigma$.
As expected
this shows strong correlation,
which applies equally to the full catalogue
satisfying $\As < \Ascut$.
The same conclusion
applies to $\fNLfold$, which itself correlates strongly with
$\fNLeq$.
The orange dashed line corresponds to the approximate
relation $\fNLsq = - 0.006 + 0.92 \fNLeq$,
from which we conclude that
the typical amplitude of $\fNLsq$
is very similar to $\fNLeq$
but just a little smaller.
The characteristic clustering of values
between
$-0.6 \lesssim \fNLeq \lesssim -0.1$
is clearly visible
(cf. the blue distribution in Fig.~\ref{fig:fNLeq-CDF}).

In the top right panel of Fig.~\ref{fig:fNL-squeezed}
we plot the shift
$\ShiftSqEq \equiv \fNLsq - \fNLeq$
between
squeezed and equilateral configurations,
expressed as a percentage of the equilateral amplitude.
The shift almost always falls between $-4\%$ and $-6\%$,
with values near $-4\%$ being favoured.
In absolute values this corresponds to
$\ShiftSqEq \approx \text{few} \times 10^{-3}$. 
In models for which the slow-roll approximation
applies throughout, we expect
the slow-roll parameters
to set the scale of the
shift~\cite{Maldacena:2002vr,Burrage:2011hd}.
For example, in the axion--quadratic model
studied in Ref.~\cite{Dias:2016rjq}
the shift computed between the same configurations
is roughly $\sim \text{few} \times 10^{-2}$,
which is comparable to $\eta = \d \ln \epsilon / \d N$.
From this point of view the typical value of $\Delta$
observed in the $\brane{3}$/$\antibrane{3}$ model is not unusually
large,
and indeed is comparable to $\eta$ in
many realizations (see Fig.~\ref{fig:usr-results}
and also the discussion of the residual
$\ShiftEqFold$ above).

It is possible that some of the more significant shifts
of $10\%$ or larger
are associated with
an
unusual change in amplitude
near equilateral configurations, which might be
expected from a `gelaton'-like or `QSFI'-like
scenario~\cite{Tolley:2009fg,Chen:2009we,Chen:2009zp,Gong:2013sma}.
Assuming the rate $0.07\%$ suggested
by McAllister et al., we would expect to observe
perhaps $\Or(10)$
QSFI-like examples in a catalogue of
this size~\cite{McAllister:2012am}.
It would be exceptionally interesting
(but numerically expensive)
to
compute exact bispectrum shapes for the trajectories
that exhibit the largest shifts between equlateral
and squeezed configurations.

\para{Shape dependence}
In general the bispectrum may depend on shape and scale
through the ratios
$k_t / \kpiv$ and $k_i / k_t$ for $i = 1, 2, 3$.
As we move from equilateral to squeezed configurations,
taking $\vect{k}_3$ to be the squeezed momentum,
the `shape' ratios
$k_1 / k_t$, $k_2 / k_t$ vary between $1/3$ and $1/2$.
Meanwhile, $k_3 / k_t$ varies between $1/3$ and $0.0025$.
Assuming the strongest dependence comes from the squeezed
momentum, the effective `shape'
running
can be written in terms of a
parameter $\nshape$~\cite{Burrage:2011hd,Byrnes:2015dub}
\begin{equation}
    \fNL \approx \fNLpiv \left( \frac{k_3}{k_t} \right)^{\nshape} ,
\end{equation}
where $\fNLpiv$ is a fiducial value taken here to be
the value of $\fNL$ on an equilateral configuration at $k_t = 3\kpiv$.
In the bottom left panel of Fig.~\ref{fig:fNL-squeezed}
we plot the distribution of $\nshape$
for our sample of squeezed configurations.
It shows very pronounced clustering near
$\nshape \approx 0.02$.
The scale dependence of the spectrum is divided
out of $\fNL$ by construction,
but it is possible that this shape dependence
is generated by the same underlying
process~\cite{Byrnes:2015dub,Kenton:2016abp}.
In particular,
the shape of the $\ns$ distribution in
Fig.~\ref{fig:code-compare-4up} is qualitatively
similar to (but not the same as) the distribution of $\nshape$
in Fig.~\ref{fig:fNL-squeezed}.
One might therefore expect the shape running to be small
when the spectrum falls in the observationally-allowed
window,
but our `squeezed' catalogue contains too
few trajectories to make a definitive statement.
It would be interesting to study the shape
running in a larger
sample.

\subsubsection{Large non-Gaussianity from rapid change of angular minima}
\label{sec:abrupt-change-minima}
In certain
very
rare examples we observe the synthesis of large non-Gaussianity,
apparently caused by abrupt shifts in the brane trajectory.
These occurrence of these rare shifts was recognized by Agarwal et al.,
who conjectured they might generate appreciable
three-point correlations~\cite{Agarwal:2011wm}.
Our results demonstrate this conjecture to be essentially correct.

\begin{figure}
    \centering
    \includegraphics[scale=1.0]{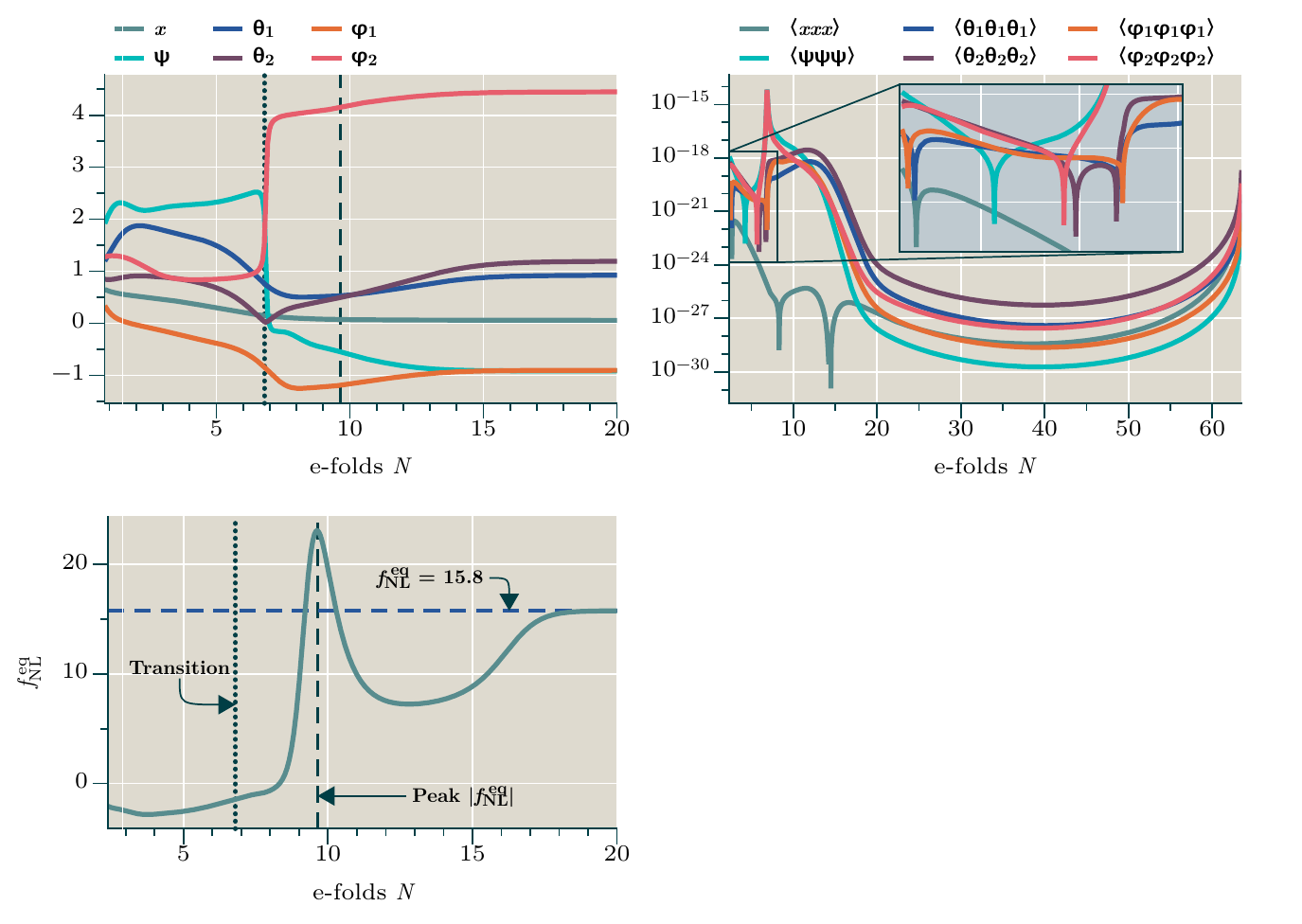}
    \caption[Example trajectory showing abrupt change between
    angular minima]
    {\label{fig:fNLVacuumChange}Example trajectory exhibiting rapid
    transition between angular minima.
    \semibold{Top left}: Evolution of background fields.
    The dotted vertical line indicates the approximate time of
    the transition.
    The dashed vertical line indicates the time at which peak
    $|\fNLeq|$ is achieved (see bottom left panel).
    \semibold{Bottom left}:
    Evolution of $\fNLeq$ on a scale $k_t/3 = \kpiv = 0.002 \, \Mpc^{-1}$.
    The axes match the top left panel, and the approximate transition
    time and peak $|\fNLeq|$ are marked in the same way.
    \semibold{Top right}: Time evolution of the dimensionless
    3-point functions
    $(k_1 k_2 k_3)^2 \langle O_1(\vect{k}_1) O_2(\vect{k}_3) O_3(\vect{k}_3) \rangle'$,
    for selected operators $O_i$.
    The prime $'$ attached to the correlation function
    denotes that the momentum-conserving $\delta$-function
    $(2\pi)^3 \delta(\vect{k}_1 + \vect{k}_2 + \vect{k}_3)$ has been
    removed.}
\end{figure}

\para{Typical evolution}
In Fig.~\ref{fig:fNLVacuumChange} we show the time evolution for a typical
trajectory exhibiting a rapid transition between
distinct angular minima.
In the top-left panel we plot the time evolution of the background
fields. The transition occurs at roughly
$N\approx 6.8$ e-folds from the initial time
(marked by the dotted vertical line)
and is characterized by rapid evolution of $\phi_2$ and $\psi$.
The motion is overdamped by Hubble friction and the fields
settle smoothly into the new vacuum.
After the transition the system approaches an adiabatic limit.
The dashed vertical line indicates the time at which maximum
$|\fNLeq|$ is attained.

In the bottom-left panel we show the time evolution of $\fNLeq$,
evaluated on the scale $k_t/3 = \kpiv = 0.002 \, \Mpc^{-1}$.
The amplitudes $\fNLfold$ and $\fNLsq$ exhibit qualitatively similar
time-dependence, although there are differences of detail.
In each of these measures,
the amplitude of 3-point correlations
exhibits rapid growth after the transition
up to a peak at $N \approx 9.6$,
followed by rapid decay.
Eventually, as the adiabatic limit is reached,
the equilateral amplitude asymptotes to $\fNLeq \sim 15.8$,
with slightly different values for the other configurations.
We plot the evolution only up to $N=20$, beyond which the
dynamics become uninteresting.
For ease of comparison
the top-left and bottom-left panels share the same horizontal scale.

This time evolution of $\fNLeq$ closely matches the behaviour
described by Elliston et al., associated with rolling between
two approximate critical points of the potential~\cite{Elliston:2011dr}.
In this case the critical points should be regarded as the angular minima
connected by the rapid transition.
Note that
Ref.~\cite{Elliston:2011dr}
considered only the case of two-field dynamics in which the trajectory
rolls from the vicinity of a critical point at which the Hessian has
one dominant negative eigenvalue
to the vicinity of a different critical point at which the Hessian has
one dominant positive eigenvalue.
However, it is plausible that similar behaviour occurs in transitions
between critical points with different numbers
of positive and negative eigenvalues.
This example suggests it may be profitable to develop
a general theory of such transitions.
For example,
the above analysis clearly invites
a connexion with the language and methods of Morse theory.
To our knowledge, however, this connexion is currently unexplored.

\para{Numerical fidelity}
The abrupt character of the transition requires that we exercise caution
before accepting numerical results.
We have verified that the solutions are stable
to adjustment of the time-sample mesh,
numerical tolerances,
and changes of stepper.
For these trajectories
we have compared the output
from our Dormand--Prince
$4^{\text{th}}$/$5^{\text{th}}$-order
stepper
with
an adaptive Bulirsch--Stoer stepper
that can work up to 8$^{\text{th}}$ order
where the solution has rapid changes or discontinuities.
So far as we can judge,
our numerical methods
appear to correctly control the solution for each correlation
function during the transition.
Further, despite the abrupt evolution of the background
fields, we have verified that the eigenvalues of the
mass matrix evolve smoothly.

In the top-right panel of Fig.~\ref{fig:fNLVacuumChange}
we plot the time evolution for a representative
sample of field-space correlation functions.
These demonstrate that the evolution is smooth
and the crucial decay during the subhorizon era is being
followed accurately. It is this era that is typically
most difficult to integrate, and conversely noise in
this region is usually a reliable indicator of numerical
problems~\cite{Dias:2015rca,Dias:2016rjq}.
For this trajectory
there are no significant subhorizon oscillations.
Other
trajectories \emph{do} exhibit them
in off-diagonal correlators $\langle 1 2 \rangle$
as a consequence of
unbalanced
phase factors $\sim \e{\im (E_1 - E_2) t}$
between states $1$ and $2$
with energies $E_1$, $E_2$.
Where we have performed spot checks on such trajectories,
the solutions are stable to the changes described above
and the oscillations appear to be smoothly resolved.

\para{Summary}
We caution that trajectories exhibiting these rapid transitions are
rare; there is no sense in which they can be regarded as typical
features of the $\brane{3}$/$\antibrane{3}$ model---although,
curiously, the single adiabatic trajectory satisfying the
(separate) WMAP7 $\WMAPsigma$ bounds on $\As$ and $\ns$
is of this type.
We observe ten
examples of this kind in our primary
catalogue for which $|\fNLeq|$ is larger than unity,
although there are certainly more examples that fall
below this arbitrary threshold.
The cases where $|\fNLeq| > 1$ correspond to $\fNLeq$
equal to
$-18.7$, $-12.3$, $-9.3$, $-8.71$, $-8.49$, $-7.31$,
$11.0$, $12.3$, $15.8$, $75.8$.
Their rarity suggests
it is unlikely that
observable signatures of these transitions could be used to constrain
the model.
Nevertheless,
despite their atypicality, these trajectories are very interesting
as examples of behaviour that has been observed in toy
models~\cite{Elliston:2011dr}, but not
(to our knowledge) in a model motivated by a purpose
\emph{other} than its ability to generate large 3-point correlations.

\subsection{`Small' ensemble comparisons}
\label{sec:small-studies}
We now consider the `small' ensembles produced by the
{\PyTransport} pipeline.
(See Table~\ref{Table:SmallStudies}.)
In general these show that observational predictions are
strikingly insensitive to many of the arbitrary choices
made in~{\S}\ref{sec:brane-potential}.
Specifically, we find very little difference
between our principal catalogue
and the `small' catalogues produced
by varying initial conditions,
$\phiUV$,
or the truncation point of the potential.
(When varying the initial conditions we do observe a
preferred non-zero initial $x$-velocity of the brane,
but this does not propagate into the final distributions.)

\para{Homogeneous model}
Significant changes do occur when dropping
contributions to the potential
sourced by the bulk flux product $\gstring |\Lambda|^2/96$.
We describe this as the `homogeneous' case
(see Table~\ref{Table:SmallStudies}),
which corresponds to setting
$\mathscr{C}_{\sumL, \sumM} = 0$.
This is an artificial test
in the sense that there is no meaningful
limit
that would justify
dropping the $\mathscr{C}_{\sumL, \sumM}$
while retaining the zero modes of the bulk Laplacian.
However, the outcome is still interesting because 
it illuminates how features of the model
arise from particular sectors of the potential.

\begin{figure}
    \centering
    \includegraphics{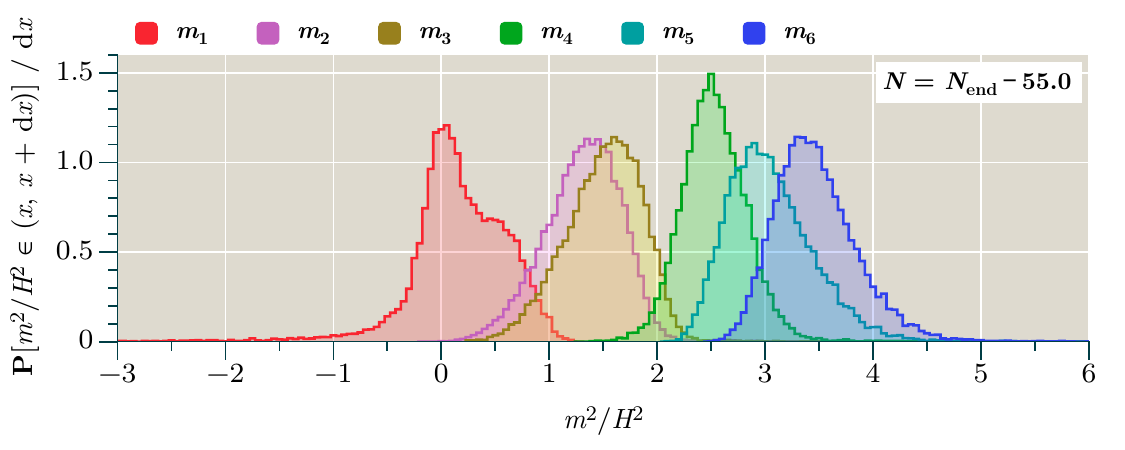}
    \caption[Mass distribution for
    `small study' with homogeneous terms only]
    {\label{fig:small-homog-massdist}Mass distribution
    with $\mathscr{C}_{\sumL, \sumM} = 0$
    (see Table~\ref{Table:SmallStudies}).
    Colour assignments
    correspond to Fig.~\ref{fig:results-mass-spectrum}.}
\end{figure}

In Fig.~\ref{fig:small-homog-massdist} we plot the mass
distribution for this catalogue, measured at
$55$ e-folds prior to the end of inflation.
In comparison with
Fig.~\ref{fig:results-mass-spectrum}
the heavy eigenstates are more nearly degenerate, and their
numerical values of $m^2/H^2$ are significantly smaller.
Meanwhile, the tachyonic state (red distribution in
Figs.~\ref{fig:results-mass-spectrum} and~\ref{fig:small-homog-massdist})
has a significantly broader tail towards
negative values.
The conclusion is apparently that the heavy masses
are dominated by contributions
sourced from the bulk fluxes.
It is this feature that causes us to speculate
that the difference in mass spectra
reported by Agarwal et al.~\cite{Agarwal:2011wm},
Dias et al.~\cite{Dias:2012nf}
and this work
may be associated with differing treatment of
these sourced modes.
Because the sourced modes also change the distribution
of the lightest mass eigenstate, it is possible
that they contribute to the difference in $\epsilon$
observed between all three studies.

\begin{figure}
    \centering
    \includegraphics{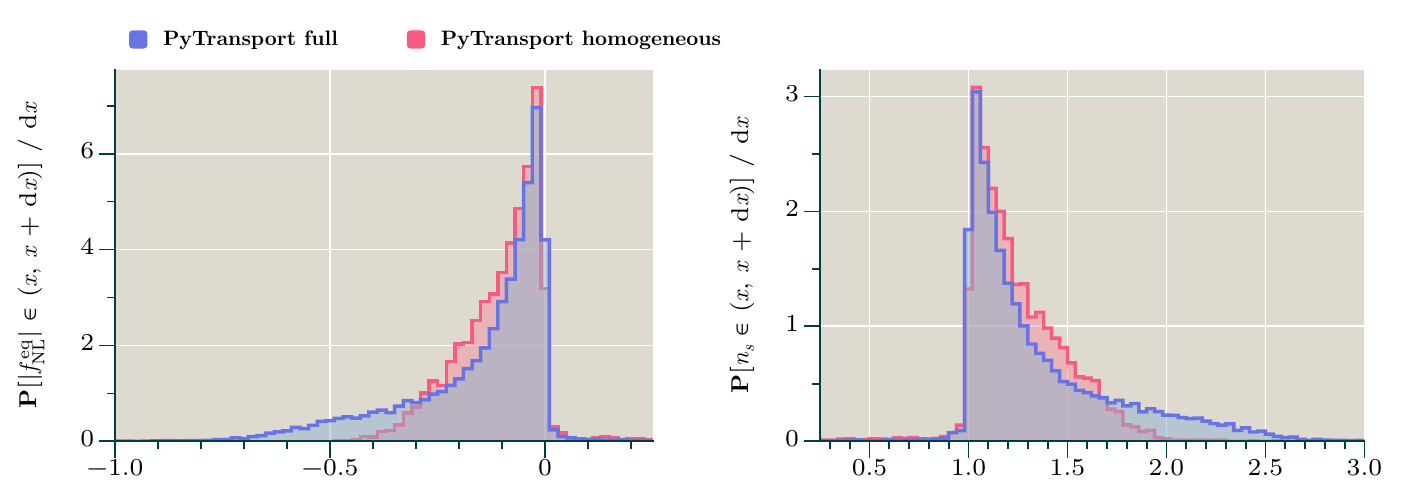}
    \caption[Change in observables for
    `small study' with homogeneous terms only]
    {\label{fig:small-homog-observables}Change in observables
    for the `homogeneous' catalogue.
    No cut is applied to $\As$.
    \semibold{Left}: distribution of $\fNLeq$
    with the full potential (blue) and in the homogeneous
    case (red).
    \semibold{Right}: distribution of $\ns$
    with the same conventions.}
\end{figure}

By itself, a change in the spectrum of heavy modes
need not imply any shift in the distribution of
observables.
Therefore in Fig.~\ref{fig:small-homog-observables}
we plot the distributions for $\fNLeq$ and $\ns$
in the homogeneous case,
without applying a cut on $\As$.
Although the peak of each distribution remains fixed,
the shape and length of the tail
is adjusted significantly.
We conclude that the impact
on observables is modest.
However, if treatment of the flux-sourced modes
is really responsible for the differences observed between
our analysis and Agarwal et al.\
or Dias et al.,
it is possible that
conclusions at the level of observables may
exhibit only limited sensitivity to these differences.

\begin{figure}
    \centering
    \includegraphics{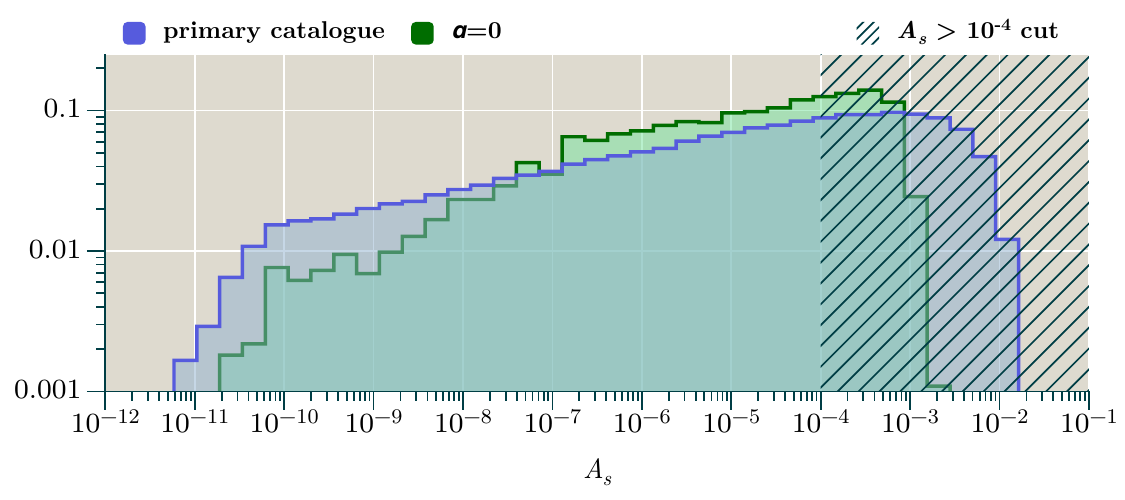}
    \caption[Distribution of $\As$ in the primary
    catalogue and `small study' with  $\alpha=0$]
    {\label{fig:small-study-As}Distribution of
    $\As$ in the primary catalogue (blue)
    and the $\alpha=0$ `small study' (green).
    The hatched region indicates the part of the
    distribution excluded by our cut $\As < 10^{-4}$.}
\end{figure}

\para{$\alpha=0$ model}
The last study is the restriction to $\alpha=0$, for which the
posterior $Q$-distribution was already discussed in
Fig.~\ref{fig:results-Q-alpha}.
This enables us to connect our analysis with
the previous studies reported
by Agarwal et al.,
Dias et al.\
and McAllister et al.,
each of which set $\alpha=0$.
We do not observe significant shifts
in the distribution of observables,
except for a small shift in the $\As$
distribution
(see Fig.~\ref{fig:small-study-As}).
This conclusion was anticipated
by Hertog \& Janssen~\cite{Hertog:2015zwh}.
On this basis, it appears that
the differences observed between our analysis
and these previous studies
should be attributed to structural differences in the
potential rather than changes in the sampling procedure.

\section{Discussion}
\label{sec:discussion}

In this paper we have revisited
the `delicate' $\brane{3}$/$\antibrane{3}$ brane inflationary
model,
and computed the amplitude of three-point correlations
for the density perturbations.
These have not previously been reported.
We find that $\fNLeq$, $\fNLfold$ and $\fNLsq$
are all highly correlated
with nearly degenerate amplitudes
close to Maldacena's single-field
prediction
$\fNL \approx -(5/12)(\ns-1)$
for $|\nt| \ll 1$~\cite{Maldacena:2002vr},
even when $|\fNL|$ is large.
For trajectories that yield observationally
acceptable CMB angular spectra $C_\ell^{XY}$
it follows that the amplitude of three-point correlations
is usually negligible.
For example, the
best-fit trajectory for the Planck2015 $TT$+$TE$+$EE$ likelihood
yields $|\fNL| \sim 5 \times 10^{-3}$.
If the restriction to the observationally
acceptable region is dropped,
the model
has a tendency to produce
scale-dependent, nonmonotonic power spectra
with corresponding $|\fNL|$ of order unity.

A small number of trajectories yield atypically large
$|\fNL|$, associated with abrupt transitions of the angular
fields between different minima.
At the extreme, we observe values $|\fNL| = \Or(100)$,
although still with unacceptable spectral behaviour.
Closer to the allowed observational window we find
one (adiabatic) trajectory of this type that yields
$\fNL = -0.749$ and $\ns = 0.94$ with an acceptable
$\As$.
However, because the spectrum is scale dependent the fit is
not as good as these values would suggest.
The occurrence of large three-point correlation amplitudes
for these trajectories was originally conjectured by
Agarwal et al.~\cite{Agarwal:2011wm}.
The large bispectra observed in our catalogue
all appear to be produced by such transitions
and not `gelaton' or `QSFI'
effects~\cite{Tolley:2009fg,Chen:2009we,Chen:2009zp}.
In our smaller catalogue of squeezed configurations
we observe some significant shifts between squeezed
and equilateral configurations that could possibly be
associated with QSFI behaviour,
although the corresponding bispectrum amplitudes do not
appear to be large.
To decide one way or the other would require full
computations of the bispectrum shape on these trajectories.
Overall, the shape usually appears close to local,
although there are exceptional cases that it would be
interesting to explore.

We find only weak dependence on most of the arbitrary choices
made during construction of the potential. In particular, we find
our observable distributions are robust to variations in
initial conditions and the field-space diameter
$\phiUV$.
We find they are also robust to changes in the truncation
point of the potential, even for observables derived from
the three-point function.
This implies that the intrinsic three-body couplings
do not play a significant role in setting the
amplitude of the bispectrum,
which is consistent with the suggestion that
the observed
large bispectra
are sourced by field-space evolution on superhorizon scales
associated with transitions between minima.
When
contributions to the potential
sourced by bulk fluxes 
are dropped we find that the mass distribution changes
significantly, but the effect on observables is much
more modest.
Variation in
the treatment of these contributions may be responsible for
the observed differences between independent analyses of
the model.
If so, there is some basis for optimism
that predictions for observables (except perhaps $r$)
might be robust.

It is sometimes suggested that $|\fNL| \sim 1$
is a generic prediction of inflationary models in which multiple
fields are active~\cite{Meerburg:2019qqi}.
This does appear to be the case for
what we could call
`type 1' models where the fields
are spectators during inflation,
but become active in the post-inflationary universe---%
as in the curvaton, inhomogeneous-end-of-inflation
and modulated reheating models~\cite{Alabidi:2010ba}.
Meanwhile, there is relatively little evidence
to guide our expectations for `type 2' models in which the fields
are active during inflation, but the evolution becomes
adiabatic before reheating.
Many toy models of this type produce negligible $|\fNL|$.
The $\brane{3}$/$\antibrane{3}$ model represents
a more realistic example of this type,
which can be attributed to the frequent emergence
of an effectively adiabatic trajectory long before
observable scales leave the horizon.
To set our expectations for the interpretation of
upcoming large galaxy surveys~\cite{Alvarez:2014vva},
it would be exceptionally interesting to know whether
the same conclusion extends to a larger class of multiple-field
models motivated by ideas in high energy physics.

In common with all prior analyses, we find the model is not
favourable observationally. Our best-fit trajectory
yields a log-likelihood $\ln \lik \approx -600$
based on Planck2015 temperature and polarization
measurements~\cite{Ade:2015xua},
but most trajectories produce spectra that are
significantly too
scale dependent---even
where the amplitude falls in
the allowed range.
Typically the bispectrum will also be
scale dependent whenever the spectrum is scale dependent.
Additionally,
there is evidence for a
characteristic
\emph{shape} running $\nshape \approx 0.02$ in the bispectrum
that may possibly be generated from the same source
as the scale dependence of the power spectrum.
It is not yet clear whether
the shape running
can be significant when the scale dependence
of the bispectrum is small---%
but, if so,
this could be an interesting observational signature.

Our software pipeline depended on development of a new interface
between {\CppTransport} and
\href{https://bitbucket.org/joezuntz/cosmosis/wiki/Home}{\Cosmosis}~\cite{Zuntz:2014csq}.
A future release of {\CppTransport} is scheduled to include
this interface, which can be used with any inflationary
model.
A similar interface to
the more recent
\href{https://bitbucket.org/joezuntz/cosmosis/wiki/Home}{\Cobaya}
framework~\cite{Torrado:2020dgo} is also planned.
These interfaces
dramatically simplify the construction of
end-to-end pipelines to constrain inflationary models
directly from data---a capability which has
been available for a long time in the collider phenomenology
community, but which has taken longer to become mainstream for
inflationary model analysis.

To mitigate the risk of implementation errors we deploy two
independent pipelines.
These show excellent agreement at the level of individual
trajectories, but
disagree in detail at the level of the entire catalogue.
The factors driving this disagreement have not yet been
identified, but we believe they relate to the sampling
implementation.
After applying the cut $\As < \Ascut$ the catalogues are brought
into statistical agreement.
Because our codes operate at tree-level,
a cut of this kind would likely be required anyway to
remove trajectories on which there is a risk
of the leading loop corrections
becoming relevant.

Despite the complexity of the model, compute times are not
prohibitive.
Samples of the two-point function can be computed in a time of order
seconds. They could perhaps be included in a parameter estimation
Monte Carlo if not too many samples are needed
to correctly predict the shape of the spectrum.
Samples of the three-point function are substantially
more expensive, with integration times of the order of
a few hundred seconds depending on the underlying hardware.
Therefore inclusion of this information in a Monte Carlo
is impractical for the $\brane{3}$/$\antibrane{3}$ model.
In simpler models the compute time is much smaller, although even
in this more optimistic case
it would be necessary to sample the bispectrum sparsely
rather than compute its full shape.
It is an interesting question whether reliable methods can
be developed to incorporate this information in
a practical likelihood calculation.

The $\brane{3}$/$\antibrane{3}$ model brings other
computational challenges.
We find that observables derived from the three-point function
can require enhanced numerical precision
for accurate computation, which is presumably a consequence
of roundoff error due to large cancellations between
the many intermediate
terms that appear in this model.
A second example is accurate computation of the
tensor spectral index $\nt$,
because the tensor spectrum is extremely flat.
Nevertheless, such accurate computations
seem worth pursuing
because of the possibility that rare trajectories
can violate the `consistency bound' $r \leq -8 \nt$
due to significant anticorrelations imprinted at
horizon exit.

\para{Comparison to previous work}
Our results are in qualitative agreement with previous
analyses of the
model~\cite{Agarwal:2011wm,Dias:2012nf,McAllister:2012am}.
However there are quantitative differences.
In particular there is
evidence for some systematic differences
between the trajectories in our catalogue
and those constructed by Agarwal et al.\
and McAllister et al.
Our typical values of $Q$ and $\epsilon$
(and hence $r$)
are somewhat different,
and our typical mass spectra have a qualitatively different structure
but similar numerical magnitudes.
There is also some evidence,
although less strong,
for similar systematic differences relative to
Dias et al.

We have
suggested these differences may be related to a different treatment
of contributions to the potential sourced by bulk fluxes.
There is also our inclusion of $\alpha$ as
a sampling parameter,
whereas Agarwal et al.,
Dias et al., and McAllister et al.\
imposed $\alpha=0$.
This seems to be responsible for some of the shift in
the (unobservable) parameter $Q$, and
also slightly broadening the distribution of $\As$.
Our typical values of $\epsilon$ and $r$ are
relatively close to those reported to Dias et al., 
although not the same.
However, conversely,
our mass spectra are in closer agreement with
McAllister et al.
We find a population of trajectories yielding red
values for $\ns$
that was not identified by previous analyses.
The observational relevance of this population
is not yet clear
because their spectra almost certainly contain
interesting features.

Despite these differences,
there are many areas of agreement between
our results and those reported by
earlier studies.
The overall shape of the mass spectrum,
the form (but not normalization)
of the $r$-vs.-$\ns$ correlation,
the occurrence of rapid transits between different minima
of the angular fields,
the presence of strong anticorrelation between field
fluctuations at horizon exit (`destructive interference' in the
language of McAllister et al.)
and possible transgressions of the `consistency bound'
are all points of agreement.
Although not discussed explicitly in~\S\ref{sec:Results},
our catalogue reproduces a similar probability of inflation
and distribution of total number of e-folds
$\Ntot$
to those reported by Agarwal et al. However, Agarwal et al.\
gave a persuasive analytic argument 
for the functional form of the distribution
of $\Ntot$,
which suggests it does not depend sensitively on
properties of the potential.

A natural question is whether our conclusions are reliable
given the apparent discrepancies between
different analyses of the model.
Ultimately this will require forensic comparison
between the separate implementations
of the $\brane{3}$-brane potential.
The most optimistic outlook is that these differences
reflect only minor divergences in construction of the potential,
against which observables may be
fairly robust, as in~\S\ref{sec:small-studies}.
The broad qualitative agreement between
all three analyses apparently does suggest there can not
be very
significant differences in the
structure of the $\brane{3}$-brane potential,
at least
when it is able to support an extended epoch of inflation.

To assist future comparison
we have attempted to give enough detail in~\S\ref{sec:construct-model}
that it will be possible for third parties to replicate
our analysis.
Also, we have made our model specification
and sampling parameters available
to download from the Zenodo open data repository.
These files are provided under
a permissive CreativeCommons attribution license.

This situation illustrates the advantages of a standardized format for
specification of inflationary models, analogous to the
\emph{Universal FeynRules Output}
format used in the collider phenomenology community
to specify the particle content, coupling constants and interactions
of a model~\cite{Degrande:2011ua}.
As inflationary models become more complex,
there seems a persuasive argument for the community to
converge on a standardized way to exchange similar
specifications.

\acknowledgments
We would like to thank
Peter Adshead,
Nishant Agarwal,
Mafalda Dias,
Anatoly Dymarsky,
Jonny Frazer
and
S\'{e}bastien Renaux-Petel for
helpful correspondence.
An early version of the {\CppTransport} interface
to {\Cosmosis} was written by Sean Butchers, whom we thank
for assistance during development of the pipeline
described in~\S\ref{sec:experimental-procedure}.
The `by hand' fits for $\nt$ using smoothed power spectra
were done in conjunction with Thomas Butler
and Jordan Ramsay--Clements.
DS acknowledges support from STFC consolidated grants
ST/P000525/1 and ST/T000473/1.
KM acknowledges support from STFC training grant ST/S505766/1.
AM acknowledges support from the University of Sussex
Junior Research Associate scheme and a Royal Astronomical Society
Undergraduate Fellowship.

\appendix
\section{Data availability}
\label{sec:data-availability}

The primary data product generated
for the work reported in this
paper is our trajectory catalogue.
We have made this available from the Zenodo
open-access repository, together with the {\CppTransport}
model file
that documents our construction of the
D3-brane potential and its parametrization.
We also include configuration files for the {\Cosmosis} pipeline
used to compute observables for each trajectory.
This pipeline depends on a modified interface between {\Cosmosis}
and the {\CLASS} Boltzmann code.
\begin{grouppanel}
    {
    \renewcommand{\arraystretch}{1.1}
    \begin{tabular}{ll}
        \semibold{License} & \href{https://creativecommons.org/licenses/by-nc/4.0/}{Creative Commons Attribution-NonCommercial 4.0 International} \\
        & \includegraphics[scale=0.12]{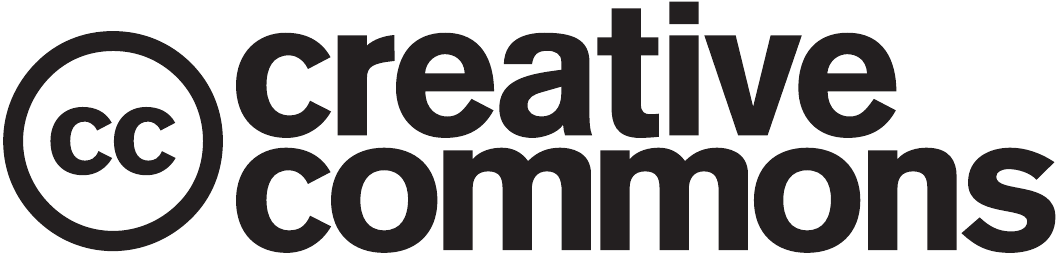}
        \includegraphics[scale=0.15]{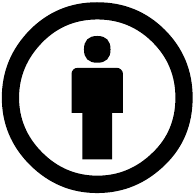}
        \includegraphics[scale=0.15]{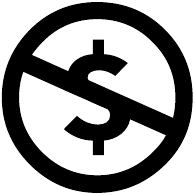} \\
        \semibold{Author} & Kareem Marzouk, Alessandro Maraio \& David Seery \\
        \semibold{DOI} & \href{https://doi.org/10.5281/zenodo.4742082}{\raisebox{-1mm}{\includegraphics[scale=0.6]{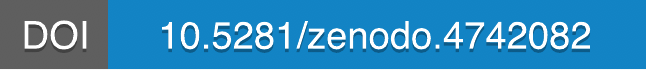}}} \\
        \semibold{Attribution} & Please cite \texttt{zenodo.org} DOI and this paper \\
        \semibold{Download} & {\small \url{https://zenodo.org/record/4742082}}
    \end{tabular}
    }
\end{grouppanel}

\para{Pipeline}
We explain in Appendix~\ref{sec:run_time}
that this model requires a custom version of {\CppTransport}.
Researchers wishing to replicate our pipeline
should therefore use
commit \href{https://github.com/ds283/CppTransport/commit/35c5ad8f200c8dc550b97145738e5f1563c011b6}{\commit{35c5ad8f}}.
This implements an interface to the {\Cosmosis} framework.
It also includes the necessary patches to ingest pre-computed expressions
for the inverse metric, the components of the connexion, 
and the Riemann tensor (also described in Appendix~\ref{sec:run_time}).
These pre-computed expressions are part of the Zenodo deposit.

The customizations in
\href{https://github.com/ds283/CppTransport/commit/35c5ad8f200c8dc550b97145738e5f1563c011b6}{\commit{35c5ad8f}}
include extensions to the model
file grammar. A new release of {\CppTransport} is planned that will include
the {\Cosmosis} interface, but the model file format should be regarded as
unstable and subject to change.
The interface targets
{\Cosmosis} 1.6 (commit
\href{https://bitbucket.org/joezuntz/cosmosis/commits/b33cd531e0da9eb2d0223015c7771ead1f9794fa?at=develop}{\commit{b33cd531e}}).

\para{Other data products}
The Planck2015 data products, including the $TT$, $TE$ and $EE$
angular power spectra (and uncertainties) used in~\S\ref{sec:Results},
can be obtained from the
\href{https://wiki.cosmos.esa.int/planckpla2015/index.php/Main_Page}{Planck Legacy Archive}.

\section{Computational requirements}
\label{sec:run_time}

It was explained in~{\S}\ref{sec:introduction}
that numerical predictions in the $\brane{3}$-brane 
model are challenging
due to the complexity of the potential.
In this Appendix we report in more detail on some of these challenges,
in the hope that such details may be useful to researchers
interested in incorporating observables derived from
three-point functions in their analyses.

\para{Translation step}
For both {\CppTransport} and {\PyTransport}
the analysis of a model is a two-step process.
First, a model description must be translated into customized
{\Cpp} code that implements the transport equations
for each correlation function.
This requires a number of computations involving
computer algebra.
Second, this code must be processed by the {\Cpp} compiler
before practical computations can be performed.
There are resource implications in each step.
Where timings are quoted, they relate to our
test machine---%
an Apple 16$^{\prime\prime}$
MacBook Pro with $16 \, \text{GB}$
and an i9-9880H running at $2.30 \, \text{GHz}$.

An initial difficulty is that the {\CppTransport} translator does
not work
successfully out-of-the-box for this model, because the {\GiNaC}
library on which it relies performs poorly when computing
the inverse conifold metric.
This is somewhat surprising given that the inversion is not
particularly difficult, but likely reflects the fact that
{\GiNaC} is not optimized for matrix operations.
Instead, we perform the inversion with {\SymPy}
and allow {\CppTransport} to read in the result of the
calculation.
It is likely that future versions of {\CppTransport}
will allow the
inverse field-space metric to be specified
explicitly as part
of the model description.
While this reduces the degree of automation in
the analysis, it allows more potent tools such
as {\SymPy} or {\Mathematica} to be used for
those parts of the calculation to which they
are best suited.

We find there are benefits to following the
same procedure
for the components of the connexion
and
the Riemann tensor.
Using {\SymPy} to pre-compute these leads to
a significant reduction in the time required
for the {\CppTransport} translator to perform
common sub-expression elimination (`CSE'),
which is needed to keep time and memory
requirements in check during the compile stage.
The happens
because the CSE
algorithm used in {\CppTransport}
does not perform well with very large expressions
due to design constraints imposed by {\GiNaC}.

With these choices, we find it takes of order
$\sim 1 \, \hours$
for {\CppTransport} to perform
translation on our test machine.
While this is not insignificant,
the procedure need only be done once.

\para{Compile step}
{\CppTransport} produces a \emph{core} \texttt{.h}
file of size $\sim 26 \, \text{Mb}$
and an MPI \emph{implementation} \texttt{.h}
file of size $\sim 800 \, \text{kb}$.
(For details, see Ref.~\cite{Seery:2016lko}.)
In future there may be some scope to reduce the size
of the core file by more intelligent pooling of
lengthy expressions.
Code of this size represents a considerable workload for
the compiler, but is still practical with
modern systems.
We find that Clang is able to compile the model
in between $10 \, \text{min}$
and $20 \, \text{min}$, with the exact time
depending on the platform and the version of Clang in use.
Memory usage peaks between $3 \, \text{Gb}$ and $4 \, \text{Gb}$.
We have tested Clang 9 and Clang 10
on Linux and macOS platforms (the latter in the guise
of Apple Clang 11 and 12).

GCC is also able to process the model, but it returns
significantly slower compile times
and has much greater resource utilization.
With GCC 10, peak memory usage was recorded in the vicinity
of $100 \, \text{Gb}$, which may be prohibitive in
a virtualized environment or on systems with slow
virtual memory.
On our test machine,
compilation did not terminate in a reasonable time, likely due to
performance issues associated with repeated paging to swap.
However, we have successfully compiled using GCC on larger machines,
including the HPC cluster used to compute our primary
catalogue.
On these machines the compile time was of order 1 hour.
The measurements reported here are not definitive,
and are intended only as a guide to relative performance.
In particular, compilation performance depends strongly on CPU
properties.

Its smaller footprint makes Clang much easier to
deploy in a resource-constrained environment such as
a virtual machine or Docker container.
Subject to this proviso, we find Docker to be a
convenient tool to deploy {\Cosmosis} and {\CppTransport}
together.

\para{Compute time}
Finally, we consider typical compute times
for samples of the inflationary two-
and three-point functions.
Although there is some variation,
computation of a single $k$-mode
of the two-point function
takes a few seconds.
The integration time for modes
of the three-point function
depends on the momentum configuration.
Equilateral and folded configurations
take a few hundred seconds,
with folded configuration
typically being between $1.5$ and $2.5$
times more expensive than the equilateral.
Squeezed configurations are most expensive.
For $\beta = 0.9$, corresponding to squeezing
$k_3 / k_t = 0.05$,
the integration is between 10 and 15 times
more expensive compared to equilateral configurations.
For $\beta = 0.95$, corresponding to squeezing
$k_3 / k_t = 0.025$,
the integration is between 10 and 20 times more
expensive compared to equilateral configurations.

\begin{figure}
    \centering
    \includegraphics{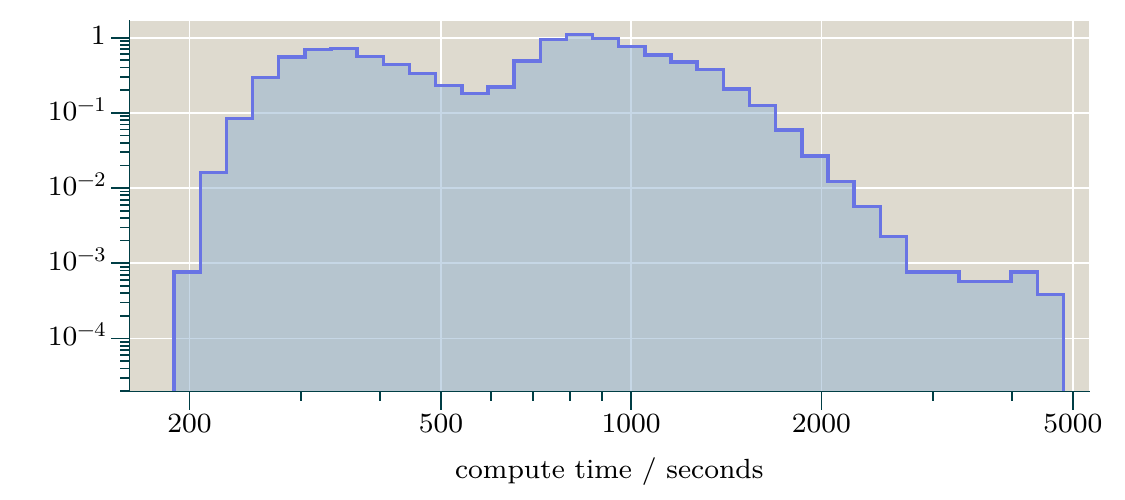}
    \caption[Distribution of integration times in the
    primary catalogue]
    {\label{fig:run_time}Distribution of integration times
    over our primary catalogue, measured in seconds.
    We have excluded
    a tail of rare trajectories up to $10^6$ seconds.
    The bimodality is due to trajectories for which we cannot
    compute the Planck2015 likelihood.}
\end{figure}

In Fig.~\ref{fig:run_time}
we plot the distribution of
compute times for the entire set of
{\cataloguesize} trajectories
in our primary catalogue
that satisfy the cut $\As < \Ascut$.
The individual calculations that contribute to this total
are listed in~{\S}\ref{sec:experimental-procedure};
note that there is no computation of
any squeezed configurations.
A majority of trajectories ($\sim 70\%$)
complete in less than $1,000 \, \seconds$,
and most ($\sim 95\%$) complete in less than $1,500 \, \seconds$.
The bimodality is due to trajectories for which we cannot
compute the Planck2015 likelihood because there are too few
total e-folds to sample the power spectrum on the
largest necessary scale (see~\S\ref{sec:experimental-procedure}).
For these trajectories we do not need to obtain a separate
sample of $\dimP_\zeta$ for use by {\CLASS},
making the calculation substantially shorter.

Although not displayed here, compute times in
{\PyTransport} are comparable to
(if generally slightly slower than)
those reported for {\CppTransport}.
Therefore, in this model, inclusion of three-point function
observables
is likely too slow for use in practical parameter
estimation using Markov Chain Monte Carlo methods.
It might be possible to compute the primordial
two-point function directly from the model, if desired,
and providing not too many sample points are used.

\section{Transport equations for the spectral index}
\label{sec:transport_spectral}

In this Appendix we briefly explain the transport
computation of the scalar and tensor spectral indices.
The superhorizon version of this calculation was given
in Ref.~\cite{Dias:2011xy},
and extended to subhorizon scales in
Ref.~\cite{Dias:2015rca}.
The discussion given here follows Ref.~\cite{Dias:2015rca}
with some refinements.

\subsection{Scalar spectral index}
First consider computation of the spectral index for $\zeta$.
As explained in~{\S}\ref{sec:computational-issues},
we start from the two-point function~\eqref{eq:scalar-2pf}
on phase space,
and define the `spectral matrix' $\tensor{n}{^{\pidx{A}}^{\pidx{B}}}$
to satisfy~\cite{Dias:2011xy}
\begin{equation}
    \label{eq:spectral-matrix-def}
    \tensor{n}{^{\pidx{A}}^{\pidx{B}}}(k)
    =
    \frac{\d \tensor{\Sigma}{^{\pidx{A}}^{\pidx{B}}}(k)}{\d \ln k} .
\end{equation}
The $\zeta$ two point function~\eqref{eq:Pzeta-def}
can be expressed at tree level in 
terms of $\tensor{\Sigma}{^{\pidx{A}}^{\pidx{B}}}$ using
\begin{equation}
    P_\zeta = \tensor{N}{_{\pidx{A}}} \tensor{N}{_{\pidx{B}}}
    \tensor{\Sigma}{^{\pidx{A}}^{\pidx{B}}}
    + \Or(\text{$1$ loop}) ,
\end{equation}
where the coefficients $\tensor{N}{_{\pidx{A}}}$ may be obtained
from standard perturbation theory or the separate universe
formula, and are momentum-independent when the mode $k$ is more
than a few e-folds outside the
horizon~\cite{Sasaki:1995aw,Elliston:2014zea}.
It follows that
\begin{equation}
    \frac{\d P_\zeta}{\d \ln k}
    \approx
    \tensor{N}{_{\pidx{A}}}
    \tensor{N}{_{\pidx{B}}}
    \tensor{n}{^{\pidx{A}}^{\pidx{B}}} ,
\end{equation}
from which the result~\eqref{eq:spectral-index-spectral-matrix}
for $\ns$ can be obtained.

\para{Transport equation}
A transport equation for $\tensor{n}{^{\pidx{A}}^{\pidx{B}}}$ can be obtained by
differentiating the transport equation for $\tensor{\Sigma}{^{\pidx{A}}^{\pidx{B}}}$.
This yields
\begin{equation}
    \label{eq:transport-nab}
    \frac{\d \tensor{n}{^{\pidx{A}}^{\pidx{B}}}}{\d N}
    =
    \tensor{u}{^{\pidx{A}}_{\pidx{C}}} \tensor{n}{^{\pidx{C}}^{\pidx{B}}}
    + \tensor{u}{^{\pidx{B}}_{\pidx{C}}} \tensor{n}{^{\pidx{A}}^{\pidx{C}}}
    + \frac{\d \tensor{u}{^{\pidx{A}}_{\pidx{C}}}}{\d \ln k} \tensor{\Sigma}{^{\pidx{C}}^{\pidx{B}}}
    + \frac{\d \tensor{u}{^{\pidx{B}}_{\pidx{C}}}}{\d \ln k} \tensor{\Sigma}{^{\pidx{A}}^{\pidx{C}}}
    + \Or(\text{$1$ loop}) .
\end{equation}
The result is a coupled system of ordinary differential
equations for $\tensor{\Sigma}{^{\pidx{A}}^{\pidx{B}}}$
and $\tensor{n}{^{\pidx{A}}^{\pidx{B}}}$.
For the definition of the `$u$-tensor'
$\tensor{u}{^{\pidx{A}}_{\pidx{B}}}$, see Dias et al.~\cite{Dias:2015rca,Dias:2016rjq}.
When the transport equation~\eqref{eq:transport-nab}
is expressed with $N$ as the independent variable, it can
be written
\begin{equation}
    \tensor{u}{^{\pidx{A}}_{\pidx{B}}} =
    \left(
        \begin{array}{c@{\hspace{1em}}c}
            0 & \delta^A_B \\
            {\mathcal{M}^A}_B & -(3-\epsilon) \delta^A_B
        \end{array}
    \right) ,
\end{equation}
where $A$, $B$ are species labels
corresponding to the phase space indices $\pidx{A}$, $\pidx{B}$.
Recall that
these indices run over the fields \emph{and} momenta.
Therefore,
when a phase space
index such as $\pidx{A}$ runs over its entire range
the corresponding species label $A$ runs over its range twice.
The matrix $\tensor{\mathcal{M}}{^A_B}$ is an effective mass matrix,
\begin{equation}
    \label{eq:scalar-effective-mass}
    \tensor{\mathcal{M}}{^A_B} =
        - \frac{k^2}{a^2 H^2} \tensor{\delta}{^A_B}
        - \frac{\tensor{M}{^A_B}}{H^2} ,
\end{equation}
where $\tensor{M}{^A_B}$ is the species
mass matrix~\eqref{eq:mass-matrix}
appearing in the Lagrangian.
Eq.~\eqref{eq:scalar-effective-mass} shows that
the $k$-dependence of $\tensor{u}{^{\pidx{A}}_{\pidx{B}}}$
is exponentially suppressed on superhorizon scales.
However, in this case it cannot be neglected.
In the formula
$P_\zeta \approx \tensor{N}{_{\pidx{A}}} \tensor{N}{_{\pidx{B}}}
\tensor{\Sigma}{^{\pidx{A}}^{\pidx{B}}}$
we evaluate $\tensor{N}{_{\pidx{A}}}$ only far outside the horizon where
$k^2/(a^2 H^2)$ is
entirely negligible.
In~\eqref{eq:transport-nab}
we integrate over times when $k$ is comparable to or
smaller than the horizon scale.
For this regime the $k^2/(a^2 H^2)$ term
is important in fixing the correct amplitude of
$\tensor{n}{^{\pidx{A}}^{\pidx{B}}}$
at horizon exit and can not be dropped.

\para{Initial conditions}
Eq.~\eqref{eq:transport-nab} can be solved simultaneously
with the transport equation for $\tensor{\Sigma}{^{\pidx{A}}^{\pidx{B}}}$,
and future versions of {\CppTransport}
will do this by default.
(The necessary functionality is already in the
\branch{201901} branch of the
\href{https://github.com/ds283/CppTransport}{GitHub repository}
for those who wish to make use of it.)
To proceed we require suitable initial conditions.
A suitable set of initial values were given
by Dias et al.~\cite{Dias:2015rca},
but these were truncated at leading order.%
    \footnote{In Ref.~\cite{Dias:2015rca}
    the initial conditions are given for
    a different matrix
    $\tensor{\tilde{n}}{^{\pidx{A}}^{\pidx{B}}}
    = (\d / \d \ln k) (k^3 \tensor{\Sigma}{^{\pidx{A}}^{\pidx{B}}})$,
    in which the factor $1/k^3$ in the ordinary power spectrum
    is removed.
    With this definition,
    Eq.~\eqref{eq:spectral-index-spectral-matrix}
    should be adjusted so the numerical constant $4$
    appearing on the right-hand side
    becomes $1$.}
In this paper,
in an attempt to improve convergence with fewer e-folds of
subhorizon evolution,
we include subleading terms
in both the slow-roll expansion and $k/(aH)$.

A next-order expression for the
scalar two-point function
was given in Ref.~\cite{Dias:2012qy},
\begin{equation}
    \langle
        \delta X^A(\vect{k}_1, t_1)
        \delta X^B(\vect{k}_2, t_2)
    \rangle
    =
    (2\pi)^3 \delta(\vect{k}_1 + \vect{k}_2)
    \big[ \tensor{\wavefn(t_1)}{^A_I} \big]^\dag
    \tensor{\wavefn(t_2)}{^B^I} ,
\end{equation}
where the superscript `$\dag$' denotes complex conjugation,
$A$ is an index in the field tangent space at time $t_1$,
$B$ is a similar index at time $t_2$,
and indices $I$, $J$, \ldots, label the field tangent space
at an arbitrary earlier time $t_\ast < t_1, t_2$
characterized by horizon exit of a fiducial
mode $k_\ast$.
As usual, all of these indices should be raised and lowered
with the field-space metric
evaluated at the appropriate
field-space coordinates.
We are interested in the equal-time commutator
for which $t_1 = t_2 = t$,
with $t$ (corresponding to conformal time $\eta$)
not too far from horizon exit of the mode $k = |\vect{k}_1| = |\vect{k}_2|$.
To avoid the logarithm $\ln(-k_\ast \eta)$
in Eq.~\eqref{eq:next-order-wavefn}
below becoming large
at this time
we must usually choose $t_\ast$ to be
roughly comparable to $t$.
The usual choice is to take $t_\ast$ to
be the moment when $k/(aH) = 1$.
Up to next-order in the slow-roll expansion,
but without making any approximation for the explicit time
dependence,
the wavefunction matrix $\tensor{\wavefn}{^A_I}$
can be written%
    \footnote{In Ref.~\cite{Dias:2012qy} this expression was
    given for a flat field manifold.
    The generalization to a curved field-space was considered
    in Ref.~\cite{Elliston:2012ab}, in which the
    \eqref{eq:next-order-wavefn} was used in the limit
    $|k\eta| \rightarrow 0$ but this expression did not
    appear explicitly.}
\begin{equation}
\label{eq:next-order-wavefn}
\begin{split}
    \tensor{\wavefn(t)}{^A_I} \equiv
    \frac{\im}{\sqrt{2k^3}}
    H_\ast
    \tensor{\Pi}{^A_K}
    \Bigg(
        &
        (1 - \im k \eta) \e{\im k \eta} \tensor{\delta}{^K_I}
        +
        \epsilon_\ast \Big[ \ln (-k_\ast \eta) - 1 \Big]
        (1 - \im k \eta) \e{\im k \eta} \tensor{\delta}{^K_I}
        \\
        &
        \mbox{} +
        \Big(
            \epsilon \tensor{\delta}{^K_I}
            - \frac{\tensor{M}{^K_I}}{3H^2}
        \Big)_\ast
        \Big[
            2 \e{\im k \eta}
            + \im \frac{\pi}{2} ( 1 - \im k \eta) \e{\im k \eta}
            - (1 + \im k \eta) \e{-\im k \eta}
            \int_{-\infty}^{k \eta} \frac{\d z}{z} \e{2\im z}
        \Big]
    \Bigg)
\end{split}
\end{equation}
where a subscript `$\ast$' indicates evaluation
at the time $t_\ast$,
$\eta \approx \int_t^\infty \d t' / a(t')$
is the conformal time corresponding to the evaluation time $t$
of the wavefunction,
and $\tensor{\Pi}{^A_I}$
is the parallel propagator on field space
evaluated on the inflationary trajectory
connecting the field-space coordinates
at times $t$ and $t_\ast$.
The first term in brackets is lowest-order in the slow-roll
expansion.
The second and third terms exhaust the next-order corrections.
For further details of Eq.~\eqref{eq:next-order-wavefn}
we refer to the literature~\cite{Dias:2012qy}.

\para{Subhorizon limit}
We wish to use~\eqref{eq:next-order-wavefn}
to supply initial conditions for $\tensor{n}{^{\pidx{A}}^{\pidx{B}}}$
when $k$ is on modestly subhorizon scales
for which $|k\eta| \approx k/(aH)$ is in the range $5$ to $10$.
To do so,
Ref.~\cite{Dias:2015rca} retained only the
term at lowest order in slow-roll and highest order
in $|k\eta|$.
Here we retain subleading terms in both expansions.

First, Eq.~\eqref{eq:next-order-wavefn}
can be used to derive correlation functions involving
the momenta $\pi^A \equiv \d X^A/\d N$.
To do so we express $\d/\d N$ in terms of $\eta$,
up to next-order in the slow-roll expansion, using
$\d / \d N = (aH)^{-1} \d / \d \eta$.
Beginning with the unequal-time correlation function,
the $N$ derivative on each momentum operator
$\pi^A$ can be brought outside the expectation brackets,
and the result evaluated by differentiation of
the wavefunction factors~\eqref{eq:next-order-wavefn}.
The equal-time limit should be taken only after all such
differentiations have been carried out.

Second, after retaining terms in the
$\wavefn \wavefn$ product only up to next-order in
the slow-roll expansion, we differentiate with respect
to $\ln k$.
This determines the element of the spectral matrix
corresponding to each correlation function.
Note that the evaluation time $\eta$ is fixed
and does not vary between $k$-modes.
The same is true for the arbitrary evaluation time $t_\ast$:
neither quantity generates any contribution
when differentiated with respect to $k$.
Finally, we expand in inverse powers of $k/(aH)$
and retain the leading- and next-order terms.
This yields
\begin{subequations}
\begin{align}
    \label{eq:field-field-spectral-ic}
    \frac{\d}{\d \ln k}
    \langle \delta X^A \delta X^B \rangle'
    & =
    -
    \tensor{\Pi}{^A_I} \tensor{\Pi}{^B_J}
    \frac{1}{2k a^2}
    \Bigg(
        \tensor{G}{^I^J}
        +
        \frac{3}{2}
        \frac{a^2 H^2}{k^2}
        \Big(
            (2 - \epsilon) \tensor{G}{^I^J}
            - \frac{\tensor{{M_\ast}}{^I^J}}{H_\ast^2}
        \Big)
        + \cdots
    \Bigg) ,
    \\
    \label{eq:field-momenta-spectral-ic}
    \frac{\d}{\d \ln k}
    \langle \delta X^A \delta \pi^B \rangle'
    & =
    \tensor{\Pi}{^A_I}\tensor{\Pi}{^B_J}
    \frac{1}{2k a^2}
    \Bigg(
        (1-\epsilon) \tensor{G}{^I^J}
        - \frac{15}{4} \frac{a^4 H^4}{k^4}
        \Big(
            3 \epsilon \tensor{G}{^I^J}
            -
            \frac{\tensor{{M_\ast}}{^I^J}}{H_\ast^2}
        \Big)
        + \cdots
    \Bigg) ,
    \\
    \label{eq:momenta-momenta-spectral-ic}
    \frac{\d}{\d \ln k}
    \langle \delta \pi^A \delta \pi^B \rangle'
    & =
    \tensor{\Pi}{^A_I}\tensor{\Pi}{^B_J}
    \frac{k}{2H^2 a^4}
    \Bigg(
        \tensor{G}{^I^J}
        + \frac{1}{2} \frac{a^2 H^2}{k^2}
        \Big(
            3 \epsilon \tensor{G}{^I^J}
            - \frac{\tensor{{M_\ast}}{^I^J}}{H_\ast^2}
        \Big)
        + \cdots
    \Bigg) ,
\end{align}
\end{subequations}
where the notation $\langle \cdots \rangle'$ means that the
momentum-conservation $\delta$-function
$(2\pi)^3 \delta(\vect{k}_1 + \vect{k}_2)$
has been dropped.
To write
Eqs.~\eqref{eq:field-field-spectral-ic}--\eqref{eq:momenta-momenta-spectral-ic}
we have combined
terms to remove explicit dependence on $t_\ast$.
Therefore $a$, $H$ and $\epsilon$ should be
determined at the evaluation time for the correlation function,
and
at lowest order we can take
the Hubble-normalized mass matrix
$\tensor{{M_\ast}}{^I^J} / H_\ast^2$ to be evaluated at the same time.
Since $\tensor{\Pi}{^A_I} \tensor{\Pi}{^B_J} \tensor{G}{^I^J}
= \tensor{G}{^A^B}$ this means we can take all terms in
Eqs.~\eqref{eq:field-field-spectral-ic}--\eqref{eq:momenta-momenta-spectral-ic}
to be evaluated at the same time and drop
the parallel propagator factors.

\subsection{Tensor spectral index}
A similar analysis applies for the tensor spectral index. Each polarization
behaves like an exactly massless scalar field, up to normalization.
Therefore~\eqref{eq:field-field-spectral-ic}--\eqref{eq:momenta-momenta-spectral-ic}
apply, with $\tensor{M}{^I^J} = 0$.
We need only one scalar field,
and to account for the normalization of the tensor polarizations
we should take $\tensor{G}{_I_J} \rightarrow \Mp^2/2$,
$\tensor{G}{^I^J} \rightarrow 2/\Mp^2$.

We write the momentum for the tensor polarizations as
$\pi_s = \d \gamma_s / \d N$~\cite{Dias:2016rjq}.
Then it follows that
\begin{subequations}
\begin{align}
    \frac{\d}{\d \ln k}
    \langle \gamma_s \gamma_{s'} \rangle'
    & =
    - \frac{1}{\Mp^2} \frac{1}{ka^2}
    \delta_{ss'}
    \Bigg(
        1
        +
        \frac{3}{2} \frac{a^2 H^2}{k^2} (2 - \epsilon)
        + \cdots
    \Bigg) ,
    \\
    \frac{\d}{\d \ln k}
    \langle \gamma_s \pi_{s'} \rangle'
    & =
    \frac{1}{\Mp^2}
    \frac{1}{ka^2}
    \delta_{ss'}
    \Bigg(
        (1 - \epsilon)
        - \frac{45\epsilon}{4} \frac{a^4 H^4}{k^4}
        + \cdots
    \Bigg) ,
    \\
    \frac{\d}{\d \ln k}
    \langle \pi_s \pi_{s'} \rangle'
    & =
    \frac{1}{\Mp^2}
    \frac{k}{H^2 a^4}
    \delta_{ss'}
    \Bigg(
        1
        +
        \frac{3\epsilon}{2} \frac{a^2 H^2}{k^2}
        + \cdots
    \Bigg) ,
\end{align}
\end{subequations}
from which the necessary initial conditions can be extracted.

\newpage

\bibliographystyle{jcap}

\bibliography{references/cosmo_refs,references/string_refs}

\providecommand{\href}[2]{#2}\begingroup\raggedright\begin{thebibliography}{100}

\bibitem{Appelquist:1974tg}
T.~Appelquist and J.~Carazzone, \emph{{Infrared Singularities and Massive
  Fields}}, \href{https://doi.org/10.1103/PhysRevD.11.2856}{\emph{Phys. Rev. D}
  {\bfseries 11} (1975) 2856}.

\bibitem{Baumann:2014nda}
D.~Baumann and L.~McAllister, \emph{{Inflation and String Theory}}, Cambridge
  Monographs on Mathematical Physics. Cambridge University Press, 2015,
  \href{https://doi.org/10.1017/CBO9781316105733}{10.1017/CBO9781316105733},
  [\href{https://arxiv.org/abs/1404.2601}{{\ttfamily 1404.2601}}].

\bibitem{Degrande:2011ua}
C.~Degrande, C.~Duhr, B.~Fuks, D.~Grellscheid, O.~Mattelaer and T.~Reiter,
  \emph{{UFO -- The Universal FeynRules Output}},
  \href{https://doi.org/10.1016/j.cpc.2012.01.022}{\emph{Comput. Phys. Commun.}
  {\bfseries 183} (2012) 1201}
  [\href{https://arxiv.org/abs/1108.2040}{{\ttfamily 1108.2040}}].

\bibitem{Alloul:2013bka}
A.~Alloul, N.~D. Christensen, C.~Degrande, C.~Duhr and B.~Fuks,
  \emph{{FeynRules 2.0 -- A complete toolbox for tree-level phenomenology}},
  \href{https://doi.org/10.1016/j.cpc.2014.04.012}{\emph{Comput. Phys. Commun.}
  {\bfseries 185} (2014) 2250}
  [\href{https://arxiv.org/abs/1310.1921}{{\ttfamily 1310.1921}}].

\bibitem{Seery:2016lko}
D.~Seery, \emph{{CppTransport: a platform to automate calculation of
  inflationary correlation functions}},
  \href{https://arxiv.org/abs/1609.00380}{{\ttfamily 1609.00380}}.

\bibitem{Butchers:2018hds}
S.~Butchers and D.~Seery, \emph{{Numerical evaluation of inflationary 3-point
  functions on curved field space--with the transport method \& CppTransport}},
  \href{https://doi.org/10.1088/1475-7516/2018/07/031}{\emph{JCAP} {\bfseries
  1807} (2018) 031} [\href{https://arxiv.org/abs/1803.10563}{{\ttfamily
  1803.10563}}].

\bibitem{Mulryne:2016mzv}
D.~J. Mulryne and J.~W. Ronayne, \emph{{PyTransport: A Python package for the
  calculation of inflationary correlation functions}},
  \href{https://arxiv.org/abs/1609.00381}{{\ttfamily 1609.00381}}.

\bibitem{Ronayne:2017qzn}
J.~W. Ronayne and D.~J. Mulryne, \emph{{Numerically evaluating the bispectrum
  in curved field-space--with PyTransport 2.0}},
  \href{https://doi.org/10.1088/1475-7516/2018/01/023}{\emph{JCAP} {\bfseries
  1801} (2018) 023} [\href{https://arxiv.org/abs/1708.07130}{{\ttfamily
  1708.07130}}].

\bibitem{Dias:2012nf}
M.~Dias, J.~Frazer and A.~R. Liddle, \emph{{Multifield consequences for D-brane
  inflation}}, \href{https://doi.org/10.1088/1475-7516/2013/03/E01,
  10.1088/1475-7516/2012/06/020}{\emph{JCAP} {\bfseries 1206} (2012) 020}
  [\href{https://arxiv.org/abs/1203.3792}{{\ttfamily 1203.3792}}].

\bibitem{Mulryne:2009kh}
D.~J. Mulryne, D.~Seery and D.~Wesley, \emph{{Moment transport equations for
  non-Gaussianity}},
  \href{https://doi.org/10.1088/1475-7516/2010/01/024}{\emph{JCAP} {\bfseries
  1001} (2010) 024} [\href{https://arxiv.org/abs/0909.2256}{{\ttfamily
  0909.2256}}].

\bibitem{Mulryne:2010rp}
D.~J. Mulryne, D.~Seery and D.~Wesley, \emph{{Moment transport equations for
  the primordial curvature perturbation}},
  \href{https://doi.org/10.1088/1475-7516/2011/04/030}{\emph{JCAP} {\bfseries
  1104} (2011) 030} [\href{https://arxiv.org/abs/1008.3159}{{\ttfamily
  1008.3159}}].

\bibitem{Dias:2011xy}
M.~Dias and D.~Seery, \emph{{Transport equations for the inflationary spectral
  index}}, \href{https://doi.org/10.1103/PhysRevD.85.043519}{\emph{Phys. Rev.
  D} {\bfseries 85} (2012) 043519}
  [\href{https://arxiv.org/abs/1111.6544}{{\ttfamily 1111.6544}}].

\bibitem{Anderson:2012em}
G.~J. Anderson, D.~J. Mulryne and D.~Seery, \emph{{Transport equations for the
  inflationary trispectrum}},
  \href{https://doi.org/10.1088/1475-7516/2012/10/019}{\emph{JCAP} {\bfseries
  10} (2012) 019} [\href{https://arxiv.org/abs/1205.0024}{{\ttfamily
  1205.0024}}].

\bibitem{Dias:2015rca}
M.~Dias, J.~Frazer and D.~Seery, \emph{{Computing observables in curved
  multifield models of inflation---A guide (with code) to the transport
  method}}, \href{https://doi.org/10.1088/1475-7516/2015/12/030}{\emph{JCAP}
  {\bfseries 1512} (2015) 030}
  [\href{https://arxiv.org/abs/1502.03125}{{\ttfamily 1502.03125}}].

\bibitem{Dias:2016rjq}
M.~Dias, J.~Frazer, D.~J. Mulryne and D.~Seery, \emph{{Numerical evaluation of
  the bispectrum in multiple field inflation---the transport approach with
  code}}, \href{https://doi.org/10.1088/1475-7516/2016/12/033}{\emph{JCAP}
  {\bfseries 1612} (2016) 033}
  [\href{https://arxiv.org/abs/1609.00379}{{\ttfamily 1609.00379}}].

\bibitem{Weinberg:2005vy}
S.~Weinberg, \emph{{Quantum contributions to cosmological correlations}},
  \href{https://doi.org/10.1103/PhysRevD.72.043514}{\emph{Phys. Rev. D}
  {\bfseries 72} (2005) 043514}
  [\href{https://arxiv.org/abs/hep-th/0506236}{{\ttfamily hep-th/0506236}}].

\bibitem{Weinberg:2006ac}
S.~Weinberg, \emph{{Quantum contributions to cosmological correlations. II. Can
  these corrections become large?}},
  \href{https://doi.org/10.1103/PhysRevD.74.023508}{\emph{Phys. Rev. D}
  {\bfseries 74} (2006) 023508}
  [\href{https://arxiv.org/abs/hep-th/0605244}{{\ttfamily hep-th/0605244}}].

\bibitem{Seery:2010kh}
D.~Seery, \emph{{Infrared effects in inflationary correlation functions}},
  \href{https://doi.org/10.1088/0264-9381/27/12/124005}{\emph{Class. Quant.
  Grav.} {\bfseries 27} (2010) 124005}
  [\href{https://arxiv.org/abs/1005.1649}{{\ttfamily 1005.1649}}].

\bibitem{Seery:2008ms}
D.~Seery, \emph{{Magnetogenesis and the primordial non-gaussianity}},
  \href{https://doi.org/10.1088/1475-7516/2009/08/018}{\emph{JCAP} {\bfseries
  08} (2009) 018} [\href{https://arxiv.org/abs/0810.1617}{{\ttfamily
  0810.1617}}].

\bibitem{Flauger:2016idt}
R.~Flauger, M.~Mirbabayi, L.~Senatore and E.~Silverstein, \emph{{Productive
  Interactions: heavy particles and non-Gaussianity}},
  \href{https://doi.org/10.1088/1475-7516/2017/10/058}{\emph{JCAP} {\bfseries
  10} (2017) 058} [\href{https://arxiv.org/abs/1606.00513}{{\ttfamily
  1606.00513}}].

\bibitem{Seery:2009hs}
D.~Seery, \emph{{A parton picture of de Sitter space during slow-roll
  inflation}}, \href{https://doi.org/10.1088/1475-7516/2009/05/021}{\emph{JCAP}
  {\bfseries 05} (2009) 021} [\href{https://arxiv.org/abs/0903.2788}{{\ttfamily
  0903.2788}}].

\bibitem{Arkani-Hamed:2015bza}
N.~Arkani-Hamed and J.~Maldacena, \emph{{Cosmological Collider Physics}},
  \href{https://arxiv.org/abs/1503.08043}{{\ttfamily 1503.08043}}.

\bibitem{Elliston:2011dr}
J.~Elliston, D.~J. Mulryne, D.~Seery and R.~Tavakol, \emph{{Evolution of $\fNL$
  to the adiabatic limit}},
  \href{https://doi.org/10.1088/1475-7516/2011/11/005}{\emph{JCAP} {\bfseries
  1111} (2011) 005} [\href{https://arxiv.org/abs/1106.2153}{{\ttfamily
  1106.2153}}].

\bibitem{Seery:2012vj}
D.~Seery, D.~J. Mulryne, J.~Frazer and R.~H. Ribeiro, \emph{{Inflationary
  perturbation theory is geometrical optics in phase space}},
  \href{https://doi.org/10.1088/1475-7516/2012/09/010}{\emph{JCAP} {\bfseries
  1209} (2012) 010} [\href{https://arxiv.org/abs/1203.2635}{{\ttfamily
  1203.2635}}].

\bibitem{GarciaBellido:1995qq}
J.~Garcia-Bellido and D.~Wands, \emph{{Metric perturbations in two field
  inflation}}, \href{https://doi.org/10.1103/PhysRevD.53.5437}{\emph{Phys. Rev.
  D} {\bfseries 53} (1996) 5437}
  [\href{https://arxiv.org/abs/astro-ph/9511029}{{\ttfamily
  astro-ph/9511029}}].

\bibitem{Weinberg:2003sw}
S.~Weinberg, \emph{{Adiabatic modes in cosmology}},
  \href{https://doi.org/10.1103/PhysRevD.67.123504}{\emph{Phys. Rev. D}
  {\bfseries 67} (2003) 123504}
  [\href{https://arxiv.org/abs/astro-ph/0302326}{{\ttfamily
  astro-ph/0302326}}].

\bibitem{Weinberg:2004kf}
S.~Weinberg, \emph{{Must cosmological perturbations remain non-adiabatic after
  multi-field inflation?}},
  \href{https://doi.org/10.1103/PhysRevD.70.083522}{\emph{Phys. Rev. D}
  {\bfseries 70} (2004) 083522}
  [\href{https://arxiv.org/abs/astro-ph/0405397}{{\ttfamily
  astro-ph/0405397}}].

\bibitem{Weinberg:2004kr}
S.~Weinberg, \emph{{Can non-adiabatic perturbations arise after single-field
  inflation?}}, \href{https://doi.org/10.1103/PhysRevD.70.043541}{\emph{Phys.
  Rev. D} {\bfseries 70} (2004) 043541}
  [\href{https://arxiv.org/abs/astro-ph/0401313}{{\ttfamily
  astro-ph/0401313}}].

\bibitem{Rigopoulos:2003ak}
G.~I. Rigopoulos and E.~P.~S. Shellard, \emph{{The separate universe approach
  and the evolution of nonlinear superhorizon cosmological perturbations}},
  \href{https://doi.org/10.1103/PhysRevD.68.123518}{\emph{Phys. Rev.}
  {\bfseries D68} (2003) 123518}
  [\href{https://arxiv.org/abs/astro-ph/0306620}{{\ttfamily
  astro-ph/0306620}}].

\bibitem{Lyth:2004gb}
D.~H. Lyth, K.~A. Malik and M.~Sasaki, \emph{{A General proof of the
  conservation of the curvature perturbation}},
  \href{https://doi.org/10.1088/1475-7516/2005/05/004}{\emph{JCAP} {\bfseries
  05} (2005) 004} [\href{https://arxiv.org/abs/astro-ph/0411220}{{\ttfamily
  astro-ph/0411220}}].

\bibitem{Elliston:2011et}
J.~Elliston, D.~Mulryne, D.~Seery and R.~Tavakol, \emph{{Evolution of
  non-Gaussianity in multi-scalar field models}},
  \href{https://doi.org/10.1142/S0217751X11054280,
  10.1142/S2010194511001292}{\emph{Int. J. Mod. Phys.} {\bfseries A26} (2011)
  3821} [\href{https://arxiv.org/abs/1107.2270}{{\ttfamily 1107.2270}}].

\bibitem{Elliston:2014zea}
J.~Elliston, S.~Orani and D.~J. Mulryne, \emph{{General analytic predictions of
  two-field inflation and perturbative reheating}},
  \href{https://doi.org/10.1103/PhysRevD.89.103532}{\emph{Phys. Rev.}
  {\bfseries D89} (2014) 103532}
  [\href{https://arxiv.org/abs/1402.4800}{{\ttfamily 1402.4800}}].

\bibitem{Blas:2011rf}
D.~Blas, J.~Lesgourgues and T.~Tram, \emph{{The Cosmic Linear Anisotropy
  Solving System (CLASS) II: Approximation schemes}},
  \href{https://doi.org/10.1088/1475-7516/2011/07/034}{\emph{JCAP} {\bfseries
  1107} (2011) 034} [\href{https://arxiv.org/abs/1104.2933}{{\ttfamily
  1104.2933}}].

\bibitem{Zuntz:2014csq}
J.~Zuntz, M.~Paterno, E.~Jennings, D.~Rudd, A.~Manzotti, S.~Dodelson et~al.,
  \emph{{CosmoSIS: modular cosmological parameter estimation}},
  \href{https://doi.org/10.1016/j.ascom.2015.05.005}{\emph{Astron. Comput.}
  {\bfseries 12} (2015) 45} [\href{https://arxiv.org/abs/1409.3409}{{\ttfamily
  1409.3409}}].

\bibitem{Meerburg:2019qqi}
P.~D. Meerburg et~al., \emph{{Primordial Non-Gaussianity}},
  \href{https://arxiv.org/abs/1903.04409}{{\ttfamily 1903.04409}}.

\bibitem{Dvali:1998pa}
G.~R. Dvali and S.~H.~H. Tye, \emph{{Brane inflation}},
  \href{https://doi.org/10.1016/S0370-2693(99)00132-X}{\emph{Phys. Lett.}
  {\bfseries B450} (1999) 72}
  [\href{https://arxiv.org/abs/hep-ph/9812483}{{\ttfamily hep-ph/9812483}}].

\bibitem{Burgess:2001fx}
C.~P. Burgess, M.~Majumdar, D.~Nolte, F.~Quevedo, G.~Rajesh and R.-J. Zhang,
  \emph{{The Inflationary brane anti-brane universe}},
  \href{https://doi.org/10.1088/1126-6708/2001/07/047}{\emph{JHEP} {\bfseries
  07} (2001) 047} [\href{https://arxiv.org/abs/hep-th/0105204}{{\ttfamily
  hep-th/0105204}}].

\bibitem{Dvali:2001fw}
G.~R. Dvali, Q.~Shafi and S.~Solganik, \emph{{D-brane inflation}},  in
  \emph{{4th European Meeting From the Planck Scale to the Electroweak Scale
  (Planck 2001) La Londe les Maures, Toulon, France, May 11-16, 2001}}, 2001,
  \href{https://arxiv.org/abs/hep-th/0105203}{{\ttfamily hep-th/0105203}}.

\bibitem{Kachru:2003sx}
S.~Kachru, R.~Kallosh, A.~D. Linde, J.~M. Maldacena, L.~P. McAllister and S.~P.
  Trivedi, \emph{{Towards inflation in string theory}},
  \href{https://doi.org/10.1088/1475-7516/2003/10/013}{\emph{JCAP} {\bfseries
  0310} (2003) 013} [\href{https://arxiv.org/abs/hep-th/0308055}{{\ttfamily
  hep-th/0308055}}].

\bibitem{Kachru:2003aw}
S.~Kachru, R.~Kallosh, A.~D. Linde and S.~P. Trivedi, \emph{{De Sitter vacua in
  string theory}},
  \href{https://doi.org/10.1103/PhysRevD.68.046005}{\emph{Phys. Rev.}
  {\bfseries D68} (2003) 046005}
  [\href{https://arxiv.org/abs/hep-th/0301240}{{\ttfamily hep-th/0301240}}].

\bibitem{Randall:1999ee}
L.~Randall and R.~Sundrum, \emph{{A Large mass hierarchy from a small extra
  dimension}}, \href{https://doi.org/10.1103/PhysRevLett.83.3370}{\emph{Phys.
  Rev. Lett.} {\bfseries 83} (1999) 3370}
  [\href{https://arxiv.org/abs/hep-ph/9905221}{{\ttfamily hep-ph/9905221}}].

\bibitem{Randall:1999vf}
L.~Randall and R.~Sundrum, \emph{{An Alternative to compactification}},
  \href{https://doi.org/10.1103/PhysRevLett.83.4690}{\emph{Phys. Rev. Lett.}
  {\bfseries 83} (1999) 4690}
  [\href{https://arxiv.org/abs/hep-th/9906064}{{\ttfamily hep-th/9906064}}].

\bibitem{Baumann:2006th}
D.~Baumann, A.~Dymarsky, I.~R. Klebanov, J.~M. Maldacena, L.~P. McAllister and
  A.~Murugan, \emph{{On D3-brane Potentials in Compactifications with Fluxes
  and Wrapped D-branes}},
  \href{https://doi.org/10.1088/1126-6708/2006/11/031}{\emph{JHEP} {\bfseries
  11} (2006) 031} [\href{https://arxiv.org/abs/hep-th/0607050}{{\ttfamily
  hep-th/0607050}}].

\bibitem{Baumann:2007np}
D.~Baumann, A.~Dymarsky, I.~R. Klebanov, L.~McAllister and P.~J. Steinhardt,
  \emph{{A Delicate universe}},
  \href{https://doi.org/10.1103/PhysRevLett.99.141601}{\emph{Phys. Rev. Lett.}
  {\bfseries 99} (2007) 141601}
  [\href{https://arxiv.org/abs/0705.3837}{{\ttfamily 0705.3837}}].

\bibitem{Baumann:2007ah}
D.~Baumann, A.~Dymarsky, I.~R. Klebanov and L.~McAllister, \emph{{Towards an
  Explicit Model of D-brane Inflation}},
  \href{https://doi.org/10.1088/1475-7516/2008/01/024}{\emph{JCAP} {\bfseries
  0801} (2008) 024} [\href{https://arxiv.org/abs/0706.0360}{{\ttfamily
  0706.0360}}].

\bibitem{Baumann:2008kq}
D.~Baumann, A.~Dymarsky, S.~Kachru, I.~R. Klebanov and L.~McAllister,
  \emph{{Holographic Systematics of D-brane Inflation}},
  \href{https://doi.org/10.1088/1126-6708/2009/03/093}{\emph{JHEP} {\bfseries
  03} (2009) 093} [\href{https://arxiv.org/abs/0808.2811}{{\ttfamily
  0808.2811}}].

\bibitem{Baumann:2009qx}
D.~Baumann, A.~Dymarsky, S.~Kachru, I.~R. Klebanov and L.~McAllister,
  \emph{{Compactification Effects in D-brane Inflation}},
  \href{https://doi.org/10.1103/PhysRevLett.104.251602}{\emph{Phys. Rev. Lett.}
  {\bfseries 104} (2010) 251602}
  [\href{https://arxiv.org/abs/0912.4268}{{\ttfamily 0912.4268}}].

\bibitem{Baumann:2010sx}
D.~Baumann, A.~Dymarsky, S.~Kachru, I.~R. Klebanov and L.~McAllister,
  \emph{{D3-brane Potentials from Fluxes in AdS/CFT}},
  \href{https://doi.org/10.1007/JHEP06(2010)072}{\emph{JHEP} {\bfseries 06}
  (2010) 072} [\href{https://arxiv.org/abs/1001.5028}{{\ttfamily 1001.5028}}].

\bibitem{Gubser:1998vd}
S.~S. Gubser, \emph{{Einstein manifolds and conformal field theories}},
  \href{https://doi.org/10.1103/PhysRevD.59.025006}{\emph{Phys. Rev.}
  {\bfseries D59} (1999) 025006}
  [\href{https://arxiv.org/abs/hep-th/9807164}{{\ttfamily hep-th/9807164}}].

\bibitem{Ceresole:1999zs}
A.~Ceresole, G.~Dall'Agata, R.~D'Auria and S.~Ferrara, \emph{{Spectrum of type
  IIB supergravity on AdS(5) $\times$ $T^{1,1}$: Predictions on {$N=1$}
  SCFTs}}, \href{https://doi.org/10.1103/PhysRevD.61.066001}{\emph{Phys. Rev.}
  {\bfseries D61} (2000) 066001}
  [\href{https://arxiv.org/abs/hep-th/9905226}{{\ttfamily hep-th/9905226}}].

\bibitem{Ceresole:1999ht}
A.~Ceresole, G.~Dall'Agata and R.~D'Auria, \emph{{KK spectroscopy of type IIB
  supergravity on AdS(5) $\times$ $T^{1,1}$}},
  \href{https://doi.org/10.1088/1126-6708/1999/11/009}{\emph{JHEP} {\bfseries
  11} (1999) 009} [\href{https://arxiv.org/abs/hep-th/9907216}{{\ttfamily
  hep-th/9907216}}].

\bibitem{Agarwal:2011wm}
N.~Agarwal, R.~Bean, L.~McAllister and G.~Xu, \emph{{Universality in D-brane
  Inflation}}, \href{https://doi.org/10.1088/1475-7516/2011/09/002}{\emph{JCAP}
  {\bfseries 1109} (2011) 002}
  [\href{https://arxiv.org/abs/1103.2775}{{\ttfamily 1103.2775}}].

\bibitem{Mortonson:2010er}
M.~J. Mortonson, H.~V. Peiris and R.~Easther, \emph{{Bayesian Analysis of
  Inflation: Parameter Estimation for Single Field Models}},
  \href{https://doi.org/10.1103/PhysRevD.83.043505}{\emph{Phys. Rev. D}
  {\bfseries 83} (2011) 043505}
  [\href{https://arxiv.org/abs/1007.4205}{{\ttfamily 1007.4205}}].

\bibitem{Easther:2011yq}
R.~Easther and H.~V. Peiris, \emph{{Bayesian Analysis of Inflation II: Model
  Selection and Constraints on Reheating}},
  \href{https://doi.org/10.1103/PhysRevD.85.103533}{\emph{Phys. Rev. D}
  {\bfseries 85} (2012) 103533}
  [\href{https://arxiv.org/abs/1112.0326}{{\ttfamily 1112.0326}}].

\bibitem{Larson:2010gs}
D.~Larson et~al., \emph{{Seven-Year Wilkinson Microwave Anisotropy Probe (WMAP)
  Observations: Power Spectra and WMAP-Derived Parameters}},
  \href{https://doi.org/10.1088/0067-0049/192/2/16}{\emph{Astrophys. J. Suppl.}
  {\bfseries 192} (2011) 16} [\href{https://arxiv.org/abs/1001.4635}{{\ttfamily
  1001.4635}}].

\bibitem{Komatsu:2010fb}
{\scshape WMAP} collaboration, \emph{{Seven-Year Wilkinson Microwave Anisotropy
  Probe (WMAP) Observations: Cosmological Interpretation}},
  \href{https://doi.org/10.1088/0067-0049/192/2/18}{\emph{Astrophys. J. Suppl.}
  {\bfseries 192} (2011) 18} [\href{https://arxiv.org/abs/1001.4538}{{\ttfamily
  1001.4538}}].

\bibitem{Elliston:2012ab}
J.~Elliston, D.~Seery and R.~Tavakol, \emph{{The inflationary bispectrum with
  curved field-space}},
  \href{https://doi.org/10.1088/1475-7516/2012/11/060}{\emph{JCAP} {\bfseries
  1211} (2012) 060} [\href{https://arxiv.org/abs/1208.6011}{{\ttfamily
  1208.6011}}].

\bibitem{Mulryne:2013uka}
D.~J. Mulryne, \emph{{Transporting non-Gaussianity from sub to super-horizon
  scales}}, \href{https://doi.org/10.1088/1475-7516/2013/09/010}{\emph{JCAP}
  {\bfseries 1309} (2013) 010}
  [\href{https://arxiv.org/abs/1302.3842}{{\ttfamily 1302.3842}}].

\bibitem{McAllister:2012am}
L.~McAllister, S.~Renaux-Petel and G.~Xu, \emph{{A Statistical Approach to
  Multifield Inflation: Many-field Perturbations Beyond Slow Roll}},
  \href{https://doi.org/10.1088/1475-7516/2012/10/046}{\emph{JCAP} {\bfseries
  1210} (2012) 046} [\href{https://arxiv.org/abs/1207.0317}{{\ttfamily
  1207.0317}}].

\bibitem{Chen:2009we}
X.~Chen and Y.~Wang, \emph{{Large non-Gaussianities with Intermediate Shapes
  from Quasi-Single Field Inflation}},
  \href{https://doi.org/10.1103/PhysRevD.81.063511}{\emph{Phys. Rev. D}
  {\bfseries 81} (2010) 063511}
  [\href{https://arxiv.org/abs/0909.0496}{{\ttfamily 0909.0496}}].

\bibitem{Chen:2009zp}
X.~Chen and Y.~Wang, \emph{{Quasi-Single Field Inflation and
  Non-Gaussianities}},
  \href{https://doi.org/10.1088/1475-7516/2010/04/027}{\emph{JCAP} {\bfseries
  04} (2010) 027} [\href{https://arxiv.org/abs/0911.3380}{{\ttfamily
  0911.3380}}].

\bibitem{Gong:2013sma}
J.-O. Gong, S.~Pi and M.~Sasaki, \emph{{Equilateral non-Gaussianity from heavy
  fields}}, \href{https://doi.org/10.1088/1475-7516/2013/11/043}{\emph{JCAP}
  {\bfseries 11} (2013) 043} [\href{https://arxiv.org/abs/1306.3691}{{\ttfamily
  1306.3691}}].

\bibitem{Hertog:2015zwh}
T.~Hertog and O.~Janssen, \emph{{Sharp Predictions from Eternal Inflation
  Patches in D-brane Inflation}},
  \href{https://doi.org/10.1088/1475-7516/2017/04/011}{\emph{JCAP} {\bfseries
  04} (2017) 011} [\href{https://arxiv.org/abs/1512.02722}{{\ttfamily
  1512.02722}}].

\bibitem{Vilenkin:1994pv}
A.~Vilenkin, \emph{{Topological inflation}},
  \href{https://doi.org/10.1103/PhysRevLett.72.3137}{\emph{Phys. Rev. Lett.}
  {\bfseries 72} (1994) 3137}
  [\href{https://arxiv.org/abs/hep-th/9402085}{{\ttfamily hep-th/9402085}}].

\bibitem{Linde:1994wt}
A.~D. Linde and D.~A. Linde, \emph{{Topological defects as seeds for eternal
  inflation}}, \href{https://doi.org/10.1103/PhysRevD.50.2456}{\emph{Phys. Rev.
  D} {\bfseries 50} (1994) 2456}
  [\href{https://arxiv.org/abs/hep-th/9402115}{{\ttfamily hep-th/9402115}}].

\bibitem{Kim:2010ud}
S.~A. Kim, A.~R. Liddle and D.~Seery, \emph{{Non-gaussianity in axion Nflation
  models}}, \href{https://doi.org/10.1103/PhysRevLett.105.181302}{\emph{Phys.
  Rev. Lett.} {\bfseries 105} (2010) 181302}
  [\href{https://arxiv.org/abs/1005.4410}{{\ttfamily 1005.4410}}].

\bibitem{Baumann:2006cd}
D.~Baumann and L.~McAllister, \emph{{A Microscopic Limit on Gravitational Waves
  from D-brane Inflation}},
  \href{https://doi.org/10.1103/PhysRevD.75.123508}{\emph{Phys. Rev.}
  {\bfseries D75} (2007) 123508}
  [\href{https://arxiv.org/abs/hep-th/0610285}{{\ttfamily hep-th/0610285}}].

\bibitem{Klebanov:2000hb}
I.~R. Klebanov and M.~J. Strassler, \emph{{Supergravity and a confining gauge
  theory: Duality cascades and chi SB resolution of naked singularities}},
  \href{https://doi.org/10.1088/1126-6708/2000/08/052}{\emph{JHEP} {\bfseries
  08} (2000) 052} [\href{https://arxiv.org/abs/hep-th/0007191}{{\ttfamily
  hep-th/0007191}}].

\bibitem{Romans:1984an}
L.~Romans, \emph{{New Compactifications of Chiral $N=2 d=10$ Supergravity}},
  \href{https://doi.org/10.1016/0370-2693(85)90479-4}{\emph{Phys. Lett. B}
  {\bfseries 153} (1985) 392}.

\bibitem{Candelas:1989js}
P.~Candelas and X.~C. de~la Ossa, \emph{{Comments on Conifolds}},
  \href{https://doi.org/10.1016/0550-3213(90)90577-Z}{\emph{Nucl. Phys. B}
  {\bfseries 342} (1990) 246}.

\bibitem{Witten:1993yc}
E.~Witten, \emph{{Phases of N=2 theories in two-dimensions}},
  \href{https://doi.org/10.1016/0550-3213(93)90033-L}{\emph{AMS/IP Stud. Adv.
  Math.} {\bfseries 1} (1996) 143}
  [\href{https://arxiv.org/abs/hep-th/9301042}{{\ttfamily hep-th/9301042}}].

\bibitem{Klebanov:1998hh}
I.~R. Klebanov and E.~Witten, \emph{{Superconformal field theory on
  three-branes at a Calabi--Yau singularity}},
  \href{https://doi.org/10.1016/S0550-3213(98)00654-3}{\emph{Nucl. Phys. B}
  {\bfseries 536} (1998) 199}
  [\href{https://arxiv.org/abs/hep-th/9807080}{{\ttfamily hep-th/9807080}}].

\bibitem{Giddings:2001yu}
S.~B. Giddings, S.~Kachru and J.~Polchinski, \emph{Hierarchies from fluxes in
  string compactifications},
  \href{https://doi.org/10.1103/PhysRevD.66.106006}{\emph{Phys.Rev.D}
  {\bfseries 66} (2002) 106006}
  [\href{https://arxiv.org/abs/hep-th/0105097}{{\ttfamily hep-th/0105097}}].

\bibitem{Sasaki:1995aw}
M.~Sasaki and E.~D. Stewart, \emph{A general analytic formula for the spectral
  index of the density perturbations produced during inflation},
  \href{https://doi.org/10.1143/PTP.95.71}{\emph{Prog.Theor.Phys.} {\bfseries
  95} (1996) 71} [\href{https://arxiv.org/abs/astro-ph/9507001}{{\ttfamily
  astro-ph/9507001}}].

\bibitem{Nakamura:1996da}
T.~T. Nakamura and E.~D. Stewart, \emph{{The Spectrum of cosmological
  perturbations produced by a multicomponent inflaton to second order in the
  slow roll approximation}},
  \href{https://doi.org/10.1016/0370-2693(96)00594-1}{\emph{Phys. Lett. B}
  {\bfseries 381} (1996) 413}
  [\href{https://arxiv.org/abs/astro-ph/9604103}{{\ttfamily
  astro-ph/9604103}}].

\bibitem{GrootNibbelink:2001qt}
S.~Groot~Nibbelink and B.~van Tent, \emph{{Scalar perturbations during multiple
  field slow-roll inflation}},
  \href{https://doi.org/10.1088/0264-9381/19/4/302}{\emph{Class. Quant. Grav.}
  {\bfseries 19} (2002) 613}
  [\href{https://arxiv.org/abs/hep-ph/0107272}{{\ttfamily hep-ph/0107272}}].

\bibitem{Sun:2005ig}
C.-Y. Sun and D.-H. Zhang, \emph{{Non-gaussianity of general multiple-field
  inflationary models}},
  \href{https://doi.org/10.1142/S0218271806008887}{\emph{Int. J. Mod. Phys. D}
  {\bfseries 15} (2006) 1259}
  [\href{https://arxiv.org/abs/astro-ph/0510709}{{\ttfamily
  astro-ph/0510709}}].

\bibitem{Rigopoulos:2005xx}
G.~Rigopoulos, E.~Shellard and B.~van Tent, \emph{{Non-linear perturbations in
  multiple-field inflation}},
  \href{https://doi.org/10.1103/PhysRevD.73.083521}{\emph{Phys. Rev. D}
  {\bfseries 73} (2006) 083521}
  [\href{https://arxiv.org/abs/astro-ph/0504508}{{\ttfamily
  astro-ph/0504508}}].

\bibitem{Peterson:2011yt}
C.~M. Peterson and M.~Tegmark, \emph{{Testing multifield inflation: A geometric
  approach}}, \href{https://doi.org/10.1103/PhysRevD.87.103507}{\emph{Phys.
  Rev. D} {\bfseries 87} (2013) 103507}
  [\href{https://arxiv.org/abs/1111.0927}{{\ttfamily 1111.0927}}].

\bibitem{Kaiser:2012ak}
D.~I. Kaiser, E.~A. Mazenc and E.~I. Sfakianakis, \emph{{Primordial Bispectrum
  from Multifield Inflation with Nonminimal Couplings}},
  \href{https://doi.org/10.1103/PhysRevD.87.064004}{\emph{Phys. Rev. D}
  {\bfseries 87} (2013) 064004}
  [\href{https://arxiv.org/abs/1210.7487}{{\ttfamily 1210.7487}}].

\bibitem{Peter1927}
F.~Peter and H.~Weyl, \emph{Die {V}ollst\"{a}ndigkeit der primitiven
  darstellungen einer geschlossenen kontinuierlichen gruppe},
  \href{https://doi.org/10.1007/BF01447892}{\emph{Mathematische Annalen}
  {\bfseries 97} (1927) 737}.

\bibitem{Salam:1981xd}
A.~Salam and J.~Strathdee, \emph{{On Kaluza-Klein Theory}},
  \href{https://doi.org/10.1016/0003-4916(82)90291-3}{\emph{Annals Phys.}
  {\bfseries 141} (1982) 316}.

\bibitem{Ceresole:1999rq}
A.~Ceresole, G.~Dall'Agata, R.~D'Auria and S.~Ferrara, \emph{{Superconformal
  field theories from IIB spectroscopy on AdS(5) $\times$ $T^{1,1}$}},
  \href{https://doi.org/10.1088/0264-9381/17/5/311}{\emph{Class. Quant. Grav.}
  {\bfseries 17} (2000) 1017}
  [\href{https://arxiv.org/abs/hep-th/9910066}{{\ttfamily hep-th/9910066}}].

\bibitem{Gandhi:2011id}
S.~Gandhi, L.~McAllister and S.~Sjors, \emph{{A Toolkit for Perturbing Flux
  Compactifications}},
  \href{https://doi.org/10.1007/JHEP12(2011)053}{\emph{JHEP} {\bfseries 12}
  (2011) 053} [\href{https://arxiv.org/abs/1106.0002}{{\ttfamily 1106.0002}}].

\bibitem{McAllister:2016vzi}
L.~McAllister, P.~Schwaller, G.~Servant, J.~Stout and A.~Westphal,
  \emph{{Runaway Relaxion Monodromy}},
  \href{https://doi.org/10.1007/JHEP02(2018)124}{\emph{JHEP} {\bfseries 02}
  (2018) 124} [\href{https://arxiv.org/abs/1610.05320}{{\ttfamily
  1610.05320}}].

\bibitem{1959142}
\emph{{T}he {R}otation {G}roups},  in \emph{Group Theory}, E.~P. Wigner, ed.,
  vol.~5 of \emph{Pure and Applied Physics}, pp.~142--152, Elsevier, (1959),
  \href{https://doi.org/https://doi.org/10.1016/B978-0-12-750550-3.50019-7}{DOI}.

\bibitem{osti_4389568}
J.~Schwinger, \emph{On angular momentum}. 1, 1952,
  \href{https://doi.org/10.2172/4389568}{10.2172/4389568}.

\bibitem{Baumann:2011su}
D.~Baumann and D.~Green, \emph{{Equilateral Non-Gaussianity and New Physics on
  the Horizon}},
  \href{https://doi.org/10.1088/1475-7516/2011/09/014}{\emph{JCAP} {\bfseries
  09} (2011) 014} [\href{https://arxiv.org/abs/1102.5343}{{\ttfamily
  1102.5343}}].

\bibitem{Bauer:2001ig}
C.~Bauer and H.~S. Do, \emph{{One loop integrals with {XLOOPS--GiNaC}}},
  \href{https://doi.org/10.1016/S0010-4655(02)00158-3}{\emph{Comput. Phys.
  Commun.} {\bfseries 144} (2002) 154}
  [\href{https://arxiv.org/abs/hep-ph/0102231}{{\ttfamily hep-ph/0102231}}].

\bibitem{Easther:2013rva}
R.~Easther, J.~Frazer, H.~V. Peiris and L.~C. Price, \emph{{Simple predictions
  from multifield inflationary models}},
  \href{https://doi.org/10.1103/PhysRevLett.112.161302}{\emph{Phys. Rev. Lett.}
  {\bfseries 112} (2014) 161302}
  [\href{https://arxiv.org/abs/1312.4035}{{\ttfamily 1312.4035}}].

\bibitem{Liddle:2003as}
A.~R. Liddle and S.~M. Leach, \emph{{How long before the end of inflation were
  observable perturbations produced?}},
  \href{https://doi.org/10.1103/PhysRevD.68.103503}{\emph{Phys. Rev. D}
  {\bfseries 68} (2003) 103503}
  [\href{https://arxiv.org/abs/astro-ph/0305263}{{\ttfamily
  astro-ph/0305263}}].

\bibitem{Adshead:2008vn}
P.~Adshead and R.~Easther, \emph{{Constraining Inflation}},
  \href{https://doi.org/10.1088/1475-7516/2008/10/047}{\emph{JCAP} {\bfseries
  10} (2008) 047} [\href{https://arxiv.org/abs/0802.3898}{{\ttfamily
  0802.3898}}].

\bibitem{Adshead:2010mc}
P.~Adshead, R.~Easther, J.~Pritchard and A.~Loeb, \emph{{Inflation and the
  Scale Dependent Spectral Index: Prospects and Strategies}},
  \href{https://doi.org/10.1088/1475-7516/2011/02/021}{\emph{JCAP} {\bfseries
  02} (2011) 021} [\href{https://arxiv.org/abs/1007.3748}{{\ttfamily
  1007.3748}}].

\bibitem{Fergusson:2006pr}
J.~R. Fergusson and E.~P.~S. Shellard, \emph{{Primordial non-Gaussianity and
  the CMB bispectrum}},
  \href{https://doi.org/10.1103/PhysRevD.76.083523}{\emph{Phys. Rev.}
  {\bfseries D76} (2007) 083523}
  [\href{https://arxiv.org/abs/astro-ph/0612713}{{\ttfamily
  astro-ph/0612713}}].

\bibitem{Rigopoulos:2004ba}
G.~Rigopoulos, E.~Shellard and B.~van Tent, \emph{{A Simple route to
  non-Gaussianity in inflation}},
  \href{https://doi.org/10.1103/PhysRevD.72.083507}{\emph{Phys. Rev. D}
  {\bfseries 72} (2005) 083507}
  [\href{https://arxiv.org/abs/astro-ph/0410486}{{\ttfamily
  astro-ph/0410486}}].

\bibitem{Aghanim:2015xee}
{\scshape Planck} collaboration, \emph{{Planck 2015 results. XI. CMB power
  spectra, likelihoods, and robustness of parameters}},
  \href{https://doi.org/10.1051/0004-6361/201526926}{\emph{Astron. Astrophys.}
  {\bfseries 594} (2016) A11}
  [\href{https://arxiv.org/abs/1507.02704}{{\ttfamily 1507.02704}}].

\bibitem{Ade:2015xua}
{\scshape Planck} collaboration, \emph{{Planck 2015 results. XIII. Cosmological
  parameters}},
  \href{https://doi.org/10.1051/0004-6361/201525830}{\emph{Astron. Astrophys.}
  {\bfseries 594} (2016) A13}
  [\href{https://arxiv.org/abs/1502.01589}{{\ttfamily 1502.01589}}].

\bibitem{Kinney:2005vj}
W.~H. Kinney, \emph{{Horizon crossing and inflation with large eta}},
  \href{https://doi.org/10.1103/PhysRevD.72.023515}{\emph{Phys. Rev. D}
  {\bfseries 72} (2005) 023515}
  [\href{https://arxiv.org/abs/gr-qc/0503017}{{\ttfamily gr-qc/0503017}}].

\bibitem{Martin:2012pe}
J.~Martin, H.~Motohashi and T.~Suyama, \emph{{Ultra Slow-Roll Inflation and the
  non-Gaussianity Consistency Relation}},
  \href{https://doi.org/10.1103/PhysRevD.87.023514}{\emph{Phys. Rev. D}
  {\bfseries 87} (2013) 023514}
  [\href{https://arxiv.org/abs/1211.0083}{{\ttfamily 1211.0083}}].

\bibitem{Namjoo:2012aa}
M.~H. Namjoo, H.~Firouzjahi and M.~Sasaki, \emph{{Violation of non-Gaussianity
  consistency relation in a single field inflationary model}},
  \href{https://doi.org/10.1209/0295-5075/101/39001}{\emph{EPL} {\bfseries 101}
  (2013) 39001} [\href{https://arxiv.org/abs/1210.3692}{{\ttfamily
  1210.3692}}].

\bibitem{Mooij:2015yka}
S.~Mooij and G.~A. Palma, \emph{{Consistently violating the non-Gaussian
  consistency relation}},
  \href{https://doi.org/10.1088/1475-7516/2015/11/025}{\emph{JCAP} {\bfseries
  11} (2015) 025} [\href{https://arxiv.org/abs/1502.03458}{{\ttfamily
  1502.03458}}].

\bibitem{Romano:2016gop}
A.~E. Romano, S.~Mooij and M.~Sasaki, \emph{{Global adiabaticity and
  non-Gaussianity consistency condition}},
  \href{https://doi.org/10.1016/j.physletb.2016.08.025}{\emph{Phys. Lett. B}
  {\bfseries 761} (2016) 119}
  [\href{https://arxiv.org/abs/1606.04906}{{\ttfamily 1606.04906}}].

\bibitem{Chen:2013aj}
X.~Chen, H.~Firouzjahi, M.~H. Namjoo and M.~Sasaki, \emph{{A Single Field
  Inflation Model with Large Local Non-Gaussianity}},
  \href{https://doi.org/10.1209/0295-5075/102/59001}{\emph{EPL} {\bfseries 102}
  (2013) 59001} [\href{https://arxiv.org/abs/1301.5699}{{\ttfamily
  1301.5699}}].

\bibitem{Chen:2013eea}
X.~Chen, H.~Firouzjahi, E.~Komatsu, M.~H. Namjoo and M.~Sasaki, \emph{{In-in
  and $\delta N$ calculations of the bispectrum from non-attractor single-field
  inflation}}, \href{https://doi.org/10.1088/1475-7516/2013/12/039}{\emph{JCAP}
  {\bfseries 12} (2013) 039} [\href{https://arxiv.org/abs/1308.5341}{{\ttfamily
  1308.5341}}].

\bibitem{Cai:2017bxr}
Y.-F. Cai, X.~Chen, M.~H. Namjoo, M.~Sasaki, D.-G. Wang and Z.~Wang,
  \emph{{Revisiting non-Gaussianity from non-attractor inflation models}},
  \href{https://doi.org/10.1088/1475-7516/2018/05/012}{\emph{JCAP} {\bfseries
  05} (2018) 012} [\href{https://arxiv.org/abs/1712.09998}{{\ttfamily
  1712.09998}}].

\bibitem{Dias:2014msa}
M.~Dias, J.~Elliston, J.~Frazer, D.~Mulryne and D.~Seery, \emph{{The curvature
  perturbation at second order}},
  \href{https://doi.org/10.1088/1475-7516/2015/02/040}{\emph{JCAP} {\bfseries
  02} (2015) 040} [\href{https://arxiv.org/abs/1410.3491}{{\ttfamily
  1410.3491}}].

\bibitem{Wands:2000dp}
D.~Wands, K.~A. Malik, D.~H. Lyth and A.~R. Liddle, \emph{{A New approach to
  the evolution of cosmological perturbations on large scales}},
  \href{https://doi.org/10.1103/PhysRevD.62.043527}{\emph{Phys. Rev. D}
  {\bfseries 62} (2000) 043527}
  [\href{https://arxiv.org/abs/astro-ph/0003278}{{\ttfamily
  astro-ph/0003278}}].

\bibitem{Meyers:2013gua}
J.~Meyers and E.~R.~M. Tarrant, \emph{{Perturbative Reheating After
  Multiple-Field Inflation: The Impact on Primordial Observables}},
  \href{https://doi.org/10.1103/PhysRevD.89.063535}{\emph{Phys. Rev. D}
  {\bfseries 89} (2014) 063535}
  [\href{https://arxiv.org/abs/1311.3972}{{\ttfamily 1311.3972}}].

\bibitem{Hotinli:2017vhx}
S.~C. Hotinli, J.~Frazer, A.~H. Jaffe, J.~Meyers, L.~C. Price and E.~R.~M.
  Tarrant, \emph{{Effect of reheating on predictions following multiple-field
  inflation}}, \href{https://doi.org/10.1103/PhysRevD.97.023511}{\emph{Phys.
  Rev. D} {\bfseries 97} (2018) 023511}
  [\href{https://arxiv.org/abs/1710.08913}{{\ttfamily 1710.08913}}].

\bibitem{Mukhanov:1985rz}
V.~F. Mukhanov, \emph{{Gravitational Instability of the Universe Filled with a
  Scalar Field}}, {\emph{JETP Lett.} {\bfseries 41} (1985) 493}.

\bibitem{Sasaki:1986hm}
M.~Sasaki, \emph{{Large Scale Quantum Fluctuations in the Inflationary
  Universe}}, \href{https://doi.org/10.1143/PTP.76.1036}{\emph{Prog. Theor.
  Phys.} {\bfseries 76} (1986) 1036}.

\bibitem{GarciaBellido:1996qt}
J.~Garcia-Bellido, A.~D. Linde and D.~Wands, \emph{{Density perturbations and
  black hole formation in hybrid inflation}},
  \href{https://doi.org/10.1103/PhysRevD.54.6040}{\emph{Phys. Rev. D}
  {\bfseries 54} (1996) 6040}
  [\href{https://arxiv.org/abs/astro-ph/9605094}{{\ttfamily
  astro-ph/9605094}}].

\bibitem{Maldacena:2002vr}
J.~M. Maldacena, \emph{{Non-Gaussian features of primordial fluctuations in
  single field inflationary models}},
  \href{https://doi.org/10.1088/1126-6708/2003/05/013}{\emph{JHEP} {\bfseries
  05} (2003) 013} [\href{https://arxiv.org/abs/astro-ph/0210603}{{\ttfamily
  astro-ph/0210603}}].

\bibitem{Senatore:2009cf}
L.~Senatore and M.~Zaldarriaga, \emph{{On Loops in Inflation}},
  \href{https://doi.org/10.1007/JHEP12(2010)008}{\emph{JHEP} {\bfseries 12}
  (2010) 008} [\href{https://arxiv.org/abs/0912.2734}{{\ttfamily 0912.2734}}].

\bibitem{Assassi:2012et}
V.~Assassi, D.~Baumann and D.~Green, \emph{{Symmetries and Loops in
  Inflation}}, \href{https://doi.org/10.1007/JHEP02(2013)151}{\emph{JHEP}
  {\bfseries 02} (2013) 151} [\href{https://arxiv.org/abs/1210.7792}{{\ttfamily
  1210.7792}}].

\bibitem{Frazer:2013zoa}
J.~Frazer, \emph{{Predictions in multifield models of inflation}},
  \href{https://doi.org/10.1088/1475-7516/2014/01/028}{\emph{JCAP} {\bfseries
  01} (2014) 028} [\href{https://arxiv.org/abs/1303.3611}{{\ttfamily
  1303.3611}}].

\bibitem{Dias:2012qy}
M.~Dias, R.~H. Ribeiro and D.~Seery, \emph{{The \ensuremath{\delta}N formula is
  the dynamical renormalization group}},
  \href{https://doi.org/10.1088/1475-7516/2013/10/062}{\emph{JCAP} {\bfseries
  10} (2013) 062} [\href{https://arxiv.org/abs/1210.7800}{{\ttfamily
  1210.7800}}].

\bibitem{Copeland:1993zn}
E.~J. Copeland, E.~W. Kolb, A.~R. Liddle and J.~E. Lidsey,
  \emph{{Reconstructing the inflaton potential: Perturbative reconstruction to
  second order}}, \href{https://doi.org/10.1103/PhysRevD.49.1840}{\emph{Phys.
  Rev. D} {\bfseries 49} (1994) 1840}
  [\href{https://arxiv.org/abs/astro-ph/9308044}{{\ttfamily
  astro-ph/9308044}}].

\bibitem{Copeland:1993jj}
E.~J. Copeland, E.~W. Kolb, A.~R. Liddle and J.~E. Lidsey,
  \emph{{Reconstructing the inflation potential, in principle and in
  practice}}, \href{https://doi.org/10.1103/PhysRevD.48.2529}{\emph{Phys. Rev.
  D} {\bfseries 48} (1993) 2529}
  [\href{https://arxiv.org/abs/hep-ph/9303288}{{\ttfamily hep-ph/9303288}}].

\bibitem{Babich:2004gb}
D.~Babich, P.~Creminelli and M.~Zaldarriaga, \emph{{The Shape of
  non-Gaussianities}},
  \href{https://doi.org/10.1088/1475-7516/2004/08/009}{\emph{JCAP} {\bfseries
  0408} (2004) 009} [\href{https://arxiv.org/abs/astro-ph/0405356}{{\ttfamily
  astro-ph/0405356}}].

\bibitem{Burrage:2011hd}
C.~Burrage, R.~H. Ribeiro and D.~Seery, \emph{{Large slow-roll corrections to
  the bispectrum of noncanonical inflation}},
  \href{https://doi.org/10.1088/1475-7516/2011/07/032}{\emph{JCAP} {\bfseries
  07} (2011) 032} [\href{https://arxiv.org/abs/1103.4126}{{\ttfamily
  1103.4126}}].

\bibitem{Tolley:2009fg}
A.~J. Tolley and M.~Wyman, \emph{{The Gelaton Scenario: Equilateral
  non-Gaussianity from multi-field dynamics}},
  \href{https://doi.org/10.1103/PhysRevD.81.043502}{\emph{Phys. Rev. D}
  {\bfseries 81} (2010) 043502}
  [\href{https://arxiv.org/abs/0910.1853}{{\ttfamily 0910.1853}}].

\bibitem{Byrnes:2015dub}
C.~T. Byrnes, D.~Regan, D.~Seery and E.~R.~M. Tarrant, \emph{{The hemispherical
  asymmetry from a scale-dependent inflationary bispectrum}},
  \href{https://doi.org/10.1088/1475-7516/2016/06/025}{\emph{JCAP} {\bfseries
  06} (2016) 025} [\href{https://arxiv.org/abs/1511.03129}{{\ttfamily
  1511.03129}}].

\bibitem{Kenton:2016abp}
Z.~Kenton and D.~J. Mulryne, \emph{{The Separate Universe Approach to Soft
  Limits}}, \href{https://doi.org/10.1088/1475-7516/2016/10/035}{\emph{JCAP}
  {\bfseries 10} (2016) 035}
  [\href{https://arxiv.org/abs/1605.03435}{{\ttfamily 1605.03435}}].

\bibitem{Alabidi:2010ba}
L.~Alabidi, K.~Malik, C.~T. Byrnes and K.-Y. Choi, \emph{{How the curvaton
  scenario, modulated reheating and an inhomogeneous end of inflation are
  related}}, \href{https://doi.org/10.1088/1475-7516/2010/11/037}{\emph{JCAP}
  {\bfseries 11} (2010) 037} [\href{https://arxiv.org/abs/1002.1700}{{\ttfamily
  1002.1700}}].

\bibitem{Alvarez:2014vva}
M.~Alvarez et~al., \emph{{Testing Inflation with Large Scale Structure:
  Connecting Hopes with Reality}},
  \href{https://arxiv.org/abs/1412.4671}{{\ttfamily 1412.4671}}.

\bibitem{Torrado:2020dgo}
J.~Torrado and A.~Lewis, \emph{{Cobaya: Code for Bayesian Analysis of
  hierarchical physical models}},
  \href{https://arxiv.org/abs/2005.05290}{{\ttfamily 2005.05290}}.

\end{thebibliography}\endgroup

\end{document}